\documentclass[11pt]{article}

\usepackage{enumitem}
\usepackage{ytableau}
\usepackage{lipsum}
\DeclareFontFamily{U}{mathb}{\hyphenchar\font90}
\DeclareFontShape{U}{mathb}{m}{n}{
      <50> <60> <70> <80> <90> <100> gen * mathb
      <109.95> mathb10 <21> <140.4> <170.28> <200.74> <24.88> mathb12
      }{}
\DeclareSymbolFont{mathb}{U}{mathb}{m}{n}
\DeclareMathSymbol{\nabla}{7}{mathb}{"99}
\usepackage{pgf,tikz,pgfplots}
\pgfplotsset{compat=1.15}
\usetikzlibrary{arrows}
\usetikzlibrary{shapes.multipart,chains,scopes}
\usepackage{url}
\usepackage{verbatim}
\usepackage{comment,cite,comment}
\usetikzlibrary {decorations.pathmorphing}
\usetikzlibrary{decorations.pathreplacing}
\usepackage{tkz-euclide}
\usepackage{multirow}
\usepackage{phaistos}
\parskip=\medskipamount

\usepackage[margin=1in]{geometry}

\csname @addtoreset\endcsname{equation}{section}

\usepackage{hyperref}
\hypersetup{
	unicode,	% use with \texorpdfstring
	colorlinks,
	citecolor=magenta,% filecolor=black,%
	linkcolor=magenta,
	urlcolor=magenta,% pdftex
	bookmarksopen=true,
	bookmarksnumbered
	}
\usepackage{empheq}

\usepackage[T1]{fontenc}
\usepackage[utf8]{inputenc}
\usepackage{fdsymbol}
\usepackage{charter}
\usepackage{lettrine}
\usepackage{ebgaramond}
\usepackage{musicography}

\usepackage{blindtext}

\usepackage{subfiles}

\def\d{{{\rm d}}}

\def\1{1\hspace{-4pt}1}
\def\j1{\widetilde{1\hspace{-4pt}1}}

\def\bec{\begin{center}}
\def\ec{\end{center}}

\def\C{\Gamma}

\def\m{\mu}

\def\t{\tau}

\def\cL{{\cal L}}

\def\cA{{\cal A}}

\def\cR{{\cal R}}
\def\cS{{\cal S}}

\def\cH{{\cal H}}
\def\cQ{{\cal Q}}
\def\cE{{\cal E}}

\def\V{{\mathtt{V}}}
\def\C{{\mathtt{C}}}
\def\tmu{{\tilde{\mu}}}
\def\d{{\rm d}}

\usepackage{caption}
\usepackage{subcaption}

\def\be{\begin{equation}}
\def\ee{\end{equation}}
\def\bea{\begin{eqnarray}}
\def\eea{\end{eqnarray}}
\def\ba{\begin{array}}
\def\ea{\end{array}}

\definecolor{rougef}{rgb}{0.7,0,0}
\definecolor{vertf}{rgb}{0,0.6,0}
\definecolor{bleuf}{rgb}{0,0,0.9}

\def\bes{\begin{eqnarray}\begin{split}}
\def\ees{\end{split}\end{eqnarray}}

\newcommand{\bse}{\begin{subequations}}
\newcommand{\ese}{\end{subequations}}

 \usepackage{cleveref}
\crefformat{section}{\S#2#1#3}
\crefformat{subsection}{\S#2#1#3}
\crefformat{subsubsection}{\S#2#1#3}
\crefrangeformat{section}{\S\S#3#1#4 to~#5#2#6}
\crefmultiformat{section}{\S\S#2#1#3}{ and~#2#1#3}{, #2#1#3}{ and~#2#1#3}

\usepackage{pgfornament}
\usepackage{ctable}
\title{{\huge Thesis Project}}
\author{Felipe Diaz}
\date{\today}
\usepackage{mathtools}
\begin{document}
\begin{titlepage}
\setcounter{page}{1}
\begin{center}
\begin{figure}[t]
    \centering\includegraphics[width=0.2\textwidth]{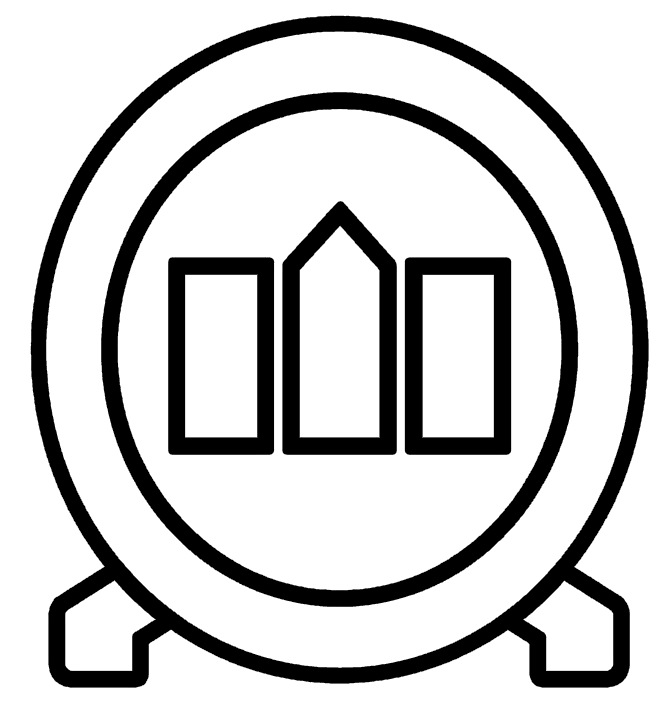}
\end{figure}
{\textsc \Huge Universidad Andr\'es Bello \\ Facultad de Ciencias Exactas}
\vskip 7pt
\rule{158mm}{0.9mm}
\vskip 7pt
%~~~~~~~~~~~~~~~~~~~~~~~~~~~~~~~~~~~
{\Huge Probing the Holographic Universe: Aspects of Entanglement Entropy Modifications}
%~~~~~~~~~~~~~~~~~~~~~~~~~~~~~~~~~~~
\vskip 0pt
\rule{158mm}{0.9mm}
\vskip 15pt
\pgfornament[width=8cm]{87}
\vskip 10pt
{\large {\underline{\it Author}}: Felipe Diaz  \\ {\underline{\it Supervisor}}: Alberto Faraggi${}^{\natural}$ \\ {\underline{\it Co-supervisor:}~Adolfo Cisterna${}^{\flat, \flat\kern-1.4pt\flat}$}}

\vskip 10pt
\underline{\Large Thesis Commission}:
\\
\vskip 10pt
{\large{Olivera Miskovic${}^{\musDoubleSharp}$\\ Rodrigo Aros${}^{\natural}$ \\Rodrigo Olea${}^{\musDoubleSharp}$ \\ Brenno Vallilo${}^{\natural}$}} 
\vskip 4pt 

{\Large{August, 2024}}
\vskip 4 pt
{\small{A thesis presented to the Universidad Andr\'es Bello in fulfillment of the thesis \\ requirement for the degree of Doctor of Philosophy in Physic}}
\vskip 10pt
{\small \it
${}^\natural$ Departamento de Ciencias F\'isicas, Universidad Andr\'es Bello, Sazi\'e 2212, Santiago, Chile \\
${}^\sharp$ Instituto de F \'isica, Pontificia Universidad Cat\'olica de Chile, Av. Vicu\~{n}a Mackenna 4860, Santiago, Chile \\
${}^\flat$ Sede Esmeralda, Universidad de Tarapac\'a, Av. Luis Emilio Recabarren 2477, Iquique, Chile \\ 
${}^{\flat\kern-1.4pt\flat}$ Institute of Theoretical Physics, Charles University, V
Hole\v{s}ovi\v{c}k\'ach 2, 180 00 Prague 8, Czech Republic \\
${}^{\musDoubleSharp}$ Instituto de F\'isica, Pontificia Universidad Cat\'olica de Valpara\'iso, Avda. Universidad 330, Curauma, Valpara\'iso, Chile}
\vskip 10pt
%\rule{170mm}{0.1mm}
{\footnotesize ${}^a$\href{mailto:f.diazmartinez@uandresbello.edu}{\texttt{f.diazmartinez@uandresbello.edu}}}
\end{center}
\end{titlepage}
\newpage
\newpage
\vspace*{\fill}
\begin{center}{\textbf{\Large Probing the Holographic Universe: \\ Aspects of
Entanglement Entropy Modifications}}

\vspace{10pt}
\emph{by} Felipe Diaz
\vspace{20pt}
%%%%%%%%%%%%%%%%%%%%%%%%%%%

%%%%%%%%%%%%%%%%%%%%%%%%%%%
\textbf{\large Abstract}
%%%%%%%%%%%%%%%%%%%%%%%%%%%
%%%%%%%%%%%%%%%%%%%%%%%%%%%

\end{center}

Entanglement entropy stands as a cornerstone in understanding the intricate interplay between quantum mechanics and information theory. In this thesis we study the influence of energy fluctuations and acceleration on entanglement entropy within the framework of Quantum Gravity. We explore three distinct scenarios wherein these factors provide insights into the underlying quantum theory. 
First, we examine the entanglement among disjoint Rindler observers within a de Sitter spacetime. Previous studies have revealed that the theory exhibits a flat entanglement spectrum. By employing a conformal description of the cosmological horizon, we introduce 1-loop corrections to the partition function of the conformal field theory which disrupt the flat entanglement spectrum, resulting in the emergence of an area law for the entanglement entropy. Remarkably, this transformation preserves the finiteness of the dimensionality of the Hilbert space previously discovered in the literature.
Next, we progress to explore the holographic entanglement entropy of field theories featuring a global $U(1)$ symmetry. We first extend the holographic prescription to include non-linearly charged theories, and study bulk thermal and quantum fluctuations. We find that the the dual theory undergoes spontaneous symmetry breaking, accompanied by a decrease in the leading-order coefficient of the $\langle TT\rangle$ correlator due to quantum corrections.  As an explicit example, we consider conformal electrodynamics coupled to three-dimensional gravity. We show how a black hole solution within this framework enables the study of two-dimensional free bosons.
Finally, we study three-dimensional hairy accelerating black holes extracting several holographic quantities. Remarkably, our analysis reveals that these solutions can be mapped to marginally multi-trace deformed thermal two-dimensional conformal field theories existing within a curved background. 
In contrast to BTZ black hole, we observe that as acceleration increases, the accessible region of the conformal boundary decreases, leading to a reduction in the entanglement entropy. This suggests a loss of information in the dual theory due to acceleration.
\vspace*{\fill}
\begin{center}
\pgfornament[width=1cm]{9}
\end{center}
\vspace*{\fill}

\newpage
\vspace*{\fill}
\paragraph{List of publications.} The thesis is based on the following publications
\begin{itemize}
    \item[\cite{Arenas-Henriquez:2022pyh}] {\sc G.~A}renas--Henriquez, {\sc F.~D}iaz, and {\sc P.~S}undell. ``\href{https://link.springer.com/article/10.1007/JHEP08(2022)261}{Logarithmic Corrections, Entanglement Entropy, and UV Cutoffs in de Sitter Spacetime}"~, JHEP {\bf 08} (2022), 261. \href{https://arxiv.org/abs/2206.10427}{[hep-th/2206.10427]}~.
    \item[\cite{Arenas-Henriquez:2022ntz}] {\sc G.~A}renas--Henriquez, {\sc F.~D}iaz, and {\sc Y.~N}ovoa. ``\href{https://link.springer.com/article/10.1007/JHEP05(2023)072}{Thermal fluctuations of black holes with non-linear electrodynamics and charged Renyi entropy}"~, JHEP {\bf 05} (2023), 072. \href{https://arxiv.org/abs/2211.06355}{[hep-th/2211.06355]}~.
    \item[\cite{Arenas-Henriquez:2023hur}] {\sc A.~C}isterna, {\sc G.~A}renas--Henriquez, {\sc F.~D}iaz, and {\sc R.~G}regory. ``\href{https://link.springer.com/article/10.1007/JHEP09(2023)122}{Accelerating Black Holes in $2+1$ dimensions: Holography revisited}"~, JHEP {\bf 09} (2023), 122. \href{https://arxiv.org/abs/2308.00613}{[hep-th/2308.00613]}~.
    \item[\cite{Cisterna:2023qhh}] {\sc A.~C}isterna, {\sc F.~D}iaz, {\sc R.~B.~M}ann, and {\sc J.~O}liva. ``\href{https://link.springer.com/article/10.1007/JHEP11(2023)073}{Exploring accelerating hairy black holes in $2+1$ dimensions: the asymptotically locally anti-de Sitter class and its holography}"~, JHEP {\bf 11} (2023), 073. \href{https://arxiv.org/abs/2309.05559}{[hep-th/2309.05559]}~.
\end{itemize}
While the thesis was being prepared, the following collaborative works were also published by the author, results from which have not been included in this thesis:
\begin{itemize}   
    \item[\cite{Caceres:2023gfa}] {\sc N.~C}aceres, {\sc C.~C}orral, {\sc F.~D}iaz, and {\sc R.~O}lea. ``\href{https://link.springer.com/article/10.1007/JHEP05(2024)125}{Holographic renormalization of Horndeski gravity}"~, JHEP {\bf{05}} (2024), 125. \href{https://arxiv.org/abs/2311.04054}{[hep-th/2311.04054]}~.
    
    \item[\cite{diaz2024fractionalI}] {\sc F.~D}iaz, {\sc C.~I}azeolla, and {\sc P.~S}undell, ``\href{https://arxiv.org/abs/2403.0228}{Fractional Spins, Unfolding, and Holography: I. Parent field equations for dual higher-spin gravity reductions}"~, {[hep-th/2403.0228]}~.
    \item[\cite{diaz2024fractionalII}] {\sc F.~D}iaz, {\sc C.~I}azeolla, and {\sc P.~S}undell, ``\href{https://arxiv.org/abs/2403.02301}{Fractional Spins, Unfolding, and Holography: II. 4D Higher Spin Gravity and 3D Conformal Dual}"~, {[hep-th/2403.02301]}~.
\end{itemize}

\vspace*{\fill}

\begin{center}
\pgfornament[width=1cm]{78}
\end{center}
\newpage
\vspace*{\fill}
\centerline{\Huge \textit{\textbf{To my mother}}} 
\vspace*{\fill}
\thispagestyle{empty}
\newpage
\vspace*{\fill}
\begin{center}\textbf{\large Acknowledgments}\end{center}

\noindent First of all, I would like to thank to my thesis supervisor, Alberto Faraggi, and co-supervisor, Adolfo Cisterna, for their guidance and unwavering patience throughout these years. From both of them, I have learn not only insights into Physics and Mathematics, but also discovered the beauty in simple aspects of life. They consistently allowed this thesis to be my own work, but steered me in the right the direction whenever he thought I needed it.

 I would like to thank to all my collaborators and friends: Gabriel Arenas--Henriquez, Jos\'e Barrientos, Nicolas Caceres, Cristobal Corral, Ruth Gregory, Carlo Iazeolla, Robert B. Mann, Julio Oliva, Rodrigo Olea, and Yerko Novoa. Without them, the papers I have been part of would not have been published. To be democratic, I would like to also thank to the "Greek gang": Giorgos Anastasiou and Ignacio Araya, both have help me and inspire through the years.  
I would like to give special thanks to Gabriel, David, and Constanza Farias with whom I have shared this entire journey. I hope that our friendship and collaboration continue to flourish for many years to come.
I extend a special acknowledgment to my collaborator, Per Sundell, whose unwavering support and guidance have been invaluable in my understanding of Higher Spin Gravity, and is a continuous source of inspiration as always remind me of the inherent beauty of Physics. Whenever I encounter difficulties or have questions about any aspect of life, he consistently finds the right words to offer guidance.

 This thesis, or any of my achievements, would have never happened without the constant support and love of my family. There are not enough words how grateful I am to you. Along this lines; I would like to thank to the "Husares” for all the funny and emotional moments through the last 15 years. You have stuck by me through so many ups and downs and I want you to know that I will always be there for you. A part of this work is yours.

% Finally, I would like to express my deepest gratitude to Gabriela. Throughout this journey, your unwavering support, understanding, and endless love have been my guiding light. Your belief in me has fueled my determination, and your encouragement has lifted me up even in the darkest of times. Thanks.

\noindent This thesis was funded by Beca Doctorado nacional (ANID) 2021 Scholarship No. 21211335, ANID/ACT210100 Anillo Grant “Holography and its applications to High Energy Physics, Quantum Gravity and Condensed Matter Systems”, and FONDECYT Regular grant No. 1210500.

\textbf{}

\vspace*{\fill}
\begin{center}
\pgfornament[width=1cm]{19}
\end{center}
\newpage
\vspace*{\fill}
\hspace*{\fill}\textit{No man is an island, \\
\hspace*{\fill}entire of itself,\\
\hspace*{\fill} every man
is a piece of the Continent,\\ 
\hspace*{\fill}a part of the main;\\
\\ 
\hspace*{\fill}if a clod be washed away by the Sea,\\
\hspace*{\fill}Europe is the less, \\
\hspace*{\fill}as well as if a promontory were,\\
\hspace*{\fill}as
well as if a manor of thy friends \\
\hspace*{\fill} or of thine
own were; \\
\\
\hspace*{\fill}any mans death diminishes me, \\
\hspace*{\fill}because I am involved in Mankind;\\
\hspace*{\fill}and therefore never send to know for whom the bell tolls;\\
\hspace*{\fill}It tolls for thee.} \\ 
\vspace{3mm}

\hspace*{\fill} John Donne 

\vspace*{\fill}
\begin{center}
\pgfornament[width=1cm]{10}
\end{center}
\newpage 

\vspace*{\fill}
\begin{center}
This page was intentionally left blank
\end{center}
\vspace*{\fill}
\thispagestyle{empty}
\newpage
\vspace*{\fill}
{\small\tableofcontents }
\vspace*{\fill}
\begin{center}
\pgfornament[width=1cm]{77}
\end{center}
\newpage
%%%%%%%%%%%%%%%%%%%%%%%%%%%%%%%%%%%%%%%%%%%%%%%%%
%%%%%%%%%%%%%%%%%%% INTROD %%%%%%%%%%%%%%%%%%%%%%
%%%%%%%%%%%%%%%%%%%%%%%%%%%%%%%%%%%%%%%%%%%%%%%%%
\section{Introduction and Motivation}

\lettrine[lines=3, depth=1]{T}{~~he} notion of entanglement is a key distinguishing feature between classical and quantum physics. In quantum field theory (QFT), entanglement refers to a phenomenon where the quantum states become correlated in such a way that the state of one particle cannot be described independently of the state of another, even when they are separated by large distances. This phenomenon arises due to the inherent non-locality of quantum mechanics and the transmission of information is fundamentally different from classical communication. Entanglement plays a crucial role in understanding various quantum phenomena. In the context of QFT, entanglement between different regions of spacetime has been of particular interest, especially in the study of black holes and the holographic principle. A measure of quantum entanglement corresponds to the \emph{entanglement entropy}, computed as the von Neumann entropy of the density matrix of the entangled subsystem and measures the amount of entanglement between different parts of a quantum system and has demonstrated a surprisingly broad spectrum of applications across various domains of physics. Its significance spans quantum information theory, condensed matter physics, general relativity, and high-energy theory, see for instance \cite{Eisert:2008ur, Latorre:2009zz, Rangamani:2016dms, Nishioka:2018khk} and reference therein. 

Entanglement plays a fascinating role in the study of quantum gravity, which seeks to unify quantum mechanics with general relativity to describe the fundamental nature of spacetime and gravity at the smallest scales. In \cite{VanRaamsdonk:2010pw, Faulkner:2013ica} it is argued that the emergence of classically connected spacetimes is related to the quantum entanglement of boundary degrees of freedom. Particularly, in the context of the AdS/CFT correspondence, knowledge of mixed states within a finite spacelike boundary region enables the complete reconstruction of a bulk region referred to as the "entanglement wedge" \cite{Czech:2012bh,Wall:2012uf,Headrick:2014cta}. This concept has proven instrumental in elucidating various phenomena associated with the duality, including quantum error correction \cite{Almheiri:2014lwa}, the entanglement of purification \cite{Espindola:2018ozt}, and the Page curve of black holes \cite{Penington:2019npb}, just to name a few. For a nice review and further references on the subject, see \cite{Harlow:2018fse}. 

The AdS/CFT correspondence \cite{Maldacena:1997re, Witten:1998qj, Gubser:1998bc} has been motivated by the desire to address longstanding questions in theoretical physics, including the nature of Quantum Gravity, the behavior of black holes, the resolution of fundamental paradoxes, and the exploration of strong coupling regimes in QFT.  offers an alternative framework for studying quantum gravity non-perturbatively by relating a gravitational theory in asymptotically locally AdS (AlAdS) spacetime to strongly coupled conformal field theory (CFT) living on its boundary, as the semiclassical regime of a weakly coupled gravitational theory is mapped onto the ultraviolet (UV) regime of a strongly coupled CFT \cite{Susskind:1998dq}. 
This duality provides a powerful tool for exploring strong coupling regimes of QFTs, which are notoriously difficult to analyze using traditional perturbative methods. Using techniques from string theory or other approaches to Quantum Gravity and exploiting the duality, one can gain insights into the non-perturbative behavior of quantum gravity by analyzing the corresponding CFT on the boundary.

One of the most important measures to test the duality is the holographic entanglement entropy. Ryu and Takayanagi conjecture \cite{Ryu:2006bv, Ryu:2006ef}\footnote{See \cite{Hubeny:2007xt} for the covariant generalization.} that the calculation of the von Neumann entropy of the reduced density matrix of the CFT is mapped to a minimization of a codimension-two surface inside AdS. More specifically, the RT proposal, later proven by Lewkowycz and Maldacena \cite{Lewkowycz:2013nqa}, states that the bipartite entanglement between a subregion $A$ and the rest of the space $B=A^{c}$ in the boundary CFT is proportional to a quarter of the area of a minimal surface $\gamma_{\rm RT}$ anchored to the CFT entangling region and extends to the boundary (see \autoref{fig:RT}). This remarkable result establishes a direct connection between the geometric properties of spacetime in the bulk gravitational theory and the quantum information encoded in the entanglement structure of the boundary CFT. Quantum corrections to the RT formula have been found in \cite{Faulkner:2013ana}, and the new quantum contributions to the holographic entanglement entropy appear as the RT surface split bulk regions generating entanglement among them, showing how the RT surface produces a bulk symmetry breaking that also appear in the field theory side \cite{Gubser:2004qj}. 

In this thesis, we use the notion of entanglement in different contexts in order to understand how quantum corrections can give information about Quantum Gravity. We first consider entanglement in de sitter (dS) spacetime which is one of the fundamental spaces describing early and late eras of our Universe, and it has been argued that Gibbons--Hawking entropy \cite{Gibbons:1976ue} appears due to entanglement between disconnected regions in the spacetime \cite{Dong:2018cuv, Arias:2019pzy, Arias:2019zug, Chandrasekaran:2022cip}. We use the two-dimensional conformal description of Killing horizons of \cite{Solodukhin:1994yz, Carlip:1999cy} and consider the 1-loop corrections found in \cite{Carlip:2000nv} which modify the Cardy entropy and introduce a logarithmic correction into the black hole entropy. We find that this modification backreacts into the Renyi entropy in a way that introduce a UV divergence into the entanglement entropy satisfying the usual area-law of entanglement \cite{Srednicki:1993im}. Moreover, we compute the Hartley entropy that accounts for the amount of states that participate in the entanglement, such that we can obtain the dimension of the Hilbert space associated with an observer within dS spacetime. We find that the the dimension of the Hilbert space is given roughly by the exponential of the GH entropy, in agreement with previous literature \cite{Banks:2000fe, Witten:2001kn}, that indicates problems in the quantization of Einstein gravity in asymptotically dS spacetime. 

We move forward and consider holographic entanglement entropy in the realms of the AdS/CFT correspondence. We consider the proposal to compute Renyi entropies when the dual CFT has a global $U(1)$ symmetry of \cite{Belin:2013uta}. We extend this notion and consider gravity coupled to non-linear electrodynamics (NLED). NLED Lagrangians were originally introduced \cite{Born:1934gh} as modifications to Maxwell's theory to solve the self-energy problem of quantum electrodynamics, but later on appeared as an important ingredient of String Theory. Moreover, considering higher-curvature terms in the IR limit of the string partition function leads to gravity coupled with NLED of the Born--Infeld type \cite{Gibbons:2000xe}. For nice reviews on NLED theories see  \cite{Gibbons:2001gy, Sorokin:2021tge}. We follow to compute the first modifications of the charged Renyi entropy by considering thermal and quantum fluctuations of the gravitational partition function. We found that the new terms appearing in the holographic entanglement entropy resemble terms appearing in field theories with a spontaneous symmetry breaking of a continuous global symmetry \cite{Metlitski:2011pr}, and the central charge of the dual theory decreases due to the small energy fluctuations. As a particular example, we consider three-dimensional gravity coupled to conformal electrodynamics \cite{Hassaine:2007py} and found that the model can be used to holographically describe charged free bosons in two dimensions. 

Finally, we focus on three-dimensional gravity, extensively used in the literature as a toy model for Quantum Gravity. Maloney and Witten \cite{Maloney:2007ud}, where they show that the sum of all known contributions to the gravitational path integral yields an infinite tower of negative energy states, potentially rendering the quantum theory non-existent. Expanding upon their findings, we identify new solutions overlooked in the Maloney-Witten sum, characterized by the presence of cosmic strings within the bulk, and show that they contribute non-trivially to the partition function already at the classical Euclidean saddle. Initially, we investigate a more generalized theory of gravity conformally coupled with a scalar field. The solutions exhibit a topological defect—cosmic strings—inducing acceleration\footnote{Recently, there have been proposals to test the astrophysical implications of such objects in four dimensions \cite{Grenzebach:2015oea, Ashoorioon:2021gjs, Ashoorioon:2022zgu }.}. Through a holographic analysis of the solutions, we demonstrate how these three-dimensional hairy accelerating black holes can serve to study marginally quartic-trace deformed two-dimensional thermal CFTs. Concluding our study, we compute the holographic entanglement entropy of the accelerating black hole under the condition of a vanishing scalar field. Notably, in a regime of small acceleration expansion, we observe a reduction in entanglement, suggesting some loss of information compared to the pure BTZ black hole counterpart.

\newpage

%%%%%%%%%%%%%%%%%%%%%%%%%%%%%%%%%%%%%%%%%%%%%%%
\vspace*{\fill}
\paragraph{Summary of results and organization of the Thesis:}
\begin{itemize}
    \item[\color{magenta}$\diamond$] In \autoref{Sec:Entanglement}, we provide a brief overview of bipartite entanglement, Renyi entropy, and their generalizations within the context of a global $U(1)$ symmetry. This section also delves into the computation of entanglement measures within a holographic framework, alongside several other key concepts that will be recurrently addressed throughout this thesis.
    \item[\color{magenta}$\diamond$] \autoref{Sec:dS} is devoted to the extension of the work \cite{Arias:2019pzy, Arias:2019zug}, where is argued that the GH entropy can be viewed as the entanglement between disconnected observers or as the entanglement entropy between the Euclidean CFTs at the past and future asymptopia. We consider the CFT description of the cosmological horizon presented in \cite{Carlip:1999cy} and consider the first quantum correction of the Cardy entropy \cite{Carlip:2000nv} and found that it is possible to break the flat entanglement spectrum previously found in the literature, but the dimension of the Hilbert space of the theory remains finite, and given by the exponential of the GH entropy.  
    \item[\color{magenta}$\diamond$] In \autoref{Sec:Fluctutations} we give a new example of how quantum fluctuations modify previous results in the holographic entanglement entropy in the context of the AdS/CFT correspondence. We consider small energy fluctuations of non linearly charged topological black holes which are dual to a thermal CFTs with a global $U(1)$ symmetry. Using the CHM map, we map the dual theory to a non-thermal CFT with an spherical entangling region. We discuss how the fluctuations modify the Renyi and entanglement entropy in a similar way that theories that suffer an spontaneous symmetry breaking of a global symmetry due to IR fluctuations of the ground state. As a particular example, we show a three-dimensional bulk model that resembles the Renyi entropy and conformal weights of the associated twist operators of a two-dimensional free charged boson. 
    \item[\color{magenta}$\diamond$] In \autoref{Sec:Acceleration}, we show how introducing acceleration into the bulk modifies the amount of entangled pairs in the dual CFT. In order to do so, we introduce a new family of accelerating solutions to three-dimensional AdS gravity coupled with a scalar field. These solutions encompass various limits of previously known black hole solutions. We conduct a brief holographic analysis of the solution and explore methods for calculating the holographic entanglement entropy. Our findings suggest that as the scalar field approaches zero, the pure gravitational solution shows a decrease in entanglement entropy with increasing acceleration. 
    \item[\color{magenta}$\diamond$] In \autoref{Sec:Conc}, we conclude with some discussion of the main results presented in this thesis and further directions.
\end{itemize}

\vspace*{\fill}
\begin{center}
\pgfornament[width=1cm]{79}
\end{center}

%%%%%%%%%%%%%%%%%%%%%%%%%%%%%%%%%%%%%%%%%%%%%%%%%
%%%%%%%%%%% Holographic Entanglement %%%%%%%%%%%%
%%%%%%%%%%%%%%%%%%%%%%%%%%%%%%%%%%%%%%%%%%%%%%%%%
\newpage
\section{Entanglement Entropy}\label{Sec:Entanglement}

Given a system decomposed into subsystems  $A$ and $B$, the subsystem $A$ can be described by the normalized density matrix $\rho_A$ which is defined by tracing over the degrees of freedom of $B$ of the total density matrix $\rho = |\Psi\rangle\langle\Psi|$~, with $|\Psi\rangle$ a pure ground state, i.e., $\rho_A = {\rm Tr}_B\rho$~, referred to as the reduced density matrix. The Hilbert space of the pure state factorizes into a direct product\footnote{See \cite{eakins2002factorization} for a discussion.}, viz, ${\cal H} = {\cal H}_A \otimes {\cal H}_{B}$~. The spectral properties of $\rho_A$ contain information about the amount of entanglement between both subsystems and it is given by the von Neumann entropy of the reduced density matrix
\begin{align}
    S_{\rm E} = -{\rm Tr}_A \rho_A \log\rho_A~.
\end{align}

The simplest example of this bipartite entanglement, appears in the two spin system. Consider a system of two particles $A$ and $B$ of spin $1/2$~, such that the Hilbert space of each particle contains two states $\cH_{A,B} = \{|0\rangle,|1\rangle\}_{A,B}$~. Using orthonormal basis ${}_{A,B}\langle i | j \rangle_{A,B} = \delta_{ij}$ for $(i,j)\in \{0,1\}$~. In this case, the total Hilbert space decompose as a tensor product of both subsystems $\cH = \cH_A \otimes \cH_B$~, give by the tensor product states $|ij\rangle = |i\rangle_A\otimes|j\rangle_B$~. 
The normalized ground state corresponds to 
\begin{align}
    |\Psi\rangle = \frac{1}{\sqrt{2}}\left(|01\rangle - |10\rangle\right)~,
\end{align}
and the reduced density matrix of a subsystem, say $A$, is
\begin{align}
    \rho_A = \frac12 \left(|0\rangle_A {}_A\langle0| + |1\rangle_A {}_A\langle1|\right) = \frac12 \begin{pmatrix}
1 & 0 \\
0 & 1 
\end{pmatrix}~.
\end{align}
Clearly $\rho_A$ is not a pure state and the von Nuemann entropy reads
\begin{align}\label{log2}
    S_{\rm E} = -{\rm Tr}_A \rho_A \log \rho_A = \log 2~,
\end{align}
showing the existence of entangled state. Such entangled state appears in the pion decay into an electron-positron system with opposite spin, due to conservation of angular momentum, separating away from each other. This shows that measuring one of the particle's spin lead to having the knowledge of the other particle, independent of the scale distance between them. This is somehow a breaking of locality which is a feature of entangled systems. Entanglement entropy quantifies the degree to which a particular state deviates from being a separable state and therefore how quantum a wave function state $|\Psi\rangle$ is.

Another interesting measure corresponds to the Renyi entropy, which consist on a non-trivial generalization of many notions of entropy while preserving additivity of independent events. 
It consist on a one-parameter generalization of the von Neumann entropy labeled by an integer $n$, refereed to as the \emph{Renyi index} or \emph{the replica parameter},
\begin{align}
    S_n = \frac{1}{1-n}\log{\rm Tr}_A \rho_A^n~,
\end{align}
such that
\begin{align}
    S_{\rm E} = \lim_{n\to1}S_n~,
\end{align}
where we have used the fact that the reduced density matrix is normalized, i.e., ${\rm Tr}_A \rho_A = 1$~, and assume an analytical continuation of $n$ to a real number.  
This entropy generalize several other notions of entropy as it contains more information about the non-trivial eigenvalues of $\rho_A$ by means of different limits of the Renyi index. Two particular important limits that we will be used, corresponds to first the min-entropy, given by the limit
\begin{align}
    S_\infty \equiv \lim_{n\to\infty}S_n = -\log \lambda_1~,
\end{align}
where $\lambda_1$ is the largest eigenvalue of the reduced density matrix, and secondly the max-entropy 
\begin{align}
    S_0 \equiv \lim_{n\to 0} S_n = \log \boldsymbol{\rm d}~,
\end{align}
where $\boldsymbol{\rm d}$ corresponds to the number of non-zero eigenvalues of $\rho_A$~, that is, the dimension of the subspace $\cH_A$ that participate in the entanglement. Finally, another interesting limit is the Collision entropy 
\begin{align}
    S_2 \equiv \lim_{n\to 2}S_n = -\log {\rm Tr}_A\rho_A^2 = \log p_2~,
\end{align}
where $p_2$ is the probability to find both subsystems being described by $\rho_A$~, i.e., the purity of the subsystem $A$~. The Collision entropy has been used to detect symmetry-protected topological states in quantum computers \cite{Azses:2020tdz}.

The Renyi entropy satisfies various inequalities between different values of the replica parameter as it is Schur-concave \cite{beck_schogl_1993,zyczkowski2003renyi}
\begin{align}\label{Rineq}
    \partial_n S_n \leq 0~,\quad \partial_n\left(\frac{n-1}{n}S_n\right)\geq 0~,\quad \partial_n\left(\left(n-1\right)S_n\right)\geq 0~,\quad \partial^2_n\left((n-1)S_n\right)\leq 0~.
\end{align}
We can give a thermal interpretation to Renyi entropy and to the inequalities \eqref{Rineq} by introducing the \emph{modular Hamiltonian} assuming that the reduced density matrix is in the image of the exponential map, viz.
\begin{align}
    \rho_A :=\exp\{-H_A\}~.
\end{align}
Then, the existence of such Hamiltonian operator implies that the Renyi entropy can be identified in terms of some thermal partition function
\begin{align}
    Z_A = {\rm Tr}_A (\rho_A^n) = {\rm Tr}_A \exp\{-\beta H_A\}~,
\end{align}
where the inverse temperature $\beta \equiv n$ is given by the replica parameter. We can then associate a free energy \label{baez2011renyi}
\begin{align}\label{FASA}
    F_A := -\frac{1}{n}\log Z_A \Rightarrow S_n = \left(1-\frac{1}{n}\right)^{-1}F_A~,
\end{align}
of the corresponding thermal system. In analogy with statistical mechanics, we can define 
\begin{align}
    E_A =-\partial_n\log Z_A~,\quad S_A = (1-n\partial_n)\log Z_A~,\quad {\cal C}_A = n^2\partial_n^2\log Z_A~,
\end{align}
being the modular energy, modular entropy, and modular capacity, respectively. Replacing the new definitions into the Renyi inequalities \eqref{Rineq}, we get
\begin{align}
    S_A\geq 0~,\qquad E_A\geq0~,\qquad {\cal C}_A\geq 0~,
\end{align}
which implies the thermal stability of the modular system. Indeed, starting with the modular system at some equilibrium temperature and then perturb the system  by dividing the equilibrium temperature by a factor of $n$~, then, the Renyi entropy measure the maximum amount of work, divided by the difference of temperature, than the system can done to move back to equilibrium at the new temperature \cite{baez2011renyi}, as can be seen in \eqref{FASA}, where the equilibrium temperature $T_0 = 1$. 
This thermal interpretation of the Renyi entropy, obtained by giving it a thermal interpretation to the reduced density matrix by means of the modular Hamiltonian, can be generalized by inserting a equilibrium temperature $T_0 \neq 1$ such that
\begin{align}
    \rho_A^n = \frac{\exp\{-nH_A/T_0\}}{Z(T_0)^n} = \frac{Z(T_0/n)}{Z(T_0)^n}~,\qquad Z(T_0) \equiv {\rm Tr}_A \exp\{-H_A/T_0\}~,
\end{align}
such that the relation between the replica parameter and temperature of the system is now 
\begin{align}
    n = T_0/T~,
\end{align}
and \eqref{FASA} becomes
\begin{align}\label{baez}
    S_n = \frac{n}{1-n}\frac{1}{T_0}\left[F(T_0) - F(T_0/n)
 \right]~.
\end{align}
We can map the Renyi entropy to the modular entropy 
\begin{align}
    S_n = \frac{n}{1-n}\frac{1}{T_0}\int_{T_0/n}^{T_0}\frac{\d m}{m^2}S_A~,
\end{align}
which is an invertible map. In fact, the Renyi entropy can be seen in terms of a fractional derivative, which are defined as
\begin{align}
    \left(\frac{\partial f}{\partial x}\right)_m := \frac{f(mx) - f(x)}{mx-x}
\end{align}
and reduces to the standard derivative when $m\to1$~. Then, the Renyi entropy and the modular free energy are related by
\begin{align}
    S_n = -\left(\frac{\partial F}{\partial T}\right)_{1/n}~,
\end{align}
indicating that the Renyi entropy corresponds to a $n$-deformation of the usual concept of entropy. 

In a general QFT, the entanglement entropy suffers from UV divergences due to the short range interaction near the the boundary of the entangling surface $\partial \Sigma$~. An usual renormalization procedure is to consider a field theory defined with a cutoff, e.g. on a lattice. In general the entanglement entropy satisfy an area law
\begin{align}
    S_{\rm E} = \alpha\frac{{\rm Area}_{\Sigma}}{\delta^{d-2}}\left(1+{\cal O}(\delta)\right)~,
\end{align}
where $\delta$ is the UV regulator, and $\alpha$ depends on the geometry in which the QFT lies. In $d= 2$~, the area law is replaced by a logarithmic divergence
\begin{align}
    S_{\rm E} = \alpha\log\frac{L}{\delta} + {\rm finite~terms}~,
\end{align}
where $L$ depends on the properties of the field theory and the entangling region geometry. 
In QFT, the two subsystems $A$ and $B$ are spatial regions separated by an entangling surface $\Sigma$~. Then, the Renyi and entanglement entropy are usually hard to compute. A prescription to obtain this quantities, consist in analytically continue the Renyi parameter to a real number such that
\begin{align}
   S_{\rm E} = -\lim_{n\to1}\frac{\log{\rm Tr}_A\rho_A^n}{n-1} = -\lim_{n\to1}\partial_n\log{\rm Tr}_A\rho_A^n~,
\end{align}
which is known as the \emph{replica trick} and is often employed in the computation of entanglement entropy in QFT. It also works in the previous example of the two particle system where now
\begin{align}
    \log {\rm Tr}_A \rho_A^n = \log 2^{1-n}~,
\end{align}
such that the replica trick gives
\begin{align}
    S_{\rm E} = -\lim_{n\to1}\partial_n \log 2^{1-n} = \log 2~,
\end{align}
in agreement with the previous calculation \eqref{log2}.  
Using the path integral representation of QFT, the trace of the $n$-th power of $\rho_A$ correspond to the partition function $Z_n$ on the $n$-fold cover of the original spacetime manifold, that is constructed by gluing $n$ copies of the original manifold along the entangling region $\Sigma$ 
 \cite{Calabrese:2004eu}. The $n$-fold cover manifold, say ${\cal C}_n$~, has a conical singularity at the entangling region $\Sigma$ with deficit angle 
 \begin{align}\label{deltan}
 \delta_n = 2\pi\left(1-\frac{1}{n}\right)~,
 \end{align}
but is invariant under the $\mathbb{Z}_n$ symmetry that shifts the modular time $\tau$ by $2\pi~ \forall n\in \mathbb{N}$~. 
 In this approach, one interchanges the computation of $\log {\rm Tr}_A\rho^n_A$ to compute a parition function in a $n$-fold cover with a branch-point, which is still far from reach, even for CFTs. These boundary conditions can be though of in terms of the insertion of a surface operator $\sigma_n$~, referred to as a \emph{twist operator}, at the branch-points (the entangling region) which intertwine $n$ copies of the field theory on a single copy of the background geometry \cite{Calabrese:2004eu, Calabrese:2005zw}. Then, the trace of the $n$-th power of the reduced density matrix is mapped to compute the vacuum expectation value of $\sigma_n$ in the tensor product of the QFT, viz.
 \begin{align}
     \langle \sigma_n \rangle = \frac{Z_n}{Z^n}~,
 \end{align}
where $Z_n$ is the parition function on the full $n-$sheeted gemoetry, such that
\begin{align}
    S_n = \frac{1}{1-n}\log\langle\sigma_n\rangle = \frac{1}{n-1}(n\log Z_1 - \log Z_n)~.
\end{align}
Then, the replica trick replaces the problem of computing $\rho^n_A$ with that of a new density matrix of a different theory that consist on $n$ copies of the original field theory. Although the idea of introducing brunch cuts by means of the twist operators immediately generalize to higher dimensions, the properties and construction of such operators is not well understood \cite{Swingle:2010jz}. In the case of the QFT to posses conformal symmetry, i.e., is a CFT, the twist operators have been understood in any dimension \cite{Hung:2014npa}, where the conformal dimension is given in terms of the energy density of the CFT in some thermal ensemble. Remarkably, the conformal dimension of the twist operators contain information of the coefficients in the leading order singularity of the stress-tensor two-point function \cite{Hung:2011nu, Hung:2014npa}. The notion of conformal dimension for this surface operators is defined in terms of the leading singularity in the $\langle T_{ij}\sigma_bn\rangle$ correlator. 
Particularly, two-dimensional CFT are very interesting theories as they have an infinite dimensional symmetry algebra \cite{Belavin:1984vu} and several examples appear in nature, see for instance \cite{Ginsparg:1993is}. In CFT$_2$, for an entanglement surface corresponding to an interval of length $L$~, the Renyi entropy reads
\begin{align}
    S_n  = \frac{c}{6}\left(1+\frac{1}{n}\right)\log\frac{L}{\delta}~,
\end{align}
where $c$ is the central charge of the theory, and the the twist operators are local primaries with scaling dimension
\begin{align}
    h_n = \frac{c}{12}\left(n - \frac{1}{n}\right)~.
\end{align}
Before delving into the properties of the twist operators, let us first briefly study some properies of entanglement entropy in CFTs.

As the entanglement entropy is a dimensionless quantities and can only depend on ratios of lenght/energy scales. In a CFT, the only lenght scales is the total system size and the UV cutoff, such that the entanglement entropy also satisfy an area law. Morevoer, for an even-dimensional CFT, the area law contains universal logarithmic divergences that encode information of the conformal anomaly appearing in the trace of the stress tensor \cite{Duff:1993wm}, by considering the scale transformations of the entanglement entropy \cite{Nishioka:2018khk} as
\begin{align}
    L \frac{dS_{\rm E}}{d L} = -\int_{\cal M} \d^dx \sqrt{g}\langle T^i{}_i\rangle + \lim_{n\to1}\partial_n\int_{{\cal M}_n}\d^dx\sqrt{g}\langle T^i{}_i\rangle~,
\end{align}
where $L$ is the lenght of the entangling region, and ${\cal M}_n$ is the $n$-fold cover of $\cal M$~. 
Then, one can solve $S_{\rm E}$ in terms of the trace of the energy momentum tensor. The entanglement entropy for an even-dimensional CFT has the structure
\begin{align}
    S_{\rm E} = \frac{\alpha_{d-2}}{\delta^{d-2}} + \frac{\alpha_{d-4}}{\delta^{d-4}} + \dots + \frac{\alpha_{2}}{\delta^{2}} + \alpha_0\log \frac{L}{\delta} + {\rm finite~terms}~,
\end{align}
where the finite terms are scheme dependent such that they do not have a physical meaning, and $\gamma_0$ is given in terms of the Weyl anomaly. 
In odd dimensions, the entanglement entropy satisfies a similar expansion but the logarithm is absent. In this case, the finite term is scheme independent and has a physical meaning. Particularly for topological field theories \cite{atiyah, wittentft2, wittentft1} the finite term corresponds to the \emph{topological entanglement entropy} \cite{Kitaev:2005dm} which is an additive constant characterizing global features of the entanglement in the ground state. 

Obtain the partition function on the $n$-fold cover manifold is a hard task even in CFTs, but when we consider an spherical entangling surface, we can compute the entanglement entropy by means of a thermal entropy. Such transformation is known as the CHM map \cite{Casini:2011kv}. 
Let us consider a constant time spherical entangled surface at constant time $S^{d-2}$ of radius $\rho = \cR$  on the Euclidean flat space
\begin{align}\label{Eflat}
    ds^2_{\rm flat} = \d t^2 + \d\rho^2 + \rho^2 \d\Omega_{d-2}^2~,
\end{align}
with $\d\Omega_{d-2}^2$ the line element of the unit $(d-2)$-sphere. Then, we can use scaling symmetry and map to distinct geometries conformally equivalent to \eqref{Eflat} to simplify the computation of $F_A$, and hence the one of $S_n$~. Particularly, mapping the flat space into an hyperbolic space $S^1\times H^{d-1}$ endowed with the metric
\begin{align}
    ds^2_{\rm hyp} = \Omega^2\left[ \d t^2 + \cR\left(\d y^2 + \sinh^2 y \d \Omega_{d-2}^2\right)\right]~,
\end{align}
by means of the complex coordinate transformation
\begin{align}\label{ftoh}
    \exp\{-y+it/\cR\} = \frac{\cR - \rho - it}{\cR + \rho + it}~.
\end{align}
where the conformal factor
\begin{align}
    \Omega = \frac{2\cR^2}{|\cR^2 - (\rho+i\tau/\cR)^2|}~.
\end{align}
The coordinate transformation \eqref{ftoh} maps the entangling surface at $\rho = \cR$ to a sphere at the asymptotic boundary of the hyperbolic geometry $y = \infty$~. The Euclidean time is periodic such that the the conformal map now defines a thermal CFT in the interior of the entangling surface with temperature
\begin{align}
    T_0 = \frac{1}{2\pi \cR}~.
\end{align}
The reduced density matrix becomes a thermal density matrix
\begin{align}\label{rhotherm}
    \rho_A = U^{-1}\rho_{\rm thermal}(T_0)U = U^{-1}\frac{\exp\{-H_A/T_0\}}{Z(T_0)}U~,
\end{align}
where $U$ denotes the unitary transformation that implements the conformal map. Therefore, after taking traces, the quantum information measures become invariant under such unitary transformation and the entanglement entropy now becomes the thermal entropy of a CFT with temperature $T_0$ on the hyperbolic space, viz.
\begin{align}
    S_{\rm E} = S_{\rm thermal}(T_0)~,
\end{align}
which is far more achievable than the computation of the partition function of a branch cover manifold.

The Renyi entropy can be computed taking the $n$-th power of \eqref{rhotherm}
\begin{align}
    \rho_A^n = U^{-1}\frac{\exp\{-nH_A/T_0\}}{Z(T_0)^n}U~,
\end{align}
which corresponds to consider a thermal ensemble with temperature
\begin{align}\label{Tneq}
    T_n = T_0/n~,
\end{align}
on the same hyperbolic geometry. 
Using \eqref{baez}, and the thermodynamic relation $S_{\rm thermal} = -\frac{\partial F}{\partial T}$ the relation between the thermal entropy and the Renyi entropy becomes
\begin{align}\label{Snthermal}
    S_n = \frac{n}{n-1}\frac{1}{T_0}\int^{T_0}_{T_0/n}\d T S_{\rm thermal}~.
\end{align}

In systems with global symmetries, the Renyi entropy does not capture the interplay of different charge sectors of the QFT. In order to extend the entanglement measures to theories with a $U(1)$ global theory, the authors of \cite{Belin:2013uta}\footnote{See \cite{Caputa:2013eka} for early computations of (holographic) entanglement entropy of theories with a $U(1)$ global symmetry.} have introduce a chemical potential into the path integral calculations of the Renyi entropy. In the grand canonical ensemble, a chemical potential $\mu$ is represented in the Euclidean path integral by a fixed background gauge potential $A_{(0)}^i$ which coupled to a spin 1 conserved current $J_i$~. As the Euclidean time period corresponds to the inverse of the system's temperature $T$, the insertion of a chemical potential appears as a non-trivial Wilson line on this thermal circle
\begin{align}
    \oint A_{(0)} = -i\mu/T~. 
\end{align}
Then, the \emph{charged Renyi entropy} is the grand canonical version of the Renyi entropy
\begin{align}\label{rhonmu}
   S_n(\mu) = \frac{1}{1-n}\log {\rm Tr}_A\rho_A^n(\mu) = \frac{1}{1-n}\log \frac{Z_n(\mu)}{Z(\mu)^n}~,\qquad \rho_A(\mu) = \rho_A \frac{\exp\{\mu Q_A\}}{n_A(\mu)}~,
\end{align}
with $n_A(\mu) := {\rm Tr}_A~\rho_A\exp\{\mu Q_A\}$~. The computation of $Z_n(\mu)$ is the same as in the uncharged case, but with a Wilson line encircling $\Sigma$, which now carries a magnetic flux\footnote{With this conventions, the chemical potential becomes a dimensionless quantity.} $-in\mu$~, and passing through all the $n$ sheets of the covering geometry.  
We can see this insertion by means of a generalized twist operator $\sigma_n(\mu)$~, which are constructed by dressing the original $\sigma_n$ with an induced Aharonov-Bohm flux through the multisheeted Riemann surface, i.e., a $(d-2)$-dimensional Dirac surface with magnetic flux $-in\mu$~. 
In the path integral formulation, the charged Renyi entropy reads
\begin{align}
    S_n = \frac{1}{n-1}\left(n\log Z(\mu) - \log Z_n(\mu)\right)~,
\end{align}
where now $Z_n(\mu)$ is the partition function in the grand canonical ensemble in the branch-cover manifold. 

In the above construction, the chemical potential is real, but the analytic continuation $\mu = i\mu_{\rm E}$, which can give rise to some singularities along the imaginary $\mu$-axis. Working with imaginary chemical potential can be useful to avoid the sign problem appearing in the lattice fermion algorithms \cite{Alford:1998sd, DElia:2002tig, DElia:2004ani}, and to to study a confinement-deconfinement phase transitions in fermionic QCD theories \cite{Roberge:1986mm, deForcrand:2002hgr} where the $U(1)$ global symmetries corresponds to the fermion number conservation and the partition function with an imaginary chemical potential has the baryonic number as the conjugate charge. For $SU(N)$ QCD, the baryon number characterize the baryon number of bound states in the deconfined phase with an integer number, such that the conjugate chemical potential must be a periodic function, while for the confined phase the partition function is perdiodic with $\mu_{\rm E}$ has period of $2\pi/N$~. 

Applying the CHM map for spherical entangling surfaces, the thermal density matrix now becomes
\begin{align}
    \rho_{\rm thermal} = \frac{\exp\{-H_A/T_0 + \mu Q\}}{Z(T_0,\mu)}=\frac{\exp\{-H_A/T_0 + i\mu_{\rm E} Q\}}{Z(T_0,\mu)}~,
\end{align}
where, as aforementioned, the chemical potential is incorporated into the thermal partition function by inserting a nontrivial Wilson loop on the Euclidean time circle
\begin{align}
    \mu_{\rm E} = \oint A_{(0)} = \int_0^{2\pi\cR} \d t_{\rm E} A_{\rm E} = 2\pi\cR A_{\rm E}~,
\end{align}
where we use that the gauge potential remains fixed along the loop. 
The Renyi entropy in terms of the partition function of the CFT on the hyperbolic geometry reads
\begin{align}
    S_n(\mu_{\rm E}) = \frac{1}{1-n}\log\frac{Z(T_0/n,\mu_{\rm E})}{Z(T_0,\mu_{\rm E})^n}~.
\end{align}
Then, the previous relations between the Renyi entropy and the thermal entropy is
\begin{align}
    S_n(\mu_{\rm E}) = \frac{n}{n-1}\frac{1}{T_0}\int^{T_0}_{T_0/n}\d T~ S_{\rm thermal}(T,\mu_{\rm E})~.
\end{align}

The insertion of twist operators give rise to non-trivial correlators with other observables of the CFT. Inserting the stress tensor at a perpendicular small distance $u$ of $\sigma_n(\mu)$, then \cite{Belin:2013uta}
\begin{align}
    \langle T_{ab}\sigma_n\rangle = -\frac{h_n}{2\pi}\frac{\delta_{ab}}{y^d}~,\qquad \langle T_{aM}\sigma_n\rangle = 0~,\qquad \langle T_{MN}\sigma_n\rangle = \frac{h_n}{2\pi}\frac{dn_M n_N - (d-1)\delta_{MN}}{y^d}
\end{align}
where $M,N\in \{0,1\}$ and $a,b \in\{2,\dots,d-1\}$, and $n_M$ is unit vector orthogonally from $\sigma_n$ to $T_{ij}$~. The leading order singularity becomes fixed up to the twist operator conformal dimensions $h_n$, which is now $\mu$ dependent as well. As shown in \cite{Hung:2011nu}, the conformal weight can be obtained in terms of the CFT energy
\begin{align}\label{hnmu}
    h_n(\mu) = \frac{n}{d-1}\frac{1}{T_0 V_{H^{d-1}}}\left(E(T_0,0) - E(T_0/n,\mu)\right)~. 
\end{align}
Moreover, Ward identities of the $U(1)$ current operator $J_i$ of the CFT satisfy \cite{Belin:2013uta}
\begin{align}
    \langle J_M \sigma_n\rangle = -\frac{ik_n(\mu)}{2\pi}\frac{\epsilon_{MN}n^N}{y^{d-1}}~, \qquad \langle J_a \sigma_n\rangle = 0~,
\end{align}
where $\epsilon_{MN}$ denotes the volume form of the two-dimensional space intersecting the spherical entanglind surface, and $k_n(\mu)$ is referred to as \emph{magnetic response} as describes the response of the current to the magnetic flux. The magnetic response can be obtained as
\begin{align}\label{magresponse}
    k_n(\mu) = 2\pi n\cR^{d-1}\rho(n,\mu)~,
\end{align}
where $\rho(n,\mu)$ is the charge density appearing in the first law in the grand canonical ensemble, $\langle J_{\rm t} \rangle = -i\rho(n,\mu)$~. 

A remarkable universal properties of the twist operators, is that the conformal weight and magnetic response satisfy\cite{Hung:2014npa}
\begin{align}
    \lim_{n\to1}\partial_\mu h_n(\mu) ={}& 2\pi^{\frac{d}{2}-1}\frac{\Gamma(d/2)}{\Gamma(d+2)}C_T~,\\
    \lim_{\mu\to0}\lim_{n\to1}\partial_\mu k_n(\mu) ={}& 4\pi^{\frac{d}{2}}\frac{\Gamma(d/2+1)}{\Gamma(d+1)}C_V~,
\end{align}
where $C_T$ and $C_V$ are the coefficients in the leading order singularities of the $\langle T T\rangle$ and $\langle J J\rangle$ correlators, respectively, viz.
\begin{align}
    \langle T_{ij}(x)T_{mn}(0)\rangle = \frac{C_T}{x^{2d}}{\cal I}_{ij,mn}~,\qquad \langle J_i(x)J_j(0)\rangle = \frac{C_V}{x^{2d-2}}I_{ij}~,
\end{align}
where
\begin{align}
    {\cal I}_{ij,mn} = I_{(ij}I_{mn)}-\frac{1}{d}\delta_{ij}\delta_{mn}~,\qquad I_{ij} = \delta_{ij} - 2\frac{x_ix_j}{|x|^2}~.
\end{align}

Another interesting quantity is the \emph{symmetry resolved Renyi entropies}. If the pure state of the full system is an eigenstate of the charge operator $[\rho,Q] = 0$ and therefore $[\rho_A,Q_A] = 0$~. Then, the reduced density matrix can be block decomposed in different charge sectors 
\begin{align}
   \rho_A = \oplus_q\Pi_q \rho_A = \oplus p(q)\rho_A(q) 
\end{align}
where $\Pi_q$ is the non local projector on eigenspace of fixed value of $q$ in the spectrum of $Q_A$~, and $p(q)$ are the probabilities of finding $q$ when measuring $Q_A$~, where we have used ${\rm Tr}_A \rho_A(q) = 1$~. Then, one would like to understand how the entanglement distributes in the different charge sectors. The von Neumann entropy now decomposes as \cite{lukin2019probing}
\begin{align}
    S_{\rm E} = \sum_q p(q)S_{\rm E}(q) - \sum_q p(q)\log p(q) = S^{c}_{\rm E} + S^f_{\rm E}~,
\end{align}
where $S_{\rm E}(q)$ is the \emph{symmetry resolved entanglement entropy} 
\begin{align}
   S_{\rm E}(q) \equiv -{\rm Tr}_A \rho_A(q)\log\rho_A(q)~. 
\end{align}
The two contributions are the \emph{configurational} entanglement entropy 
\begin{align}
    S^c_{\rm E} \equiv \sum_q p(q)S_{\rm E}(q)~,
\end{align}
measuring the total entanglement due to each sector with its corresponding weight probability, and the \emph{fluctuation} entanglement entropy
\begin{align}
    S^f_{\rm E} \equiv -\sum_q p(q)\log p(q)~,
\end{align}
which takes into account the contributions of fluctuations of the value of the charge within $A$~. Then, we can define the symmetry resolved Renyi entropy as
\begin{align}
    S_n(q) \equiv \frac{1}{1-n}\log {\rm Tr}\rho_A(q)^n~,
\end{align}
where each given charge sector $q$ can be obtained by Laplace transforming the partition function 
\begin{align}
    {\cal Z}_n(q) = \int_{-\pi}^{\pi}\frac{\d\mu_{\rm E}}{2\pi}\exp\{-iq\mu_{\rm E}\}~,\qquad Z_n(\mu_{\rm E}) \equiv {\rm Tr}_A (\rho_A^n\exp\{i Q_A \mu_{\rm E}\})~, 
\end{align}
such that
\begin{align}\label{SRSn}
    S_n(q) = \frac{1}{n-1}(n\log{\cal Z}_1(q) - \log {\cal Z}_n(q))~. 
\end{align}
Using replica methods, in \cite{ Goldstein:2017bua, xavier2018equipartition,Bonsignori:2019naz, Oblak:2021nbj} it is noticed that for several $(1+1)$-dimensional systems, the symmetry resolved Renyi entropy and entanglement entropy is the same in the different charges sectors and is proportional to area of the entangling region, i.e., the entanglement does not depend on the charge and distributes equally on each sector.

%%%%%%%%%%%%%%%%%%%%%%%%%%%%
%%%%%%%%%%%%%%%%%%%%%%%%%%%%
\subsection{Holographic Entanglement entropy }\label{SubSec:HolographyRenyi}

The AdS/CFT correspondence \cite{Maldacena:1997re, Gubser:1998bc, Witten:1988hc} states that exist dual formulations for certain quantum field theories and gravitational theories on AlAdS spacetimes. Particularly, the GPKW relation states that the supergravity approximation of the string partition function in the Euclidean regime, corresponds to the generating functional of connected diagrams of a CFT, viz.
\begin{align}\label{GPKW}
    \exp\left\{-I_{\rm grav}[g_{(0)ij},\phi_{(0)}^I]\right\} = \left\langle \exp\left\{  \int_{\partial \cal M}\d^dx \sqrt{-g_{(0)}}\left( -\frac12 g_{(0)ij}T^{ij} + \phi_{(0)}^I{\cal O}_I\right)\right\}\right\rangle_{\rm CFT}~, 
\end{align}
where $I_{\rm grav}$ is the gravitational theory with $\phi_{(0)}^I$ and $g_{(0)ij}$ being, respectively, the boundary conditions on the set of fields $\phi^I$ and the metric $g_{\mu\nu}$ of the $d+1$ dimensional bulk $\cal M$~. Then, the asymptotic values of the fields in the string theory side source gauge operators ${\cal O}^I$ at the $d$-dimensional conformal boundary $\partial \cal M$ of an AlAdS$_{d+1}$ spacetime. For instance, the asymptotic value of a bulk scalar field on an AlAdS$_{d+1}$ background sources a dual scalar operator $\cal O$ of a CFT on the fixed $d$-dimensional boundary coordinatized by $g_{(0)ij}$~. 

In this approximation, the gravitational theory are dominated in the classical regime by Einstein theory coupled to matter, viz.
\begin{align}
    I_{\rm grav} = \frac{1}{2\kappa}\int_{\cal M}\d^{d+1}x\sqrt{g}\left(R - 2\Lambda + {\rm matter}\right)~,
\end{align}
where $\kappa := 8\pi G_{d+1}$ with $G_{d+1}$ is the $(d+1)$-dimensional Newtons constant, and cosmological constant is negative, which in terms of the AdS characteristic radius
\begin{align}
    \Lambda = - \frac{d(d-1)}{2\ell^2}~.
\end{align}
the gravitational theory must be free from infrared (IR) divergences. However, in gravity theories within Anti-de Sitter (AdS) spacetimes, the unbounded volume of the hyperbolic manifold leads to long-range interaction divergences. Thus, to regularize the Euclidean action, additional boundary terms are necessary. This regularization involves introducing an IR cutoff $\delta$ and subsequently considering the IR limit in the final computation stage. This regularization technique, known as \emph{holographic renormalization}, has been extensively explored in literature such as \cite{Henningson:1998gx,deHaro:2000vlm,Skenderis:2002wp}, where the Fefferman--Graham expansion \cite{Fefferman:1984asd} of the fields is utilized to derive covariant intrinsic counterterms.

Additionally, an extra boundary term is required to ensure the well-posedness of the variational principle. For pure gravity, the action takes the form:

\begin{align}
I_{\rm ren} = I_{\rm grav} + \frac{1}{\kappa}\int_{\partial \mathcal{M}}\d^dx \sqrt{h}K + \frac{1}{\kappa}\int_{\partial \mathcal{M}}\d^dx \sqrt{h}L_{\rm ct}(h,{\cal R},\nabla{\cal R})~,
\end{align}

Here, $h_{ij}$ represents the induced metric at $r=\delta$, $K = h^{ij}K_{ij}$ denotes the extrinsic curvature with respect to the outward-pointing normal to $\mathcal{M}$, ${\cal R}$ signifies the intrinsic Ricci curvature of the boundary, and

\begin{align}
L_{\rm ct}(h,{\cal R},\nabla{\cal R}) = \frac{d-1}{\ell} + \frac{\ell}{2(d-2)}{\cal R} + \frac{\ell^3}{2(d-2)^2(d-4)}\left({\cal R}_{ij}{\cal R}^{ij} - \frac{d}{4(d-1)}{\cal R}^2\right)+\dots~,
\end{align}

The second term appears only if $d\geq 3$, the third term if $d\geq 4$, and higher-order terms in the curvature emerge in higher dimensions.

\begin{figure}[h!]
    \centering
    \includegraphics{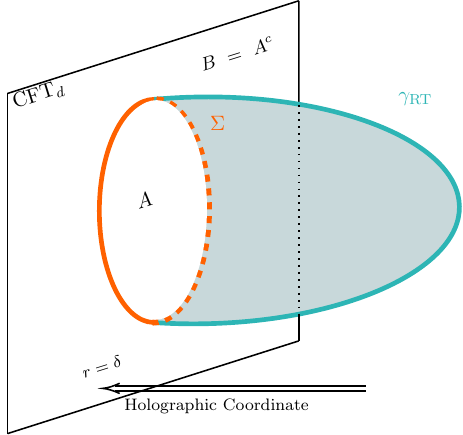}
    \caption{The RT minimal surface extends to the bulk of AdS and is its boundary coincides with the entangling region.}
    \label{fig:RT}
\end{figure}

For entanglement entropy, Ryu and Takanagi (RT) \cite{Ryu:2006bv, Ryu:2006ef} conjectured that the holographic entanglement entropy of a CFT can be obtained as
\begin{align}
    S_{\rm E} = \frac{{\rm Area}_{\gamma_{\rm RT}}}{4G_{d+1}}~,
\end{align}
where $\gamma_{\rm RT}$ is a bulk codimension-two minimal surface that is anchored at the entangling surface, see \autoref{fig:RT}.

The RT conjecture was proven by Lewkowycz and Maldacena \cite{Lewkowycz:2013nqa}, has been widely studied. For further details, see, for instance, \cite{Rangamani:2016dms, Nishioka:2018khk}  and references therein. Then, the computation of entanglement on the dual CFT corresponds to minimization problem (modulo boundary conditions) at the AdS bulk. 
One can also use the path integral representation of the computation of entanglement by means of the GPKW relation \eqref{GPKW} to find the entanglement entropy of a CFT with a gravity dual theory as\footnote{For a discussion of renormalization of holographic entanglement entropy, see \cite{Anastasiou:2019ldc} and refernce therein.}

\begin{align}
    S_{\rm E} = \lim_{n\to1}\partial_n\left(I_{\rm ren}[{\cal M}_n]-n I_{\rm ren}[{\cal M}]\right)~.
\end{align}
The formula above needs a solution ${\cal M}_n$ whose boundary $\partial{\cal M}_n$ is the $n$-fold cover manifold with conical deficit at the entangling surface used in the replica trick. The solution ${\cal M}_n$ must be such that $\partial {\cal M}_n = {\cal C}_n$ and that the modular time extends to the bulk. In order to have a such bulk solution, we need to assume that the bulk solution 
is invariant under the replica symmetry that shifts the modular time by $2\pi$~. The entangling surface corresponds to the fixed loci of the $\mathbb{Z}_n$ action on ${\cal C}_n$~, and our assumptions implies that this action extends to a codimension-two hypersurface $\gamma_{\rm RT}^n$ that is anchored to the entangling region $\Sigma$ at the boundary of the n-fold cover ${\cal M}_n$~. 
The extended bulk replica symmetry allows to define the orbifold
\begin{align}
    \widetilde{\cal M}_n = {\cal M}_n/\mathbb{Z}_n~,
\end{align}
that generates a conical singularity at the RT surface ${\gamma}_{\rm RT}$ with deficit angle given by \eqref{deltan} respects to a rescaled modular time $\widetilde{\tau} = \tau/n$~. 
We can introduce the bulk-per-replica action 
for the orbifold
\begin{align}
    \widetilde{I}[\widetilde{\cal M}_n]=I[{\cal M}]/n~,
\end{align}
such that
\begin{align}\label{rtI}
    S_{\rm E} = \lim_{n\to1} \partial_n \left[n\left(\widetilde{I}[\widetilde{\cal M}_n] - I[{\cal M}]\right)\right] = \partial_n \widetilde{I}[\widetilde{\cal M}_n]\big\rvert_{n=1}~.
\end{align}
In \cite{Lewkowycz:2013nqa} it is shown that the bulk-per-replica action contains terms that are proportional to the RT surface area, reproducing the RT conjecture. Notice that $\widetilde{I}[\widetilde{\cal M}_n]$ has contributions from the conical singularities generated by the orbifold at the RT surface. The orbifold $\widetilde{\cal M}_n$ is made by gluing $n$ replicas obtained by cutting ${\cal M}$ along a codimension-one hypersurface which ends on a codimension-two hypersurface $\Sigma$~. When the replicas are glued together in $\widetilde{\cal M}_n$, one gets a conical singularity on $\Sigma$~. In \cite{Fursaev:1995ef, Fursaev:2013fta} it is shown that 
\begin{align}
    \int_{\widetilde{\cal M}_n}\d^{d+1}x \sqrt{g}R = \frac{1}{n}\int_{\cal M}\d^{d+1}x\sqrt{g}R + 4\pi\left(1-\frac{1}{n}\right){\rm Area}_\Sigma + {\cal O}\left((n-1)^2\right)~.
\end{align}
Applying the result to Einstein gravity \cite{Dong:2016fnf}, 
\begin{align}\label{repI}
    \widetilde{I}[\widetilde{\cal M}_n] = I_{\rm grav}[\widetilde{\cal M}_n] + \frac{1}{4G_{d+1}}\left(1-\frac{1}{n}\right){\rm Area}_{\gamma^n_{\rm RT}}~,
\end{align}
where 
\begin{align}
{\rm Area}_{\gamma_{\rm RT}^n} = \int_{\gamma_{\rm RT}^n}\d^{d-1}x \sqrt{\gamma}~,    
\end{align}
with $\gamma_{ab}$ the induced metric at the extended RT surface. In \cite{Dong:2016fnf}, the second term in \eqref{repI} was interpreted as the action of a codimension-two cosmic brane, with tension $(1-1/n)/4G_{d+1}$~. Then, \eqref{rtI} reproduces on-shell the RT conjecture as ${\rm Area}_{\gamma_{\rm RT}^1} = {\rm Area}_{\gamma_{\rm RT}}$~. Indeed, the surface ${\rm Area}_{\gamma_{\rm RT}^1}$ is automatically minimal, as in the $n\to1$ limit the cosmic brane tension vanishes and the surface in this limit is simply a solution of the fields equations for the area functional $\delta {\rm Area}_{\gamma_{\rm RT}^1}=0$ given a bulk background~. For an arbitrary replica parameter the action \eqref{repI} can be used to get the holographic Renyi entropy and the associated modular system in terms of the cosmic brane area, and the Renyi inequalities \eqref{Rineq} are mapped to constraints on the area of the cosmic brane \cite{Nakaguchi:2016zqi} as, modulo field equations,
\begin{align}
    \partial_n \widetilde{I}[\widetilde{\cal M}_n] = \frac{{\rm Area}_{\gamma_{\rm RT}^n}}{4 G_{d+1}}\frac{1}{n^2}~,
\end{align}
reproducing the RT entanglement entropy in the $n\to1$ limit.

As previously discussed, the case of spherical entangling region, $\Sigma = S^{d-1}$~, there is a simplification on the CFT side due to the CHM map. We can obtain the holographic entanglement entropy using the RT formulation. Let us consider the Euclidean AdS$_{d+1}$ spacetime in Poincar\'e coordinates and polar coordinates for the transverse section, viz.,
\begin{align}
    ds^2 = \frac{\ell^2}{z^2}\left(\d z^2 + \d t_{\rm E}^2 + \d r^2+r^2\d \Omega^2_{d-2} \right)~,
\end{align}
where $z=0$ corresponds to the conformal boundary, $t_{\rm E}$ to the Euclidean time. Then, at fixed time, there is a sphere of radius $r=R$ at the conformal boundary. This shape is defined by all points $y_a$~, with $a = 1 ,\dots , d-2$~, such that $r \leq R$~. The induced metric of the RT surface
\begin{align}
    h_{ab} = g_{\mu\nu}\frac{\partial x^\mu}{\partial y^a}\frac{\partial x^\nu}{\partial y^b}~,
\end{align}
such that
\begin{align}
{\rm Area}_{\gamma_{\rm RT}} = \ell^{d-1}{\rm Vol}_{S^{d-2}}~{\rm min}\int^{R}_0 \d r L_{\rm sp}=\ell^{d-1}{\rm Vol}_{S^{d-2}}~{\rm min}\int^{R}_0 \d r \frac{r^{d-2}}{z^{d-1}}\sqrt{1+\left(\partial_r z\right)^2}~.
\end{align}
Minimization of this area corresponds to solve the Euler--Lagrange equations with $r$ treated as the Hamiltonian time,
\begin{align}
    \frac{\partial L_{\rm sp}}{\partial z} - \frac{\d }{\d  r}\left(\frac{\partial L_{\rm sp}}{\partial z'}\right) = 0~.
\end{align}
Although the problem has no conservation laws, an exact solution to field equations was found in \cite{Ryu:2006bv}
\begin{align}
r^2+z^2=R^2~,
\end{align}
for any dimension. Introducing $y = 1+r/R$ and a cutoff $\delta$~,the area becomes
\begin{align}
    {\rm Area}_{\gamma_{\rm RT}} = \ell^{d-1}{\rm Vol}_{S^{d-2}}~{\rm min}\int^{1}_{\delta/R}~ \d y\frac{(1-y^2)^{\frac{d-3}{2}}}{y^{d-2}}~,
\end{align}
which is divergent near $y=0$~. Expanding around this singular point, the leading order contribution to the entanglement entropy reads
\begin{align}
    S_{\rm E} = \frac{\ell^{d-1}}{4G_{d+1}}\left(\frac{R^{d-2}{\rm Vol}_{S^{d-2}}}{(d-2)\delta^{d-2}} + \dots\right)~,
\end{align}
which shows that the leading UV divergence in the entanglement entropy is proportional to the area of the entangling region.
We can obtain the same result using the CHM map. Conformally mapping the CFT with an spherical entangling region to a CFT in the hyperbolic disk of radius $\cal R$ at temperature $T_0=(2\pi {\cal R})^{-1}$~,  the Renyi entropy can be computed through \eqref{Snthermal} \cite{Hung:2011nu}. The holographic dual of this configuration corresponds to a black hole with an hyperbolic horizon with Hawking temperature $T_0$~.   
Eq. \eqref{Snthermal} imposes a constraint on the horizon radius as the temperature depends on $r_h$. The black hole line element
\begin{align}
    ds^2 = -f(r) \frac{\ell^2}{{\cal R}^2} \d t^2 + \frac{\d r^2}{f(r)} + r^2 \d\Sigma_{d-1}^2 ~,
\end{align}
where $\d \Sigma_{d-1}^2$ is the line element of the $(d-1)$-hyperbolic plane, and at $r\to\infty$~, the boundary metric is conformal to
\begin{align}
    ds_{(0)}^2 = -\d t^2 + {\cal R}^2 \d \Sigma_{d-1}^2~,
\end{align}
which is exactly the metric in hich the CFT lies after applying the CFT map. The metric lapse function reads
\begin{align}
    f(r) = \frac{r^2}{\ell^2} - 1 - \frac{m}{r^{d-2}}~,
\end{align}
where $m$ is an integration constant that can be related with the black hole mass \cite{Belin:2013uta}. We can rewrite the metric function in terms of the horizon radius as
\begin{align}
f(r) = \frac{r^2}{\ell ^2}-1 + \left(\frac{r_h}{r}\right)^{d-2}\left(1-\frac{r_h^2}{\ell^2}\right)~,
\end{align}
and the horizon's temperature 
\begin{align}
    T = \frac{T_0}{2}\ell f'(r_h) = \frac{T_0}{2x}\left(dx^2 - (d-2) \right)~,
\end{align}
where
\begin{align}
    x := \frac{r_h}{\ell}~.
\end{align}
The Wald entropy \cite{Wald:1993nt} of this black hole is simply given by the famous horizon's area formula
\begin{align}
S_{\rm thermal} = \V \left(\frac{r_h}{\ell}\right)^{d-1}~,\qquad \V := \frac{2\pi}{\kappa} \ell^{d-1} V_\Sigma~,
\end{align}
where 
\begin{align}
V_\Sigma \sim \frac{\Omega_{d-2}}{d-2}\frac{{\cal R}^{d-2}}{\delta^{d-2}}+\dots~,
\end{align}
is the horizon's hyperbolic volume that must be regularized with an IR cutoff $\delta$ (see \autoref{Sec:Fluctutations} for further details). Then, the Renyi entropy \eqref{Snthermal} from the point of view of the bulk becomes
\begin{align}
    S_n = \frac{n}{n-1}\int^{x_1}_{x_n} \d x~S(x)\partial_x T(x)~,
\end{align}
where $x_n$ is the largest solution to 
\begin{align}\label{xn}
T(x_n) = T_0/n~, 
\end{align}
which in terms of the black hole temperature reads
\begin{align}
    dnx_n^2 - 2x_n - n(d +2) =0 \Longrightarrow x_n =  \frac{1}{dn}\left(1+\sqrt{1+d(d-2)n^2}\right)~.
\end{align}
Then, the holographic Renyi entropy becomes \cite{Hung:2011nu}
\begin{align}
    S_n = \frac{\V}{2}\frac{n}{n-1}\left(x_1^{d-2}-x_n^{d-2} + x_1^d -x_n^d\right)~,
\end{align}
with limits
\begin{align} 
S_0 ={}& \frac{\V}{2}\left(\frac{2}{d}\right)^d\frac{1}{n^{d-1}}~, \\ S_{\rm E} ={}& \frac{\V}{2}\frac{(d-2)x_1^{d-2}}{dx_1-1}\left(1+\frac{dx_1^2}{d-2}\right)~, \label{Slimits}\\S_\infty  ={}& \frac{\V}{2}\left(x_1^{d-2}-x_\infty^{d-2} + x_1^d - x_n^d\right)~,
\end{align}
with
\begin{align}
    x_1 = \frac{d-1}{d}~,\qquad x_\infty = \sqrt{\frac{d-2}{d}}~.
\end{align}

As can be seen from \eqref{Slimits}, the entanglement entropy has the same values as the one obtained throughout the RT formula.

If the CFT has a global $U(1)$ symmetry, then we would need to charge the black hole such that the gauge symmetry in the bulk is mapped to the global symmetry at the boundary. In \autoref{Sec:Fluctutations} we consider such scenario, and generalize to NLED theories.

%%%%%%%%%%%%%%%%%%%%%%%%%%%%%%%%%%%%%%%%%%%%%%%%%
%%%%%%%%%%%%%%%%%% de Sitter %%%%%%%%%%%%%%%%%%%%
%%%%%%%%%%%%%%%%%%%%%%%%%%%%%%%%%%%%%%%%%%%%%%%%%
\newpage
\section{Non-flat Entanglement Spectrum in de Sitter}\label{Sec:dS}

Our Universe is expanding at an accelerating pace \cite{SupernovaSearchTeam:1998fmf, SupernovaCosmologyProject:1998vns}. The small value of the (positive) cosmological constant is responsible for the positive vacuum pressure which prevents an internal observer to have access to the full spacetime establishing a \emph{cosmological horizon} (also known as \emph{Rindler horizon}). The horizon area sets the size of the observable Universe, and acts as a barrier that prevents retrieve of information and observer does not have access to data at the spatial slice ${\cal I}^+$ at future infinity. An spacetime with such observable constrains, is characterized by the Friedmann--Lemaitre--Robertson--Walker (FLRW) metric with a time-dependent scale factor describing the expansion.
Current observations indicate that our Universe may have go through an inflationary epoch short after the Big Bang. During the inflation era, the Universe exponentially expanded cooling the spacetime drastically and ends with the reheating period. If the current accelerating pace does not stop, our universe will have a similar exponential expansion period at late times. This indicating that our universe can be asymptotically described at early and late times, by a FLRW with exponential scale factor. Such spacetime is known as de Sitter space (dS) and is a fundamental solution of Einsteins equations in Cosmology. 

In \cite{Gibbons:1976ue}, using Euclidean Quantum Gravity methods, it is shown that dS space has thermal properties that resemble the one of black holes. The cosmological horizon has a temperature and a conjugated entropy. An universal formulation of the microscopic description of this thermal feature is still lacking and is one of the fundamental problems in Quantum Gravity. 
The most consisting framework to study Quantum Gravity is String theory. Although is a theory which consistently unifies all fundamental interactions, due to the large gap between the electroweak scale and the string scale is hard to understand and test the phenomenological implications of the theory. Moreover, there are a myriad amount of vacua in the theory \cite{Bousso:2000xa, Douglas:2003um, Susskind:2003kw}, which makes the theory rich enough to have a plethora of different applications, for instance, through the AdS/CFT correspondence, but also makes hard to distinguish which consistent Quantum Gravity theories can be constructed from String Theory.
This is a central question in Quantum Gravity, and the \emph{swampland program} (see for instance \cite{Vafa:2005ui, Ooguri:2006in, Agmon:2022thq}) is a web of far-reaching and consequential conjectures and criteria to catalogue which UV complete gravitational theories appearing consistently as effective low-energy theories of String theory.
Particularly, due to the difficulty of constructing a stable dS vacuum solution \cite{Hull:1998vg, Maldacena:2000mw, Kachru:2003aw, Balasubramanian:2005zx, Westphal:2006tn,Hertzberg:2007wc,Hertzberg:2007wc,Covi:2008ea,Cicoli:2015ylx, Danielsson:2018ztv}, it has been argued \cite{Ooguri:2018wrx,Kachru:2003aw} that dS spacetime\footnote{See \cite{Montero:2022ghl} for a recent discussion on supersymmetric Ads spacetimes and the string swampland.} belongs to the swampland\footnote{See \cite{Geng:2019bnn,Geng:2019zsx} for a discussion on the implications of the dS/dS holographic bound \cite{Alishahiha:2004md} on the string swampland.}.  Banks \cite{Banks:2000fe} argued using M-theory methods that the Hilbert space of Quantum Gravity in assymptotically dS spacetime has finite dimension $N_{\cal H}$ given by the exponential of the quarter of the area of the cosmological horizon, viz.
\begin{align}
    N_{\cal H} = \exp\left\{\frac{{\rm Area}_{\cal H}}{4G_{d+1}}\right\}~.
\end{align}
This indicates that Einsteins classical theory in dS space can not be quantized and there must exist a more fundamental theory in which the cosmological constant is determined. 

In this section, we will study the problem of the boundedness of the dimensionality of the Hilbert space using the notion of entanglement in dS developed in \cite{Arias:2019zug, Arias:2019pzy}.  

\subsection{Deforming de Sitter Spacetime}
Four-dimensional de Sitter spacetime (dS), is the maximally symmetric solution to vacuum Einstein equations with positive cosmological constant $\Lambda = 3/\ell^2$ given by the induced metric on the hyperboloid\footnote{In this section, we use the symbol $\ell$ to denote the dS radius, which was previously employed for AdS. This is because the dS scenario is exclusively addressed here, without any reference to AdS spacetime.}

\begin{align}\label{hyperboloid}
    X_M X^M = \ell^2~,\qquad M = 0,1,\dots,4~,
\end{align}
in five-dimensional Minkowski spacetime
\begin{align}
    \d s^2 = -\d X_0^2 + \sum_{i=1}^4 \d X_i^2~.
\end{align}
The global geometry has an isometry group $O(1,3)$ and can be viewed as a foliation along a global time, whose space-like leaves are given by round 3-spheres, together with two conformal boundaries at space-like infinities ${\cal I}^\pm$. The intrinsic line elment can be coordinatized as
\begin{align}
    \d s^2 = -\d T^2 + \ell^2\cosh^2(T/\ell)\d\Omega_3^2~,
\end{align}
where $T \in (-\infty,\infty)$ is the global time coordinate, and $\d\Omega_3^2$ is the metric of the round, unit 3-sphere.  
The global dS spacetime corresponds to a FLRW universe with positivily curved spatial slice and an exponential scale factor. The exponential inflation prevents light rays to propagate to the whole spacetime. Therefore, a static observer (often referred to as a Rindler observer) has access only to a sub-region of dS referred to as the Rindler wedge or Rindler patch, and is enclosed by a cosmological (Killing) horizon. 

\begin{figure}
    \centering
    \includegraphics{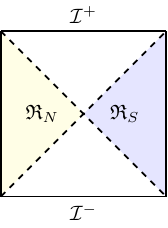}
    \caption{Penrose diagram of the extended dS space with two causal diamonds ${\mathfrak{R}}_N$ and ${\mathfrak{R}}_S$. Here $\mathcal{I}^+$ and $\mathcal{I}^-$ corresponds to the past and future conformal infinities of the global geometry, respectively.}
    \label{fig:rindler}
\end{figure}

Treating dS path integral in the Euclidean regime, Rindler patches are mapped into round $S^4$ upon Wick rotating the oberver time coordinate, and the Euclidean field theory has, respectively, an associated Hawking temperature \cite{Figari:1975km} and conjugated GH entropy\cite{Gibbons:1977mu}
\begin{align}\label{SGH}
    T_{\rm H}  = \frac{1}{2\pi\ell}~,\qquad S_{\rm GH} = \frac{A_{\cal H}}{4G_4}~,
\end{align}
where ${{\rm Area}_{\cal H}} = 4\pi\ell^2$ is the area of the cosmological horizon. 
Remarkably, the Euclidean thermodynamics of the static patch of dS spacetime resembles the one of black holes, although these are observer-dependent quantities. 
Black hole entropy has nowadays been understood as to account for the number of the black hole quantum states, in contrast with dS spacetime, where such interpretation is still not achieved as there is no notion of microstates associated with the cosmological horizon\footnote{See \cite{Maldacena:1998ih} for a mechanical statistical interpretation of three-dimensional dS entropy using a $SL(2,\mathbb{C})$ Chern--Simons formulation of 3d gravity.}. Then, the underlying quantum theory and the interpretation of the GH entrpoy remains an open puzzle (see \cite{Banks:2000fe,Witten:2001kn, Balasubramanian:2001rb} for further discussions).
In the case of AdS, the holographic proposal for entanglement entropy makes even a further understanding of the BH entropy as it is recovered by the holographic entanglement when the Ryu--Takayanagi surface coincides with the black hole horizon. 
Similarly, one could try to understand the GH entropy in terms of entanglement between gravitons inside and outside the static patch. Several approaches to understand dS entanglement (with and without holography) have been proposed \cite{Dong:2018cuv, Arias:2019pzy, Arias:2019zug, Narayan:2019pjl, Shaghoulian:2022fop, Chandrasekaran:2022cip ,Cotler:2023xku}, founding a flat entanglement spectrum\footnote{This issue also appears in the AdS/CFT holographic entanglement entropy in the leading order of the $1/G_{d+1}$ expansion, and has been solved by summing over possible RT surfaces in the bulk \cite{Dong:2018seb}.} (see \cite{Banks:2022irh} for a discussion) indicating a maximally mixed density matrix that becomes proportional to the identity operator, such that the Renyi entropy becomes constant by means of the Renyi index and equaling the GH entropy. 
We will consider horizon's fluctuations using CFT methods to generate a non-flat Renyi spectrum and study how it affects the dimensionality of the dS Hilbert space. 
In order to do so, we will follow the approach of \cite{Arias:2019pzy, Arias:2019zug} (see \cite{Diaz:2019khq, Diaz:2019khq} for further details and applications), where the entanglement between disconnected observers has been computed by considering an extending coordinate patch that contains both causally disconnected Rindler observers, and consider the reduced density matrix of one of them. This is achieved by letting $\mathbb{Z}_n$ act on the extended patch producing an orbifold with two antipodal conical defects, such that the original smooth dS geometry is achieved in the $n\to 1$ limit, referred to as the \emph{tensionless limit}. 
The extended static patch is a fibration over $S^2$ with fibers along the equator given by points, and in the northern and southern hemispheres given by radially extended Rindler patches and can be coordinatized by 
\begin{align}\label{dsextended}
    ds^2 = \cos^2\theta \d \Sigma^2 + \ell^2 \d\Omega_2^2~, 
\end{align}
where
\begin{align}\label{dSigma}
    \d\Sigma^2 = -f(r)\d t^2 + \frac{\d r^2}{f(r)}~,\qquad f(r)=1-\frac{r^2}{\ell^2}~,
\end{align}
corresponds to the two-dimensional extended Rindler observer line elements. The embedding coordinates that satisfy the hyperboloid constraint \eqref{hyperboloid} and renders the induced metric to \eqref{dsextended} are
\begin{align}\label{embe}
    X_0 =\sqrt{\ell^2-r^2}\cos\theta\sinh(t/\ell)~,\qquad X_1 = \sqrt{\ell^2-r^2}\cos\theta\cosh(t/\ell)~,\end{align}\vspace{-10mm}
    \begin{align}\nonumber 
    X_2 = r\cos\theta~,\qquad  X_3=\ell\sin\theta\cos\phi~,\qquad X_4 = \ell\sin\theta\sin\phi~,
\end{align}
with
\begin{align}
   t\in \mathbb{R}^1~,\qquad r \in (-\ell,\ell)~,\qquad \theta \in [0,\pi)~,\qquad \phi \in [0,2\pi)~,
\end{align}
covering both statics patches with a single coordinate system as shown in \autoref{fig:rindler}.

\begin{figure}
    \centering
    \includegraphics{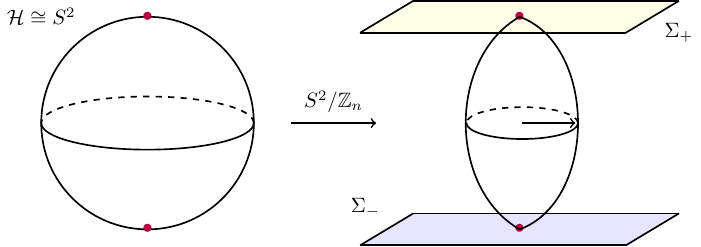}
    \caption{The horizon's spherical topology deforms to a spindle after the orbifold procedure. The set of fixed points of the $\mathbb{Z}_n$ action corresponds to $\Sigma_\pm$ at $\theta=0,\pi$, respectively. }
    \label{fig:orbi1}
\end{figure}

Let us now consider the orbifold
\begin{align}
    {\rm dS}_n := {\rm dS}/\mathbb{Z}_n~, 
\end{align}
coordinatized by
\begin{align}
    \d s^2 = \cos^2\theta \d\Sigma^2 + \ell^2\left(\d\theta^2+\sin^2\theta\frac{\d\phi^2}{n^2}\right)~,
\end{align}
with $n$-fold branched cover with a natural $\mathbb{Z}_n$ action, and two set of fixed points forming a pair $(\Sigma_+,\Sigma_-)$ of antipodal sub-manifolds of codimension-two, referred to as \emph{defects}, each with induced metric given by \eqref{dSigma}. The orbifolding of the horizon is illustrated in \autoref{fig:orbi1}

The ${\mathbb{Z}}_n$ action generates $\delta^2$-function singularities in the Riemann curvature tensor \cite{Fursaev:1995ef} generating a conical deficit $2\pi(1-1/n)$ along the horizon's azimuthal angle. 
|14Well-posedness of the variational principle requieres to supplement the gravity action with extra terms localized at the defects, providing a singular stress tensor cancelling singularities appearing in the bulk Einstein tensor, such that field equations are everywhere unambiguous. 
In the case of dS gravity, the Einstein--Hilbert action
\begin{align}
    I_{\rm EH} := \frac{1}{16\pi G_4}\int_{\rm dS}\d^4x \sqrt{-g}\left(R+\frac{6}{\ell^{2}}\right)~,
\end{align}
must be supplemented with the action of defects, that can be dressed by fixed Nambu--Goto actions with a $n$-dependent tensions, i.e.,
\begin{align}
    I_{\rm n} = \sum_{\xi = \pm}I_{\xi}~,\qquad I_\xi := T_{n_\xi}\int_{\Sigma_\xi}\d^2s \sqrt{-\sigma_\xi}~,
\end{align}
where $\Sigma_\pm$ correspond to the defect geometries endowed with metrics $\sigma_{\pm}$ and  
\begin{align}
    T_{n_\xi} := -\frac{1}{4G_4}\left(1-\frac{1}{n_\xi}\right)~,
\end{align}
the tensions. In the case at hand $n_+  = n =n_-$~. 
Then,
\begin{align}
    I_{\rm{dS}_n} := I_{\rm EH} + I_{\rm n}~,
\end{align}
becomes well-defined with a smooth tensionless limit $n\to1$ which recovers Einstein--Hilbert action in dS spacetime. 
The orbifolded double Rindler wedge has a cosmological horizon given by a bifurcated Killin horizon $S^2$-topology at $r = \pm \ell$, with a $n$-independent surface gravity matching the Hawkin dS temperature, but now the conjugated entropy has an extra $1/n$ factor respect to the GH entropy, i.e.,
\begin{align}
    S_{{\rm GH}, n} = \frac{1}{n}S_{\rm GH}~. 
\end{align}
In \cite{Arias:2019pzy}, each Rindler observer is assigned with a $n$-independent modular Hamiltonian $H_{\cal R}$ with modular temperature $T= 1/n$ conjugate to $H_{\cal R}$ such that the reduced density matrix $\rho_{\cal R} := \exp\{-H_{\cal R}\}$ gives the thermal partition function
\begin{align}
    Z_{\cal R} := {\rm tr}_{H_{\cal R}}\rho_{\cal R}^n = {\rm tr}_{H_{\cal R}}\exp\{-n H_{\cal R}\}~, 
\end{align}
computed on the $n$-fold branched cover of the Euclideanized orbifold
\begin{align}
    S_n^4 = S^4/\mathbb{Z}_n~.
\end{align}
We can now use the Calabrese--Cardy formula \cite{Calabrese:2004eu}
\begin{align}
    Z_{\cal R} = \frac{Z[S_n^4]}{Z^n[S^4]}~,
\end{align}
where $Z[S_n^4]$ is computed on the $n$-fold, branched, Euclidean cover geometry of a single observer, which is constructed by gluing together $n$ copies of the spacetime along the entangling Region $\partial({\cal R}_+ \cap {\cal R}_-)$~, which yields a branched geometry with a smooth metric with natural $\mathbb{Z}_n$ action whose ramification surface equals the entangling surface, see \cite{Arias:2019pzy, Arias:2019lzk} for details. Such maneuver is often referred to as the Replica trick. 
Then, we can compute the Renyi entropy as
\begin{align}
    S_n = \frac{1}{1-n}\left(\log Z[S_n^4]-n\log Z[S^4] \right)~,
\end{align}
which assuming locality of the gravity action and considering the first order of the Euclidean saddle approximation we get
\begin{align}
    S_n = 2S_{\rm GH}~,\qquad \forall n\geq 1~,
\end{align}
which is $n$-independent, indicating a flat entanglement spectrum. The entanglement limit $n\to1$ recovers the original smooth geometry and the GH entropy, indicating that the entropy of the cosmological horizon measures the entanglement between disconnected Rindler observers in the undeformed dS spacetime. 
The extension of the static patch allows to understand the entanglement in terms of the non-trivial topology of dS, and interpret the Rindler observer quantity in terms of a modular system. Then, the Renyi entropy limits, see \autoref{Sec:Entanglement}, give information of the quantum theory of a single Rindler observer.
In \cite{Arias:2019zug}, an effective description of the defect quantum theory has been given in terms of the Liouville CFT with modular temperature $T=n^{-1}$ whose Cardy entropy has been matched with the theories modular free energy.

\subsection{Near-Horizon Symmetry Algebra }
As shown by Carlip\footnote{See also \cite{Solodukhin:1998tc} for a related early result.} \cite{Carlip:1999cy}, the constraint algebra of General Relativity may acquiere a computable central extension when the spacetime manifold has a boundary. Morevoer, when the boundary corresponds to a Killing horizon, the assymptotic symmetry algebra corresponds to a single copy of the Virasoro algebra, indicating a CFT$_2$ description of Killing horizons. 
Then, one can use the Cardy entropy \cite{Cardy:1986ie} to acount for the number of microstates associated with a Killing horizon modulo computing the Virasoro zero modes eigenvalues. 
The validity of the Cardy entropy for Killing horizons remains an open problem as the underlying CFT may be chiral\footnote{Similarly, the Kerr/CFT correspondence uses the Cardy entropy for a CFT described in terms of a single copy of the Virasoro algebra, reproducing the correct BH entropy \cite{Guica:2008mu}, see \cite{Bredberg:2011hp, Compere:2012jk} for a discussion on the validity of the Cardy entropy in the Kerr/CFT correspondence.}. 
We follow to assign the $n$-deformed Rindler patch classes of boundary conditions at the cosmological horizon inducing Virasoro modules with a finite computable central charge when $n>1$~. In order to compute the associated asymptotic charges, we use the spacetime covariant Barnich--Brandt formalism \cite{Barnich:2001jy} which allows to compute conserved charges ${\mathfrak Q}$ which are integrable and finite for all suitable boundary conditions. In the case of General Relativity, an exact variation of the charges obeys
\begin{align}
    \delta{\mathfrak Q}_\zeta [g] =  \int_{S} k_\zeta^{\rm Einstein} [\delta g,g]~,
\end{align}
where $S$ is a codimension-two surface of integration, $\zeta$ 
 is an asymptotic Killing vector\footnote{A Killing vector $\zeta$ generates isometries and preserves a given metric structure by ensuring that the Lie variation along this vector field vanishes when acting on the metric tensor, expressed as ${\cal L}_\xi g = 0$.
An asymptotic Killing vector does not preserve the full metric but rather maintains the asymptotic structure. An asymptotic Killing vector field $\zeta$ is defined by
\begin{align}
{\cal L}_\zeta g = {\cal O}(\delta g)~,
\end{align}
where the metric $g$ follows specific boundary conditions, and $\delta g$ represents metric fluctuations allowed by these conditions.

The set of asymptotic vector fields gives rise to an asymptotic symmetry group, formed by the quotient of allowed diffeomorphisms by trivial ones, along with associated well-defined surface charges.}, $g$ is the background metric with an associated fluctuation $\delta g$, and\footnote{Tensor indices are raised and lowered with the background metric $g_{\mu\nu}$~.}
\begin{align}
    k_\zeta^{\rm Einstein}[h,g] = -\frac{1}{8\pi G}\frac{1}{4}\d x^\alpha \wedge \d x^\beta \epsilon_{\alpha\beta\mu\nu} \Big[& \zeta^\nu\nabla^\mu h^\rho{}_\rho - \zeta^\nu\nabla_\rho h^{\rho\mu} + \zeta_\rho \nabla^\nu h^{\mu\rho} \nonumber\\ &+\frac12 h^\rho{}_\rho\nabla^\nu\zeta^\mu - h^{\nu\sigma}\nabla_\rho\zeta^\mu + \frac12h^{\rho\nu}\left(\nabla^\mu\zeta_\rho + \nabla_\rho\zeta^\mu\right)\Big]~,
\end{align}
is the covariant two-form surface charge for Einstein gravity\footnote{There is an ambiguity in define the surface charge form coming from an extra freedom appearing in the definition of the presymplectic structure of the gravitational theory \cite{Iyer:1994ys}. Then, it is possible to use a different definition for charges leading to similar results in the definition of the asymptotic symmetry algebra such as \cite{Lee:1990nz, Abbott:1981ff, Silva:2002jq}.}. Extending the Lie bracket algebra of charges\footnote{Up to central charges, the Lie bracket algebra of charges $[\mathfrak{Q}_m,\mathfrak{Q}_n]:={\cal L}_{\xi_m}\mathfrak{Q}_n$ and the Lie bracket algebra of vector fields are isomorphic \cite{Brown:1986nw}.} to a Dirac bracket algebra of the conserved charges represent the asymptotic symmetry algebra up to a central term of the form \cite{Barnich:2001jy}
\begin{align}
    \left\{\mathfrak{Q}_\zeta,\mathfrak{Q}_{\zeta'}\right\} = \mathfrak{Q}_{[\zeta,\zeta']} + \int_{S}k_\zeta[{\cal L}_{\zeta'}g,g]~,
\end{align}
where $[\cdot,\cdot]$ represents the Lie bracket of vector fields and ${\cal L}_\zeta g$ is the Lie derivative of the metric tensor along the asymptotic Killing vector. 
Due to the variational principle, the surface charge two-form is a potential of the Noether three-form which is closed on the entire static patch, including both defects, such that the central extension accounts for those asymptotic degrees of freedom that extend smoothly into the entire geometry. 
If the algebra associated with the set of diffeomorphisms preserving boundary conditions contains at least one sub-algebra isomorphic to the de Witt algebra
\begin{align}
    i\left[\xi_p,\xi_q\right] = (p-q)\xi_{p+q}~,\qquad \{p,q\} \in \mathbb{Z}~,
\end{align}
the associated Dirac bracket algebra corresponds to the Virasoro algebra (upon suitable definition of the Virasoro operators in terms of the conserved charges).
In the case of dS spacetime, there is a set of asymptotic Killing vectors preserving the cosmological horizon structure \cite{Carlip:1999cy, Silva:2002jq}, which are of the form 
\begin{align}\label{AKVs}
    \zeta_p^\mu = T_p \chi^\mu + R_p\rho^\mu~,
\end{align}
where $\chi^\mu:= (\partial_t)^\mu$ is a null Killing vector field and
\begin{align}
    \rho_\mu = -\frac{1}{2\kappa}\nabla_\mu \chi ^2~,\qquad \kappa^2 := -\lim_{\chi^2\to0}\frac{1}{4\chi^2}\nabla_\rho \chi^2 \nabla^\rho\chi^2 = \frac{1}{\ell^2}~, 
\end{align}
is a vector which is normal to the horizon with vanishing norm at the horizon. Moreover, $R_p$ and $T_p$ are coordinate-dependent function which are obtained by requiring that $\chi^2$ remains null at $r = \ell$~, and have the form \cite{Silva:2002jq} 
\begin{align}
    R_p = \frac{\alpha}{\ell\kappa}i p T_p~,\qquad T_p = -\frac{\ell}{\alpha}\exp\{i p(\phi - \alpha t/\ell)\}~,
\end{align}
where $\alpha$ is an undetermined real number. Then, the asymptotic Killing vectors satisfy the algebra of diffeomorphisms of the circle
\begin{align}
    i[\zeta_p,\zeta_q]^\mu = (p-q)\zeta^\mu_{p+q}~,
\end{align}
and preserve the structure of the cosmological horizon. 
The symmetry algebra generated by the asymptotic Killing vectors \eqref{AKVs} is centrally extended by 
\begin{align}
    \int_{\cal H} k_{\zeta_p}[{\cal L}_{\zeta_q}g,g] = -i\frac{\ell^2}{4n\alpha G_4}\left(\alpha^2p^3 + 2p\right)\delta_{p+q,0}~,
\end{align}
with ${\cal H} \subset dS_n$ the orbifolded horizon manifold. Promoting Dirac brackets to quantum commutators $\{\cdot,\cdot\} \to \frac{1}{i }[\cdot,\cdot]$ and defining the Virasoro quantum operators
\begin{align}
    L_p := \mathfrak{Q}_p + \frac{3\ell^2}{8G_4}\frac{\alpha}{n}\delta_{p,0}~,
\end{align}
yields a chiral Virasoro algebra 
\begin{align}
    [L_p,L_q] = (p-q)L_{p+q} + \frac{\ell^2 }{4 G_4 }\frac{\alpha}{n}p(p^2-1)\delta_{p+q,0}~,
\end{align}
with non-trivial central charge
\begin{align}
    c_n = 12i\lim_{r\to \ell} \mathfrak{Q}_{\zeta_p}[{\cal L}_{\zeta_{-p}}g,g]\rvert_{p^3} = \frac{3\ell^2}{ G_4}\frac{\alpha}{n}~.
\end{align}
Considering the horizon field theory as a thermal chiral CFT, the corresponding Virasoro partition yields the Cardy entropy
\begin{align}
    S_{\rm C}^{(0)} = 2\pi\sqrt{\frac{1}{6}c_n \Delta} = \frac{\pi^2}{3}c_n T_c~,
\end{align}
where $T_c$ is the temperature of chiral modes near the horizon given by $T_c=(2\pi)^{-1}$ \cite{Bunch:1978yq}, and $\Delta = L_0$ the associated eigenvalue of the Virasoro zero mode which has been eliminated by means of
\begin{align}
    \frac{\partial S_{\rm C}}{\partial \Delta} = \frac{1}{T_c} \Rightarrow \Delta = \frac{\pi^2}{6}c_n T_c^2~.
\end{align}
Viewing the orbifolding procedure as a computation in a quantum theory using replicas, the Cardy entropy is identified as the modular free energy $F_n$ of the underlying degrees of freedom of the orbifolded geometry near the horizon region. 
In order to have a maximally entangled state, let us use the arbitrariness of $\alpha$ such that we rescale it as
\begin{align}
    \alpha  = 2\eta(n-1)~,\qquad \eta >0
\end{align}
where $\eta$ remains a finite undetermined constant.
With this redefinition, the central charge becomes
\begin{align}\label{Centraln}
    c_n = \eta\frac{6\ell^2}{  G_4}\left(1-\frac{1}{n}\right)~,
\end{align}
and using the Baez relation \eqref{baez}, the resulting the Renyi entropy becomes $n$-independent and proportional to the GH entropy, i.e.,
\begin{align}
    S_n = \eta S_{\rm GH}~.
\end{align}
We get, that the particular rescaling of $\eta$ leads to the possibility to recover the known result for the flat entanglement spectrum in dS. 
Then, the deformed geometry recovers the smooth geometry in the entanglement limit, whose thermal properties can be interpreted as being due to the entanglement between disconnected causal diamonds encoded into the two-dimensional chiral CFT at the cosmological horizon. 
In what follows, we will consider horizon's fluctuations and break the flat entanglement spectrum and analyze the different limits of the Renyi index. 

\subsection{Quantum Fluctuations, UV cutoffs, and the Finite Hilbert Space}
Many computations of black hole entropy relies on CFT methods \cite{Strominger:1996sh, Solodukhin:1998tc, Carlip:1999cy, Carlip:2000nv}, as previously discussed, Carlip has shown how to use them for arbitrary Killing vector, particularly in Cosmological horizons. Furthermore, Carlip obtained the logarithmic corrections associated with the Cardy entropy \cite{Carlip:2000nv} by considering the steepest descedent methods beyond the first order contribution of a two-dimenional CFT partition function. The result applies for any unitary CFT in the Cardy regime \cite{Hartman:2014oaa}, and relies on the modular invariance of the parition function, whose first-order quantum corrections is admissible if the central charge $c>> \Delta$, which is indeed satisfied for chiral modes in the Bunch--Davies vacuum \cite{Arias:2019zug}.
Then, the density of state of the chiral two-dimensional CFT with central charge $c$ and quantum corrections become
\begin{align}
    \rho(\Delta) \sim \left(\frac{c}{94\Delta^3}\right)^{\frac14}\exp\left\{2\pi\sqrt{\frac{1}{6}c\Delta}\right\} \longleftrightarrow \rho(T_c)\sim \frac{c}{144}(S_{\rm C})^{-\frac32}\exp\{ S_{\rm C}\}~.
\end{align}
Applying the result for the deformed dS geometry, we get
\begin{align}
    S_{\rm C} \sim S_{\rm C}^{(0)} - \frac32 \log S_{\rm C} = \left(1-\frac{1}{n}\right)\eta S_{\rm GH} - \frac32 \log \left[\left(1-\frac{1}{n}\right)^{\frac13}\eta S_{\rm GH}\right]~,
\end{align}
modulo constants. The Renyi entropy now becomes
\begin{align}
    S_n = \eta S_{\rm GH} - \frac32 \left(\frac{n}{n-1}\right)\log\left[\left(1-\frac{1}{n}\right)^{\frac13}\eta S_{\rm GH}\right]~,
\end{align}
which is no longer constant beyond the semi-classical order, breaking the flat spectrum due to the quantum corrections. 
The Reni entropy now has limits
\begin{align}
    S_{\rm E} ={}& \lim_{n\to1}S_n = -\frac32 \lim_{n\to1}- \frac{n}{n-1}\log\left[\left(1-\frac{1}{n}\right)^{\frac13}\eta S_{\rm GH}\right]~, \\ S_{\infty} ={}& \lim_{n\to \infty} S_n = \eta S_{\rm GH} - \frac32 \log \eta S_{\rm GH}~, \\ 
    S_0 ={}& \lim_{n\to 0}S_n = \eta S_{\rm GH}~. 
\end{align}
The entanglement entropy has the corresponding GH entropy as a first finite contribution, but acquires an extra divergent piece coming from the quantum corrections. The divergence can be regularized by introducing a cutoff $\delta$. The entanglement entropy can now be rewritten as
\begin{align}
    S_{\rm E} = \eta \frac{{{\rm Area}_{\cal H}}}{\delta^2} + \eta \frac{{{\rm Area}_{\cal H}}}{4  G_4}~,
\end{align}
corresponding to the divergent area law of entanglement entropy \cite{Srednicki:1993im}, where $\delta$ reads as
\begin{align}
    \delta^2 = \frac{2}{3}\eta {{\rm Area}_{\cal H}}\left\{\left(1-\frac{1}{n}\right)\log\left[\left(1-\frac{1}{n}\right)^\frac13 \frac{\eta {{\rm Area}_{\cal H}}}{4  G_4}\right]\right\}^{-1} \longleftrightarrow \left(1-\frac{1}{n}\right) = \frac{\delta^2}{2\eta {{\rm Area}_{\cal H}}}W_0\left(\frac{128 ^3G_4^3}{\eta^2{{\rm Area}_{\cal H}}^2\delta^2}\right)~,
\end{align}
where $W_0(x)$ is the Lambert $W$ function (or product logarithm). As we can see, in the tensionless limit 
\begin{align}
    \lim_{n\to 1}\delta = 0~,
\end{align}
indicating that the cutoff can be identified with the UV cutoff appearing in the standard QFT description literature of entanglement entropy \cite{Srednicki:1993im}. 
In this case, the bulk UV cutoff is related with the orbifold parameter which corresponds to the Renyi index due to the replica trick, and also gets a physical interpretation in terms of the energy scale of the theory. 
Similar results can be found in \cite{Solodukhin:1994yz} for asymptotically flat black holes, where one considers a conical singularity associated with the time coordinate, indicating a quantum fluctuation in the Euclidean black hole thermodynamics approach. The conical singularities modify the black hole temperature and therefore the conjugate entropy, generating two extra logarithmic divergences; one associated with UV divergences of the field theory, and a second one associated with interactions near the tip of the cone, where both regulators have been matched by hand. In our case, we have a single regulator associated with both, UV divergences and conical defects. 

The other interesting limits of Renyi entropy correspond to the min-entropy $S_\infty$ and the Harley entropy $S_0$. 
The min-entropy takes the form
\begin{align}
    S_\infty = -\log \lambda_1~,
\end{align}
where $\lambda_1$ is the largest eigenvalue of the reduced density matrix, which in the case at hand becomes
\begin{align}
    \lambda_1 = \left(\eta S_{\rm GH}\right)^{\frac32} \exp\{-\eta S_{\rm GH}\}~,
\end{align}
which due to the large value of the GH entropy, we get $\lambda << 1$, which is consistent with the fact that the min-entropy represent the smallest value of the Renyi entropy. 
Particularly, the Harley entropy $S_0$ yield information of the dimensionality of the reduced density matrix \cite{Hung:2011nu}, namely
\begin{align}
    S_0 = \log {\boldsymbol{\d}}~,
\end{align}
where $\boldsymbol{\d}$ corresponds to the number of non-vanishing elements of $\rho_{\cal R}$~. In the construction  before, this limits corresponds to moving the cutoff to the infrared leading to the properties of dS spacetime in the IR. The Harley entropy is then the dimensionality of the Hilbert space associated with a single Rindler observer, in our case is finite and given by the exponential of the GH entropy, i.e.,
\begin{align}
    \boldsymbol{\d} = \exp\{\eta S_{\rm GH}\}~.
\end{align}
As aforementioned, this controversial results have been found in previous literature \cite{Banks:2000fe, Jacobson:2022jir,Lu:2024tgj} indicating that the quantization becomes highly non-trivial as the dS isometry grup does not has non-trivial finite unitary representation and can not act faithfully on the Hilbert space, see \cite{Witten:2001kn} for further discussions. 

%%%%%%%%%%%%%%%%%%%%%%
\subsection{Codimension-two dS Holography}
\begin{figure}[h!]
    \centering
    \includegraphics{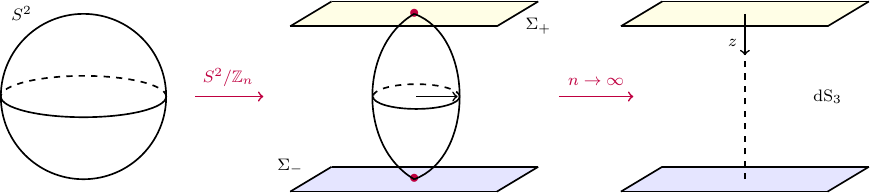}
    \caption{The large $n$ limit shrinks the spindle radius to a single transverse direction between both defects, with a geometry diffeomorphic to global dS$_3$~.}
    \label{fig:largen}
\end{figure}

The limit $n\to \infty$ is rather subtle in the geometry, as showing in \cite{Arias:2019zug}, the orbifold dS$_n$ reduces to global three-dimensional dS spacetime (dS$_3$) and the defects are sent to its conformal buondaries ${\cal I}_3^\pm$~. 
This can be seen directly in the embedding coordinates \eqref{embe}, which upon taking the large $n$ limit become
\begin{align}
    X_0 ={}& \sqrt{\ell^2-r^2}\sinh(t/\ell)\cos\theta~,\\  X_1 ={}&\sqrt{\ell^2-r^2}\cosh(t/\ell)\cos\theta~,\\ X_2 ={}& r\cos\theta \\ X_3 ={}& \ell\sin\theta~,\\ X_4 ={}& 0~,
\end{align}
which satisfying the hyperboloid constraint, the resulting induced line elemnt reads
\begin{align}
    ds^2 = \cos^2\theta \d\Sigma_2^2 + \ell^2\d\theta^2~.
\end{align}
Further taking a double Wick rotation $\t\to i\ell\hat\varphi~, \theta \to i \hat{\tau}/\ell$, the geometry becomes the one of dS$_3$ in global coordinates, which can be seen explicitly by taking $r = \ell\cos\hat\phi$~, viz.,
\begin{align}
    ds^2 = -\d \hat\tau^2 + \cosh^2(\hat{\tau}/\ell)\d\Omega_2^2~,\quad \hat{\tau} \in \mathbb{R}/\{\pm\infty\}~,\quad \hat{\varphi} \in [0,2\pi]~,\quad \hat\phi \in [0,\pi]~,
\end{align}
with $\hat\varphi$ and $\hat\phi$ coordinates of the unit, round two-sphere with line element $\d\Omega_2^2 = \d\hat\phi + \sin^2\hat\phi\d\hat\varphi^2$~. The limit is illustrated in \autoref{fig:largen}.

As can be seen, the defects are now mapped to the conformal infinities ${\cal I}^\pm$ of the global dS$_3$ spacetime. As aforementioned, in \cite{Arias:2019zug}, consistently truncating the classical theory to $\Sigma_\xi$ in terms of Liouville field theory for each defect, whose associated central charge reprouces \eqref{Centraln} when $\eta = 1/2$. Upon dimensionally reduce the Newton's constant, the sum of central charges reproduces the Strominger central charge for dS$_3$ obtained in \cite{Strominger:2001pn} using a proposed dS/CFT correspondence based on the analytical continuation of the AdS radius, i.e.,
\begin{align}
    c(\Sigma_+ \cup \Sigma_-) \stackrel{n\to\infty}{\longrightarrow} c({\cal I}_3^+ \cup {\cal I}_3^-) = \frac{3\ell}{2  G_3}~,
\end{align}
where $G_3 = G_4/4\ell$~, which using the Cardy formula reproduces the three-dimensional GH entropy by taking the large $n$ limit of the modular free energy of the total space 
\begin{align}
    S_{\rm C}^{(0)} \stackrel{n\to\infty}{\longrightarrow} \frac{\pi\ell}{2  G_3}~,
\end{align}
which equals the entropy of three-dimensional dS space \cite{Maldacena:1998ih}.
If we further use the logarithmic corrections found by Carlip, the entropy on the defects now becomes
\begin{align}
    S_{\rm C} \stackrel{n\to\infty}{\longrightarrow} \frac{\pi\ell}{2  G_3} - 3\log \frac{\pi\ell}{2  G_3}~,
\end{align}
matching the quantum corrected entropy of three-dimensional de Sitter found in \cite{Anninos:2020hfj,Chandrasekaran:2022cip}.
In this limit, the dual field theory becomes non-thermal, as the modular temperature $T=n^{-1}$ vanishes in the limit, in agreement with the intepretation of \cite{Klemm:2002ir} as dS$_3$ entropy corresponds to the Liouville momentum (see \cite{Arias:2019zug}), avoiding the problems mentioned in \cite{Dyson:2002nt} about the thermal nature of the dual CFT.

%%%%%%%%%%%%%%%%%%%%%%%%%%%%%%%%%%%%%%%%%%%%%%%%%
%%%%%%%%%%%%% Thermal fluctuations %%%%%%%%%%%%%%
%%%%%%%%%%%%%%%%%%%%%%%%%%%%%%%%%%%%%%%%%%%%%%%%%
\newpage
\section{Fluctuations and Holographic Charged Renyi entropy}\label{Sec:Fluctutations}

In the previous section, we discussed how exploring quantum corrections' influence on entanglement and Renyi entropy can help in understanding Quantum Gravity within dS space. Now, we will proceed to examine thermal and quantum fluctuations in the context of the gauge/gravity duality.
We will consider field theories with a global $U(1)$ symmetry. In order to do so, we use the formalism developed in \cite{Belin:2013uta} to study charged CFTs and the corresponding holographic duals. We extended the proposal by considering NLED theories in the bulk and consider quantum and thermal fluctuations on the string theory side which modifies the holographic Renyi entropy. 

\subsection{Holographic Non-linearly Charged Renyi Entropy}
As shown in \autoref{Sec:Entanglement}, the computations of entanglement entropy can be extended to CFTs with a global $U(1)$ charge \cite{Belin:2013uta}. In the case of spherical entangling surfaces, the CHM map allows to identify the entanglement entropy as the thermal entropy of an AlAdS charged hyperbolic black hole in the grand canonical ensemble. 

We will extend the result and include non-linearly charged theories by considering Einstein--Hilbert AdS gravity coupled to an arbitrary non linear electrodynamics (NLED) Lagrangian density ${\cal L}({\cal S})$ that depends on the Lorentz invariant scalar ${\cal S} = F_{\mu\nu}F^{\mu\nu}$~, where $F_{\mu\nu}$ is the curvature of the $U(1)$ gauge field $A_\mu$~, i.e.,
\begin{align}
    I = \int_{\cal M}\d ^{d+1}x\sqrt{-g}\left[\frac{1}{2\kappa}\left(R + \frac{d(d-1)}{\ell^2}+{\cal L}({\cal S})\right)\right]~.
\end{align}
 The field equations read
\begin{align}\label{eomned}
    E^\mu{}_\nu := R^\mu{}_\nu - \frac12 \delta^\mu{}_\nu\left(R+\frac{d(d-1)}{\ell^2}\right) - \kappa T^\mu{}_\nu = 0~, \qquad E^\mu := \nabla_\nu\left(F^{\mu\nu}\frac{\partial {\cal L}}{\partial \cal S}\right) = 0~,
\end{align}
where the energy-momentum tensor for the NLED coupling is
\begin{align}
    T^\mu{}_\nu = \delta^\mu{}_\nu \cL - 4\frac{\partial \cL}{\partial \cS}F^{\mu\lambda}F_{\nu\lambda}~.
\end{align}

Following \cite{Miskovic:2010QSR}, we solve the equations of motion for a static ansatz
\begin{align}
    ds^2 = -f(r)\frac{\ell^2}{\cR^2}\d t^2 + \frac{\d r^2}{f(r)} +r^2\gamma_{mn}\d y^m \d y^n~,
\end{align}
where $\gamma$ is the metric of a $(d-1)$ surface that can be either spherical, flat, or hyperbolic. The normalization $\cR$ is used such that when the transverse section has negative curvature, the boundary metric becomes conformally equivalent to 
\begin{align}
    ds^2 = -\d t^2 + \cR^2 \d H^2_{d-1}~
\end{align}
with $\d H_{d-1}^2$ denotes the line element of the $(d-1)$ hyperbolic plane with unit curvature.  
The gauge field ansatz
\begin{align}
    A_\mu = \left(\frac{\ell}{\cR}\phi(r)-\frac{\mu}{2\pi\cR}\right)\delta^t{}_\mu~,\qquad {\rm where}\quad \mu \equiv 2\pi \ell \phi(r_h)~,
\end{align}
is the chemical potential chosen such that the gauge potential vanishes at the black hole horizon defined by $f(r_h) = 0$~, generates an electric field $E(r) = -\partial_r\phi(r)$ and the associated field strength
\begin{align}
    F_{\mu\nu} = \frac{2\ell}{\cR}E(r)\delta^{[t}{}_{\mu}\delta^{r]}{}_{\nu} \Rightarrow \cS = -2E^2~.
\end{align}
Plugging the ansatz into the field equations we found
\begin{align}
    E^t{}_t = E^r{}_r = \frac{d-1}{2r^2}\left(rf'(r)+(d-2)(f(r)-k) -d\frac{r^2}{\ell^2}\right)-\kappa T^r{}_r = 0~,
\end{align}
that is solve for
\begin{align}
    f(r) = \frac{r^2}{\ell^2}-\frac{m}{r^{d-2}}+\frac{\cQ}{r^{d-2}}+k~,
\end{align}
where $m$ is an integration constant that can be related with the black hole mass \cite{Miskovic:2010ConCharges}, $k = 0,1,-1$ is the curvature of the transverse section, and 
\begin{align}
    \cQ(r) = \frac{2\kappa}{d-1}\int^r_\infty \d u~u^{d-1}T^r{}_r(u)~,
\end{align}
can also be related with the Noether charge associated to the $U(1)$ gauge symmetry. 
The metric function can be rewritten as
\begin{align}
    f(r) = \frac{r^2}{\ell^2} + k + \frac{\cQ}{r^{d-2}}-\left(\frac{r_h}{r}\right)^{d-2}\left(\frac{r_h^2}{\ell^2} +k +\frac{\cQ(r_h)}{r_h}  \right)~,
\end{align}
the black hole temperature is
\begin{align}
    T = \frac{T_0}{2}\ell f'(r_h) = \frac{T_0}{2}\frac{\ell}{r_h}\left[d\frac{r_h^2}{\ell^2} 
+ (d-2)k + \frac{2\kappa}{d-1}r_h^2 T^r{}_r(r_h) \right]~,
\end{align}
such that extremality occurs at
\begin{align}
    T^r{}_r(r_h) = -\frac{d-1}{2\kappa r_h^2}\left(\frac{dr_h^2}{\ell^2}+(d-2)k\right)~,
\end{align}
and the Wald entropy corresponds to
\begin{align}
    S^{(0)} = \V\left(\frac{r_h}{\ell}\right)^{d-1}~,\qquad \V \equiv \frac{2\pi}{\kappa}\ell^{d-1}V_\Sigma~,
\end{align}
where $V_\Sigma$ is the horizon's volume, that for the hypberbolic case $(k=-1)$
\begin{align}
    V_{H^{d-1}} =  \int_{H^{d-1}}\d \Sigma_{d-1} \sim  \begin{cases} \frac{\pi^{\frac{d-2}{2}}}{(d-2)\Gamma(d/2)}\frac{\cR^{d-2}}{\delta^{d-2}} + \dots~, & \mbox{if } d>2 \\[10pt] 2\log\left(\frac{2\cR}{\delta}\right)~, & d=2 \end{cases}
\end{align}
is divergent, regularized by inserting a cutoff $\delta$~. This shows how the universal divergences of the Renyi entropies are encoded on the volume of the horizon \cite{Hung:2011nu}. 
Moreover, integrating the gauge field equation $E^t = 0$ one gets a generalized Gauss law for NLED
\begin{align}
    \partial_r\left(r^{d-1}E(r)\frac{\partial \cL}{\partial \cS}\right) \Rightarrow E(r)\frac{\partial \cL}{\partial \cS}\Big\rvert_{\cS = -2E^2} = -\frac{\tilde q}{r^{d-1}}~,
\end{align}
where $\tilde{q}$ corresponds to the black hole electric charge. 

Now that the thermodynamic quantities have been defined, one can show that for an arbitrary NLED the Smarr relation is not always fulfilled \cite{Balart:2017dzt}, but it does satisfies \cite{Miskovic:2010QSR} the quantum statistical relation between the renormalized Euclidean action and the Gibbs free energy 
\begin{align}
    I^{\rm E} = \beta F = {\cal E} - TS - \frac{\mu}{2\pi\cR}\rho
\end{align}
and the first law
\begin{align}\label{fistlawcharged}
  \d {\cal E} = T\d S + \frac{\mu}{2\pi\cR}\d \rho~,
\end{align}
where $\rho$ is the charge density and  $\cE = \frac{(d-2)V_\Sigma}{2\kappa} m +\frac{1}{2}\left(1+(-1)^d\right)E_{\rm Casimir}^k$ the black hole energy that has a vacuum energy contribution when $d$ is odd of the form
\begin{align}
    E_{\rm Casimir}^k = (-k)^{\frac{d}{2}}\frac{\ell^{\frac{d}{2}-1}}{\kappa}V_\Sigma~.
\end{align}

Another quantities that will be relevant in the holographic analysis of the Renyi entropy and the associated quantum corrections is the black hole specific heat 
\begin{align}
    {\cal C} := \left(\frac{\partial E}{\partial T}\right)_{T} = T \left(\frac{\partial S^{(0)}}{\partial T}\right)_T = T(d-1)\V x^{d-2}\frac{\partial x}{\partial T}~,
\end{align}
where $E$ is the energy of the system and $T$ the equilibrium temperature, and in the second equality we have use the first law of black hole thermodynamics. In the NLED case can be written in terms of the stress-energy tensor as
\begin{align}\label{calC}
   {\cal C} = (d-1)S^{(0)}\left[ \frac{dx^2 + (d-2)k + \frac{2\ell^2\kappa}{d-1}x^2 T^r{}_r(x)}{dx^2-(d-2)k + \frac{2\ell^2\kappa}{d-1}x^2\left( T^r{}_r(x) + x\partial_x T^r{}_r(x)\right)} \right]~,
\end{align}
where
\begin{align}
    x := \frac{r_h}{\ell}~.
\end{align}
Thermal stability of the solution implies that
\begin{align}
    T^r{}_r(x) + x\partial_x T^r{}_r(x) \geq -\frac{d-1}{2\kappa\ell^2}\left(d- \frac{k(d-2)}{}  \right)~,
\end{align}
with all terms evaluated at $\cS = -2E^2$~. Additionally, the weak energy condition $\rho = -T_{\mu\nu}u^\mu u^\nu \geq 0$~, with $u^\mu$ a time-like vector, implies
\begin{align}
    T^r{}_r = T^t{}_t = {\cal L} +4\frac{\partial \cal L}{\partial \cS}E^2 \geq 0~.
\end{align}
Then
\begin{align}
    \partial_x T^r{}_r(x) \geq -\frac{d-1}{2\kappa\ell^2 x}\left(d - \frac{(d-2)k}{x^2} + \frac{2\kappa\ell^2}{d-1}T^r{}_r\right)~,
\end{align}
which for $k=-1$ the right-hand-side of the inequality becomes negative.

We can now compute the holographic quantities as explained in \autoref{SubSec:HolographyRenyi} throughout the CHM map, and see how the NLED Lagrangian modifies respect to the pure Maxwell case \cite{Belin:2013uta}. We express the NLED contributions to the boundary data in terms of the $rr$- components of the stress tensor at the deformed horizon $x_n$ defined through \eqref{xn}.

The conformal weight of the twist operators \eqref{hnmu}
\begin{align}
    h_n(\mu) = \pi n\left(\frac{\ell}{\kappa}\right)^{d-1}\left[x_n^{d-2}\ell^{d-2}\left(1-x_n^2\right) - \frac{\cQ(x_n,\mu)}{\ell x_n^{d-2}}\right]~,
\end{align}
and the magnetic response \eqref{magresponse} can be obtained through the first law \eqref{fistlawcharged} as
\begin{align}
    \rho(x_n,\mu) = -4\left(\frac{\ell}{\cR}\right)^{d-2}x_n^{d-1}\left(E\frac{\partial \cL}{\partial \cS}\right)\Big\rvert_{\cS =-2E^2, ~x = x_n}~,
\end{align}
and the central charge of the theory can be obtained by a using a series expansion of the structure coefficients around $n=1$~. The horizon radius is deformed such that $x_n$ corresponds to the largest solution to $T(x_n,\mu) = T_0/n$~, which in the NLED case becomes
\begin{align}
    n x_n^2\left(d + \frac{2\kappa\ell^2}{d-1}T^r{}_r(x_n,\mu)\right) - 2x_n + kn(d-2) = 0~,
\end{align}
which usually can not be solved analytically and relies on numerical methods, see for instance \cite{Dey:2016pei} for the Born--Infeld case. Similar issues appear for the chemical potential which needs to be solved as we are working in the grand canonical ensemble. 

The holographic nonlinearly charged Renyi entropy reads
\begin{align}\label{SnCHMmuzero}
    S_n^{(0)} = \frac{n}{n-1}\frac{\V}{2}\left[ x^{d-2} + x^d + \left(\cQ(x,\mu) +\frac{2\kappa\ell^2}{d-1} \int \d x~x^d\partial_x T^r{}_r(x,\mu) \right)\Bigg\rvert^{x_1}_{x_n}\right]~.
\end{align}
The Renyi inequalities \eqref{Rineq} can be checked using 
\begin{align}
    \partial_n S_n^{(0)} = -\frac{\partial x_n}{\partial n}\frac{\partial S_n}{\partial x}~,
\end{align}
whith
\begin{align}
    \frac{\partial x_n}{\partial n} = \frac{d-2 -x_n^2\left(d+ \frac{2\kappa}{d-1}T^r{}_r(x_n)\right)}{4nx_n\left[d + \frac{2\kappa}{d-1}\left(T^r{}_r(x_n) + \tfrac{1}{2}x_n\partial_{x_n}T^r{}_r(x_n)\right)\right] - 2}~.
\end{align}
Finally, the Renyi index can be computed in terms of black hole parameters as
\begin{align}
    n = \frac{2}{x}\left(dx^2 - (d-2) + \frac{2\kappa \ell^2}{d-1}x^2 T^r{}_r(x)\right)^{-1}~,
\end{align}
and can be used to analyse the dual CFT stability and phase transitions \cite{Belin:nindex, Fang:2016ehk}.

Now that we have define the holographic nonlinearly charged Renyi entropy, we move forward and consider thermal and quantum fluctuations on the solution and discuss how modifies the entanglement entropy and structure constants of the dual theory to interpret the correction from the CFT perspective. 

%%%%%%%%%%%%%%%%%%%%%%%%%%%%%
%%%%%%%%%%%%%%%%%%%%%%%%%%%%%
%%%%%%%%%%%%%%%%%%%%%%%%%%%%%
\subsection{Logarithmic corrections}

Several different approaches to Quantum Gravity have shown \cite{Das:2001ic,Carlip:1999cy,Carlip:2000nv,Kaul:2000kf,Cai:2009ua,Sen:2012dw,Sen:2012kpz,Pathak:2016vfc,Bobev:2023dwx} that black hole entropy gets modified by terms of the form $-\C \log({\rm Area})$~, such that the black hole entropy becomes\footnote{The same holds for dS space \cite{Anninos:2020hfj}.}
\begin{align}\label{Slog}
    S^{(1)} = S^{(0)} - \C \log S^{(0)}+\dots~, 
\end{align}
where
\begin{align}
    S^{(0)} = \frac{2\pi}{\kappa}{\rm Area}_{\rm Hor}~,
\end{align}
is the Bekenstein-Hawking entropy that is proportional to a quarter of the horizon's area. 

In \cite{Das:2001ic}, it is shown that this logarithmic corrections are universal and appear by small statistical fluctutations around equilibrium points. 
In the canonical ensemble, the partition function in terms of the inverse of the temperature $\beta = 1/T$ and the sysetem's energy $E$
\begin{align}
Z = \int \d E \rho(E)\exp\{-\beta E\}~,
\end{align}
where $\rho(E)$ is the density of states and can be obtained, for fixed $E$~, by an inverse Laplace transformation
\begin{align}
    \rho(E) = \frac{1}{2\pi i}\int_{\gamma -i\infty}^{\gamma+i \infty} \d \beta~\exp\{\log Z + \beta E\} =  \frac{1}{2\pi i}\int_{\gamma -i\infty}^{\gamma+i \infty} \d \beta~\exp\{\beta S(E)\} ~,
\end{align}
where the exact entropy function of the system 
\begin{align}
    S = \log Z + \beta E~.
\end{align}
The complex integration can be performed by the method of steepest descent around the saddle point $\beta_0$~, i.e.,
\begin{align}
    S_0' \equiv \left(\frac{\partial S}{\partial\beta}\right)\Big\rvert_{\beta_0} = 0~, 
\end{align}
which is given by the inverse of the equilibrium temperature of the system
\begin{align}
    E = \langle E\rangle =  -\left(\frac{\partial }{\beta}\log Z\right)\Big\rvert_{\beta_0}~. 
\end{align}
If we consider the next-to-leading order in an expansion around the equilibrium point, i.e.,
\begin{align}
    S = S_0 + \frac12(\beta-\beta_0)S_0'' + \dots~,
\end{align}
the density of states can be computed order by order, and the microcanonical entropy at equilibrium becomes 
\begin{align}
    S = \log \rho = S_0 -\frac12 \log S_0'' + \dots~,
\end{align}
which contains logarithmic corrections. One can rewrite the expression in the logarithm in terms fluctuations of energy noticing that
\begin{align}
    S_0'' = \langle E^2\rangle - \langle E\rangle^2 = {\cal{C}} T^2~,
\end{align}
such that in the canonical ensemble there is an universal logarithmic modification of the entropy by means of consider small thermal fluctuations in the partition function of the form
\begin{align}\label{canoSlog}
    S = S_0 - \frac12 \log {\cal C}T^2~,
\end{align}
and moreover, $\cal C$ is proportional to $S_0$~, see \eqref{calC} such that one expects a general correction proportional to $\log S_0$~. 

In the grand canonical ensemble, the partition function with $K$ fixed chemical potentials for a continuous distributions of the conserved quantities
\begin{align}\label{Zgc}
    Z = \prod_{i=0}^K \int_0^\infty \d q_i dE~\rho(N_i,E)\exp\{-\beta(E-\mu_iq_i)\}~,
\end{align}
where $N_i$ are the charges conjugates to $\mu_i$~, $\rho$ is the density of states, and $E$ the possible energies of the system. 
Laplace transforming \eqref{Zgc}, the density of states in the saddle point 
\begin{align}\label{rhozero}
    \rho(q_i,E) = \frac{1}{(2\pi)^{\frac23}{(\rm det}D_{A B})^{\frac12}}Z(\beta,\mu_i)\exp\{\beta (E-\mu_iN_i)\}~,
\end{align}
and 
\begin{align}
    D_{A B} = \frac{\partial^2}{\partial\chi^A\partial\chi^B}{\rm log}Z~,
\end{align}
where $\chi^A = \{-\beta \mu_i,\beta\}$ are the thermodynamic quantities that are considered n the thermal partition function. Then, the thermal entropy corresponds to
\begin{align}
    S = \log\rho - \frac12 \log {\rm det}D_{AB}~.
\end{align}
Notice that ${\rm det}D_{AB} >0$ is a necessity a condition for thermal stability, and correspond to the black hole's response coefficients. In the canonical ensemble one recovers \eqref{canoSlog} as
\begin{align}
    {\rm det}D_{AB} = {\cal C}T^2~,
\end{align}

Allowing for fluctuations of the particle number in the grand canonical ensemble, the coefficients of $D_{AB}$ are related with fluctuations of the thermodynamic quantities. Considering only a single charge~, in \cite{Mahapatra:2011si}, it is shown that ${\rm det}D_{AB}$ correspond to the square of the standard deviation between the product of the conserved quantities and that it can be reexpressed as
\begin{align}
    {\rm det}D_{AB} = -T^4{\cal C}\left(\frac{\partial N}{\partial \lambda}\right)_T - T^3\mu - T^4\left(\frac{\partial N}{\partial T}\right)^2_\mu~,
\end{align}
with $\lambda = -\mu/T$ and $N$ the conjugated charge to the chemical potential. 
One can further relax the condition of that the spectrum of the charges is continuous by including a Jacobian factor $J$ that appears in taking the continuous limit  of the partition function \cite{Gour:2003jj}. In this case, the Jacobian modifies the statistical entropy as
\begin{align}
    S = \log\rho - \frac12\log{\rm det}D_{AB} + \log J~.
\end{align}
The Jacobian factor is contingent on the quantum spectrum of the black hole, thus necessitating prior knowledge of the underlying Quantum Gravity theory. However, it is widely acknowledged that this field is not yet fully comprehended.

In \cite{Gour:2003jj}, it is shown that, assuming that the spectrum of each parameter is linear in the quantum numbers in which are measure and that they are independent of each other, the entropy for the AdS-Reissner--Nordstrom black hole acquires a logarithmic correction with $\C = 1$ in \eqref{Slog}
independent of the dimension. This is an example of how the $\C$ coefficient depends on the approach to Quantum Gravity. Moreover, such corrections (although with different values of $\C$) also appear when quantum corrections (see for instance \cite{Fursaev:1994te,Solodukhin:1994yz, Kaul:1998xv, Kaul:2000kf,Carlip:2000nv,Aros:2010jb,Sen:2012dw}) into the gravity theory are considered. Then, the coefficient in front of the logarithmic corrections seem to be fundamental to understand how Gravity can be understood as a field theory. 
In this subsection, we will see how the coefficient in the logarithmic corrections play a role in the holographic dictionary and how one can extract information of the quantum theory of Gravity if the AdS/CFT correspondence is assumed\footnote{See \cite{Mukherji:2002de} for an example on how the gauge/gravity duality can be used to restrict or predict the value of $\C$~.}. 
In order to do so, we extend the proposal of \cite{Mahapatra:2016iok} for the canonical ensemble and we include the logarithmic corrections of the black hole entropy into \eqref{Snthermal}. Some comments on a holographic interpretation for the corrections is given at the end of the section. 

The thermal/quantum fluctuations back reacts into the Renyi entropy as
\begin{align}\label{Sngen}
    S_n ={}& \left(1-\frac{1}{n}\right)^{-1}\frac{1}{T_0}\int^{x_1}_{x_n}\d x\left[S^{(0)} - \C \log \left(\V x^{d-1}\right)\right]\partial_x T(x,\mu)\nonumber \\ \nonumber ={}& S_n^{(0)} - \left(1-\frac{1}{n}\right)^{-1}\frac{\C}{T_0}\int^{x_1}_{x_n}\d x~\log \left(\V x^{d-1}\right) \\ \nonumber ={}& \frac{n}{n-1}\left\{\frac{\V}{2}(x^d + x^{d-2}) - \frac{\C}{2x}\left[(d-1)(2-d(1-x^2) + (2-d(1+x^2))\log\V x^{d-1}\right]\right\}\Big\rvert_{x_n}^{x_1} \\  +{}&  \frac{n}{n-1}\left(\frac{\kappa\ell^2}{d-1}\right)\int^{x_1}_{x_n}\d x\left(\V x^{d-1} - \C\log\V x^{d-1}\right)\left(T^r{}_r + x\partial_x T^r{}_r\right)~,
\end{align}
with all quantities evaluated at fixed chemical potential prior to integration. As can be seen from \eqref{Sngen}, the uncharged limit can be smoothly taken and the last term in \eqref{Sngen} vanishes, and the remaining terms correspond to universal terms appearing from the Einstein--Hilbert action. 
The Renyi inequalities \eqref{Rineq} are satisfied if the thermal entropy is positive \cite{Pastras:2014oka} as can be seen from
\begin{align}
    \partial_n \left(\frac{n-1}{n}S_n\right) = \frac{1}{n^2}S~,
\end{align}
such that, the Renyi entropy inequalities are satisfied if $\C \leq \frac{S^{(0)}}{\log S^{(0)}}$~. Usually, $\C$ is of order unity, such that if $\C \geq 0$ and $S^{(0)}\geq 0$~, then the bound can be violated if the black hole has a small area, i.e., if the black hole is microscopic \cite{Scardigli:1999jh}. Such configurations can appear at the last epoch of a black hole lifetime as Hawking radiation makes the black hole to evaporate and fluctuations start to dominating overt he classical saddle. However, such black holes seems to be thermal unstable \cite{Jizba:2009qf} and the description of statistic entropy and fluctuations is yet not fully understood. One would need to further consider the backreaction of quantum fluctuations into the geometry itself and how the thermodynamic quantities become modified. 
Then, holographically the Renyi inequalities are mapped to thermal stability of the black hole solution.

Let us for a moment consider only the gravitational sector. The conformal weights of the twist operators are modified as
\begin{align}
    h_n(\mu=0) ={}& \frac{n}{(d-1)}V_{H^{d-1}}^{-1}\int^{x_1}_{x_n}\d x~T(x)\partial_x S(x) \nonumber \\ ={}& \pi n\left(\frac{\ell}{\kappa}\right)^{d-1}\left(x_n^{d-2} -x_n^d\right) - \frac{n}{2}\frac{\C}{V_{H^{d-1}}}\frac{(dx_n + d -2)(1-x_n)}{x_n}~,
\end{align}
with 
\begin{align}
    x_n(\mu=0) = \frac{1}{dn}\left(1+\sqrt{1-2dn^2 + d^2 n^2}\right)~.
\end{align}
Then, the leading singularity of the two-point function of the stress tensor becomes modifies by the logarithmic corrections as
\begin{align}
    \partial_n h_n\rvert_{n=1}=\frac{1}{d-1}\left(2\pi\left(\frac{\ell}{\kappa}\right)^{d-1} - \frac{\C}{V_{H^{d-1}}}\right)~.
\end{align}
The second term corresponds to a small correction of the central charge of the dual theory due to the fluctuations considered in the black hole thermodynamics. If $\C>0$~, which is a requirement for the dual black hole to satisfy the generalized second law \cite{Das:2001ic}, this term can be though to as an IR correction of the central charge which is in agreement with the $c-$ and $F-$ theorems in $d=2$ and $d=4$, respectively \cite{Zamolodchikov:1986gt, Cardy:1988cwa}. Extending the result to non-trivial chemical potential, we get
\begin{align}
    h_n(\mu) = \pi n\left(\frac{\ell}{\kappa}\right)^{d-1}(x_n^{d-2} - x_n^d) - \frac{n}{2}\frac{\C}{V_{H^{d-1}}}\frac{(dx_n+d-2)(1-x_n)}{x_n} - \frac{n\mu}{2\pi\cR}\frac{\rho(x_n,\mu)}{d-1}~,
\end{align}
where $x_n = x_n(\mu)$~. In the next subsection, we will use the results in a particular example where the computations are analytical such that we can give some holographic interpretation to the corrections.

%%%%%%%%%%%%%%%%%%%%%%%%%%%%%
%%%%%%%%%%%%%%%%%%%%%%%%%%%%%
%%%%%%%%%%%%%%%%%%%%%%%%%%%%%
\subsection{Holographic Two-dimensional Free Bosons}\label{SubSec:ConfBosons}

In \cite{Belin:2013uta}, it is shown how to compute holographically the charged Renyi entropy of a CFT with a spherical entangled surface. For $d=2$~, one should consider the charged BTZ black hole. Nonetheless, such solution is charged by
\begin{align}
    \cQ = - \frac{q^2}{2}\log\left(\frac{r}{\ell \sqrt{m}}\right)~,
\end{align}
and the solution does not asymptotes to global AdS, indicating a breaking of conformal invariance at the boundary \cite{Klebanov:1999tb}\footnote{In \cite{Perez:2015jxn} it is shown that the asymptotic symmetry algebra of the canonical generators corresponds to a direct sum of ${\mathfrak{u}}(1)$ and two copies of the Virasoro algebra if one relaxes the Brown--Henneaux boundary conditions.}. Instead, let us consider three-dimensional conformal electrodynamics \cite{Hassaine:2007py} 
\begin{align}
{\cal L} = \lambda|\cS|^{\frac{3}{4}}~.
\end{align}
with $\lambda$ some real coupling constant.
The associated traceless energy-momentum tensor
\begin{align}
    T^\mu{}_\nu = \lambda|\cS|^{\frac34}\left(\delta^\mu{}_\nu - 3\frac{F^{\mu\rho}F_{\nu\rho}}{F_{\alpha\beta}F^{\alpha\beta}}\right)~,
\end{align}
with 
\begin{align}
    T^r{}_r = -\frac{q^2}{2r^3}~,
\end{align}
produces a Coulomb-like electric field
\begin{align}
    E(r) \propto \frac{q^2}{r^2}~.
\end{align}
The theory contains a black hole solution if $\lambda^2 = \frac{128\sqrt{2}}{27}\kappa q$ \cite{Cataldo:2000we} with metric function
\begin{align}\label{fcataldo}
    f(r) = \frac{r^2}{\ell^2}-m + \frac{\tilde{q}^2}{r}~,
\end{align}
which is, indeed, an AlAdS black hole with curvature singularity at the origin, as can be seen from
\begin{align}
    R_{\mu\nu\rho\sigma}R^{\mu\nu\rho\sigma} = 12\left(\frac{1}{\ell^4} + \frac{\tilde{q}^4}{2r^6}\right)~.
\end{align}

\begin{figure}[h!]
    \centering
    \includegraphics{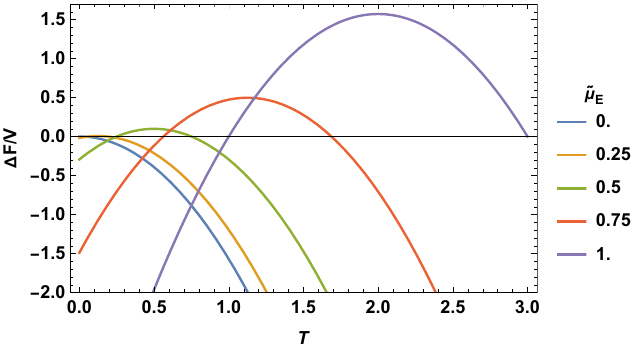}
    \caption{Difference of the density of free energies of the black hole \eqref{fcataldo} and thermal AdS for various values of $\tmu_{\rm E}$~. }
    \label{fig:hptrans}
\end{figure}

In the grand canonical ensemble, the charge is related to the chemical potential as
\begin{align}
    q = -\frac{2x\kappa}{3\pi}\mu
\end{align}
and the horizon's temperature simplifies to
\begin{align}
    T(x,\mu) = T_0(x-\tmu^2)~,\qquad \tmu^2:= \left(\frac{\kappa}{3\pi}\right)^2 2\ell \mu^2~,
\end{align}
which imposes $x\geq \tmu^2$ for real chemical potential. The specific heat 
\begin{align}
    {\cal C} = \frac{T}{T_0}\V~,
\end{align}
is always positive, ensuring the thermal stability of the solution in the grand canonical ensemble. 
We can solve \eqref{Tneq} analytically as
\begin{align}
    x_n = \frac{1}{n} + \tmu^2~,
\end{align}
and is always positive for any $\tmu$~. This adds the possibility to have a well-defined Wick rotated chemical potential $\mu_E^2 = (i\mu)^2$ in the dual theory. Nonetheless, to have a positive horizon radius in the Wick rotated version we need $n\tmu_E^2 \leq 1$ for fixed $n$~. For two-dimensional chiral bosons, such upper bound for the imaginary chemical potential appears as a restriction to avoid singularities in the free energy functional \cite{Belin:2013uta}. We will see later that this system also reproduces the Renyi entropies and twist operators of such theory. 
As a final comment on the thermodynamics of the solution, we notice that using an imaginary chemical potential, there is a possibility of having a Hawking--Page phase transition \cite{Hawking:1982dh}. This can be seen by comparing the Euclidean action of the solution $I^{\rm E}_{\rm C}$ with the thermal AdS Euclidean action $I^{\rm E}_{\rm TAdS}$ and using the relation of the action with the free energy $I^{\rm E} = \beta F$ as shown in \cite{Miskovic:2010QSR}, we get
\begin{align}
    \Delta I = I_{\rm C}^{\rm E} - I_{\rm TAdS}^{\rm E} = -S^{(0)}\left(\frac{x-2\tmu_{\rm E}^2}{x+\tmu^2_{\rm E}}\right) = -\frac{\pi \V}{2 T}\left(\tmu^2_{\rm E}- T\right)\left(3\tmu^2_{\rm E}-T\right),
\end{align}
showing a change in the sign such that a phase order phase transition between thermal AdS and the black hole at two different critical temperatures $T_c^{(1)} = \tmu^2_{\rm E}$ and $T_c^{(2)} = 3\tmu^2_{\rm E}$~. This are known as reentrant phase transitions as $T_c^{(2)} > T_c^{(1)}$~, see \autoref{fig:hptrans}. This is radically different from the pure BTZ solution which shows no phase transition due to a mass gap between the vacuum and the solution.

Moving forward, we can now compute the required holographic quantities for this example. The magnetic response
\begin{align}
    k_n(\tmu_{\rm E}) = \frac{8 \pi  \cR}{\ell}  \left(1-n \tmu_{\rm E}^2\right)\tmu_{\rm E} ~,
\end{align}
and the 
conformal weight of the twist operators for imaginary chemical potential
\begin{align}\label{hnConf}
    h_n(\mu_{\rm E}) = \frac{\pi\ell}{\kappa}\left[\left(n-\frac{1}{n}\right) + n \tmu_{\rm E}^4\right]~,
\end{align}
such that
\begin{align}
    \lim_{n\to1}\lim_{\mu\to0}\partial_n h_n(\mu_{\rm E}) = \frac{2\pi\ell}{\kappa}~.
\end{align}
The conformal weight \eqref{hnConf} behaves just as the ones found in \cite{Belin:2013uta} for a two-dimensional charged conformal bosons \cite{Belin:2013uta}. Indeed, the holographic Renyi entropy for an imaginary chemical potential reads
\begin{align}
    S_n^{(0)} = \frac{\pi\ell}{\kappa}\left[\frac12\left(1+\frac{1}{n}\right) - \tmu_{\rm E}^2 \right]~,
\end{align}
 which matches the CFT computation for a free two-dimensional boson with an imaginary chemical potential $\mu_{\rm C}$ \cite{Belin:2013uta} if one identifies $\mu_{\rm E}^2 = \frac{|\mu_{\rm C}|}{2\pi}$~. Then, in both cases, the decreasing of the Renyi entropy is dictated by the absolute value of the chemical potential, indicating the duality between both theories. The Renyi entropy limits are
\begin{align}
    S_{\rm E} = \V(1-\tmu^2_{\rm E})~, \qquad S_0 = \frac{\V}{2n}~,\qquad S_\infty = \frac{\V}{2}\left(1+2\tmu^2_{\rm E}\right)~,
\end{align}
showing that $S_0$ is independent of the charge, and for large chemical potential
\begin{align}
    \lim_{\tmu_{\rm E}\to\infty} S_n^{(0)}(\tmu_{\rm E}) = -\V\tmu^2~,
\end{align}
that is, indeed, $n$-independent at the leading order. The Renyi entropy inequalities \eqref{Rineq} can be now checked analytically
\begin{align}
    \partial_n S_n^{(0)} = -\frac{\V}{2n}~,\qquad \partial_n\left(\left(n-1\right)S_n^{(0)}\right) = \frac{\V}{2}\left(1+\frac{1}{n^2} - 2\tmu^2_{\rm E}\right)~, \\ \partial_n\left(\frac{n-1}{n}S_n^{(0)}\right)\frac{\V}{n^2}(1-n\tmu_{\rm E}^2)~,\qquad \partial^2_n((n-1)S_n^{(0)}) = -\frac{\V}{n^3}~,
\end{align}
which are identically satisfied when we return to the real chemical potential. Now, in the imaginary case, we have constrains between the Renyi index and the chemical potential, such as $n\tmu_{\rm E}^2 \geq 1$~, which is satisfied by considering a positive black hole temperature. Also notice that $\mu_{\rm E}\partial_{\mu_{\rm E}}S_n^{(0)} = 2\V \mu^2_{\rm E} \geq 0$~. 

Let us now consider thermal fluctuations computed by assuming a linear dependence between the thermodynamic quantities and the associated quantum numbers, just as in \cite{Gour:2003jj}. We get logarithmic corrections with coefficient $\C = 1/2$~, which are valid in the regime
\begin{align}
    {\cal C}(\tmu_{\rm E}) \geq \frac{2\tmu_{\rm E}^2}{T}\frac{\V \ell}{2\pi \cR S^{(0)}}~,
\end{align}
imposing a bound on the chemical potential.

\begin{figure}[t!]
\begin{center}
  \includegraphics[scale=0.42]{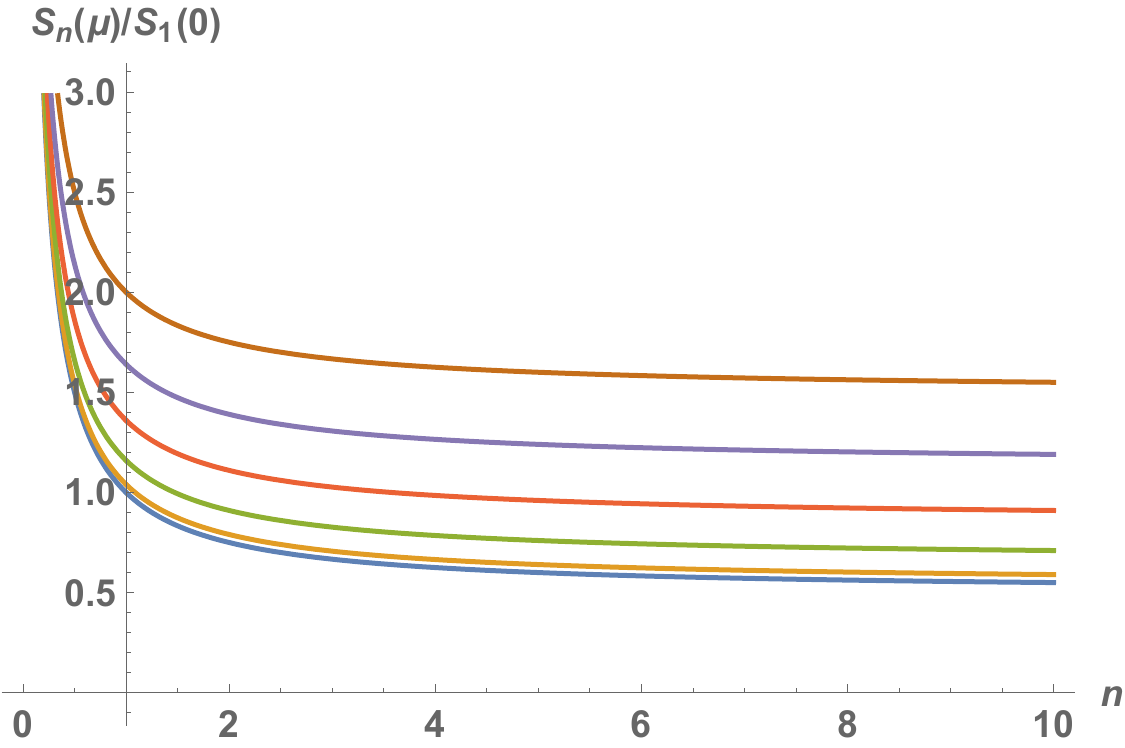}
  \includegraphics[scale=0.42]{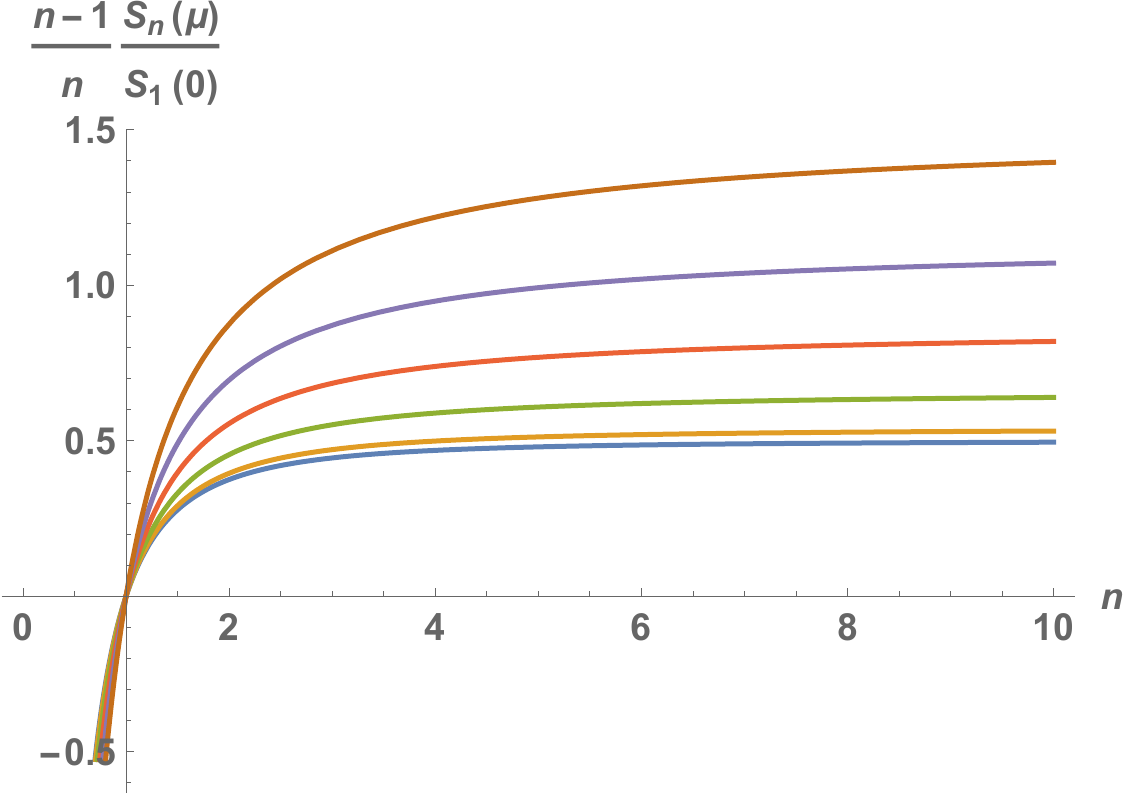}
  \captionof{figure}{Corrected Renyi entropy as a function of $n$ on the left panel, and $\frac{n}{n-1}\frac{S_n(\mu)}{S_1(0)}$ on the right panel both normalized by the zero-charge limit of the entanglement entropy $S_1(0)$,  for different values of the chemical potential with $\V = 10^{40}$~. From bottom to top, the curves corresponds to $\tmu = 0,0.2,0.4,0.6,0.8,1$. }
  \label{fig:c1}
  \end{center}
\end{figure}

\begin{figure}[t!]
\begin{center}
  \includegraphics[scale=0.4]{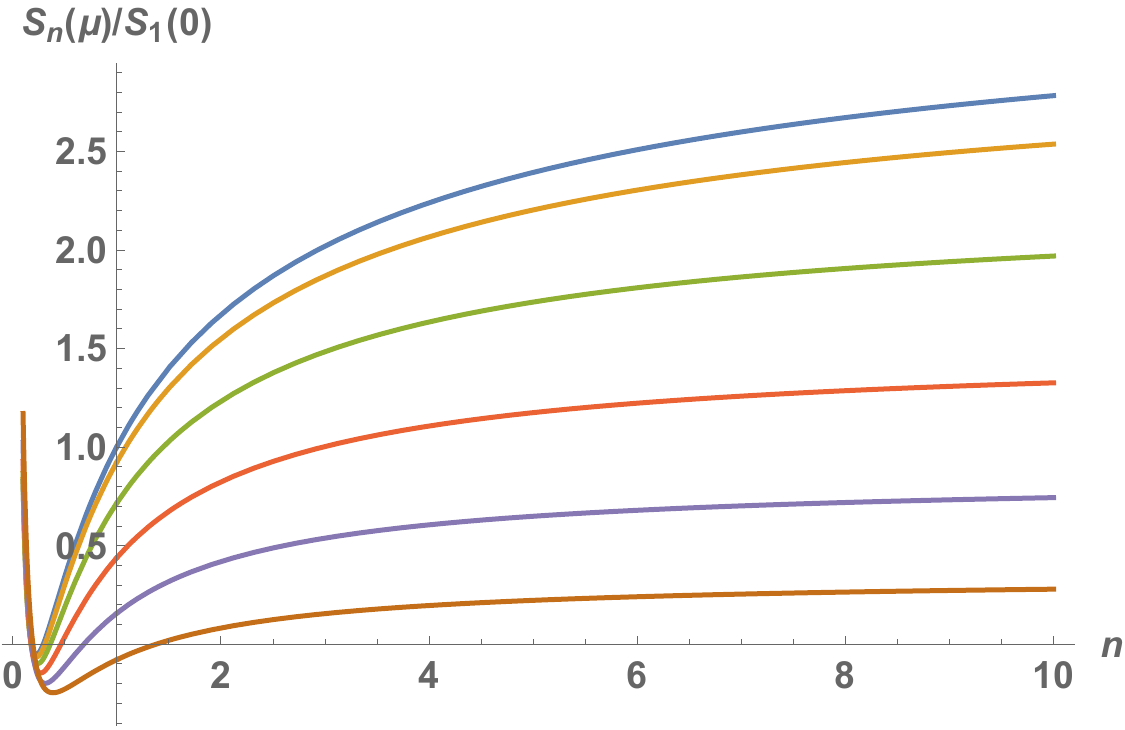}
  \includegraphics[scale=0.4]{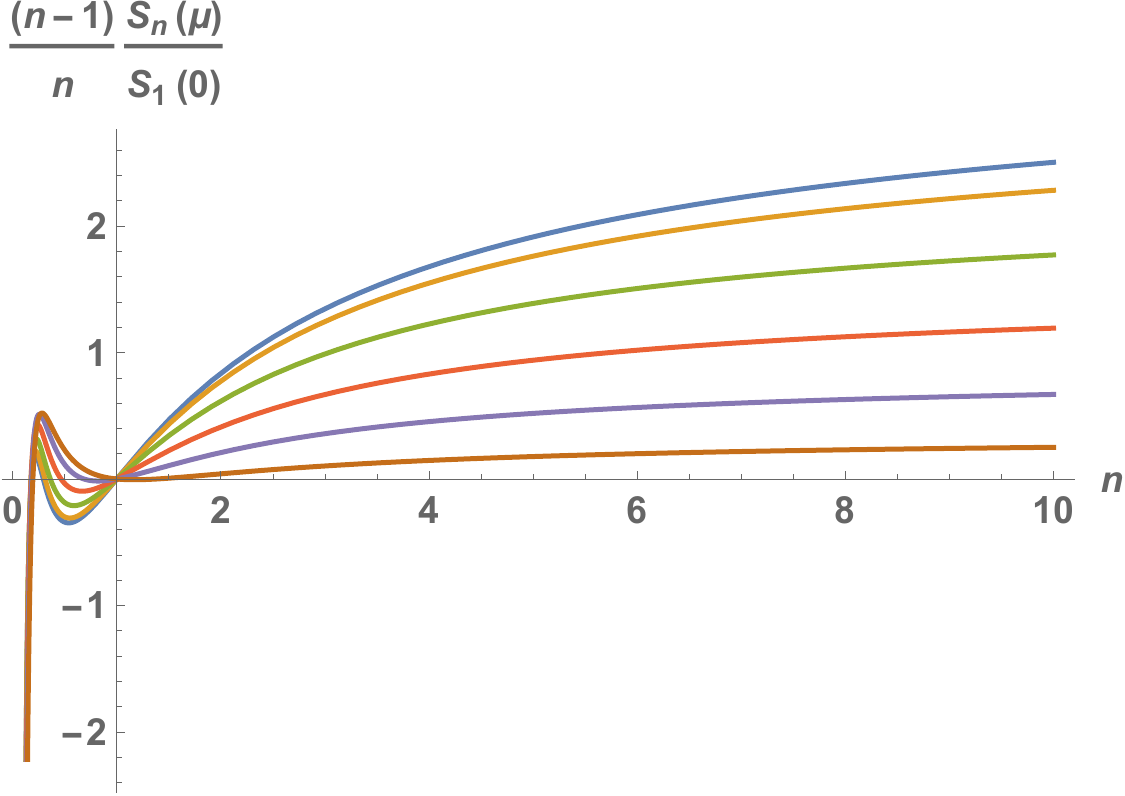}
  \captionof{figure}{Corrected Renyi entropy for three-dimensional gravity coupled to Coulomb sources with logarithmic corrections as a function of $n$ normalised by $S_1(0)$ on the left panel, and $\frac{n}{n-1}\frac{S_n(\mu)}{S_1(0)}$ on the right panel,  for different values of $\tmu$ with a UV cutoff so that $\V = 1$~. From top to bottom, the curves corresponds to $\tmu = 0,0.2,0.4,0.6,0.8,1$. }
  \label{fig:c4}
  \end{center}
\end{figure}

The corrected Renyi entropy becomes
\begin{align}
    S_n = \frac{n}{n-1}\left[\frac{\V}{2}\left(x^2_1 - x_n^2\right) +\C\left(x_1(1+\log\V x_1) - x_n(1-\log\V X_n\right)\right]~,
\end{align}
with $\C=1/2$~, with limits
\begin{align}
    S_{\rm E} = \frac{\V}{2n} - \C\log\frac{\V}{n}~, \qquad S_0 =\V(1-\tmu_{\rm E}^2) - \C \log(\V(1-\tmu^2_{\rm E}))~,\nonumber \\ S_\infty = \left(\frac{\V}{2}(x_1+x_\infty) + \C\right)(x_1-x_\infty) - \C\left(x_1\log \V x_1 + x_\infty \log\V x_\infty\right)~,
\end{align}
where $x_1 = 1-\tmu_{\rm E}^2$ and $x_\infty = -\tmu^2_{\rm E}$~. The large $\tmu_{\rm E}$ yields to
\begin{align}
    \lim_{\tmu_{\rm E}\to\infty}S_n(\tmu_{\rm E}) = -\V\tmu_{\rm E}^2 - \C\log\V \tmu_{\rm E}^2~,
\end{align}
which is again independent of $n$ at the leading order, while the Renyi inequalities now become
\begin{align}
    \partial_n S_n ={}& - \frac{(n-1)(\V(n-1)+2n\C) + 2\C(1-\tmu^2_{\rm E})\log\left(\frac{n}{n-1}\left(1-\tmu_{\rm E}^2\right)\right)}{2n^2(n-1)^2}~, \\ \partial_n((n-1)S_n) ={}&\frac{\V}{2}\left(1+n^{-2}-2\tmu_{\rm E}^2\right) - \C\left[\frac{1-n}{n} +\tmu_{\rm E}^2\log\left(\frac{1/n-\tmu^2_{\rm E}}{1-\tmu_{\rm E}^2}\right) + \log\left(\V(1-\tmu^2_{\rm E})\right) \right]~, \\ \partial_n^2((n-1)S_n) ={}& -\frac{1}{n^3}\left(\V - \frac{n\C}{1-n\tmu_{\rm E}^2}\right)~, \\ \partial_n\left(\frac{n-1}{n}S_n\right) ={}& \frac{\V}{n^2}(1-n\tmu^2_{\rm E}) - \frac{\C}{n^3}\log\left(\frac{\V}{n}(1-n\tmu^2_{\rm E}\right)~, \\ \tmu_{\rm E}\partial_{\tmu_{\rm E}} ={}& 2\tmu_{\rm E}^2\left(\V + \frac{\C}{n-1}\log\frac{1-n\tmu^2_{\rm E}}{n-n\tmu^2_{\rm E}}\right)~,
\end{align}
which can not be checked analytically. Nonetheless, it is easy to notice that if $\C$ is smaller compared with $\V$ the inequalities are satisfied. This can be seen explicitly comparing \autoref{fig:c1} and \autoref{fig:c4}

\begin{figure}[t!]
\begin{center}
  \includegraphics[scale=0.40]{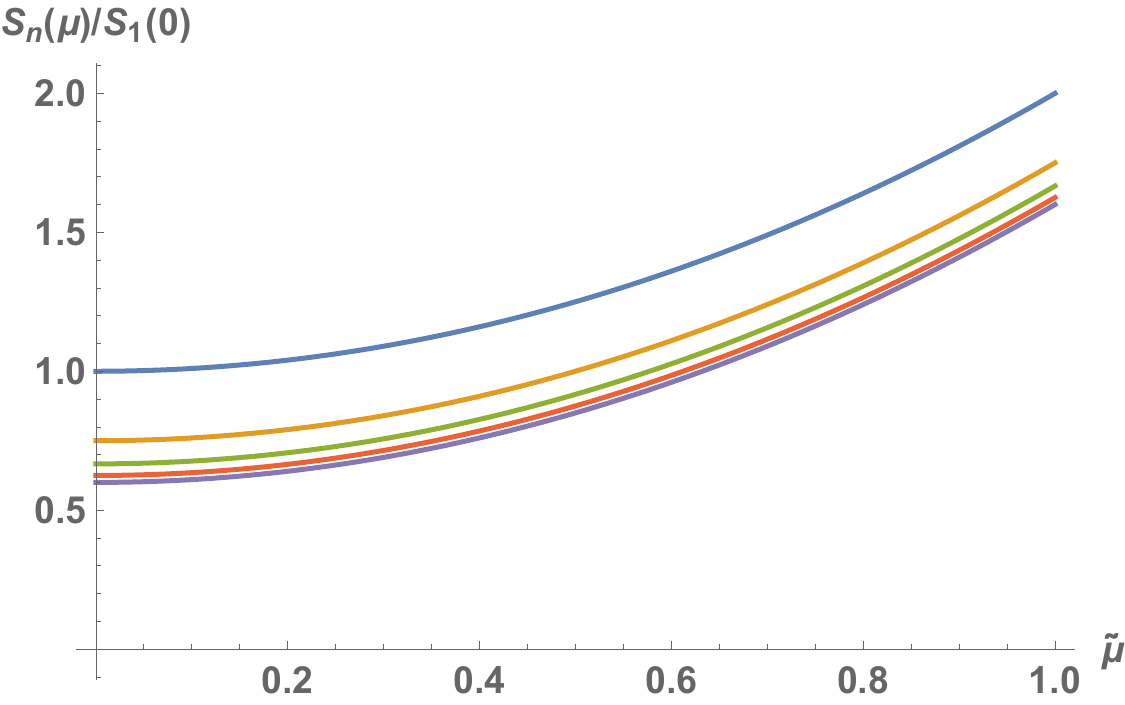}
  \includegraphics[scale=0.4]{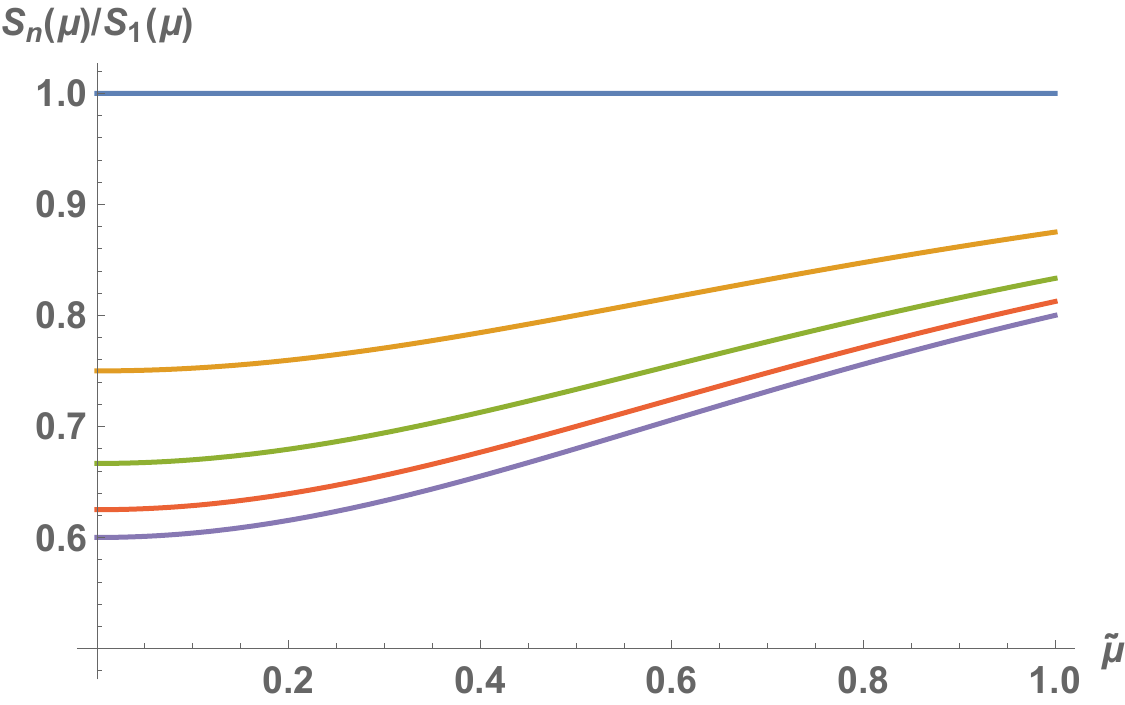}
  \captionof{figure}{Corrected charged Renyi entropy for three-dimensional gravity coupled to conformal electrodynamics with a UV cutoff such that $\V = 10^{40}$ as a function of the chemical potential normalized by the entanglement entropy of the uncharged limit $S_1(0)$ on the left panel, and normalised by the entanglement entropy $S_1(\mu)$ on the right panel. From top to bottom, the curves corresponds to $n = 1,2,3,4,5$~.}
  \label{fig:c2}
  \end{center}
\end{figure}

\begin{figure}[t!]
\begin{center}
  \includegraphics[scale=0.4]{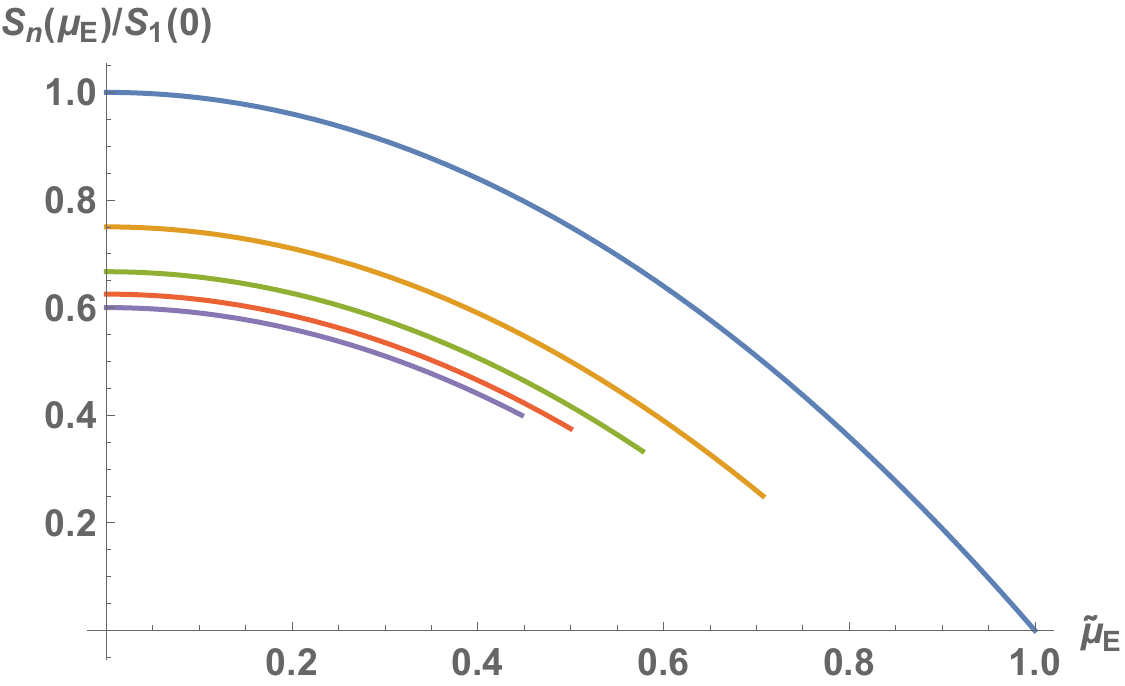}
  \includegraphics[scale=0.4]{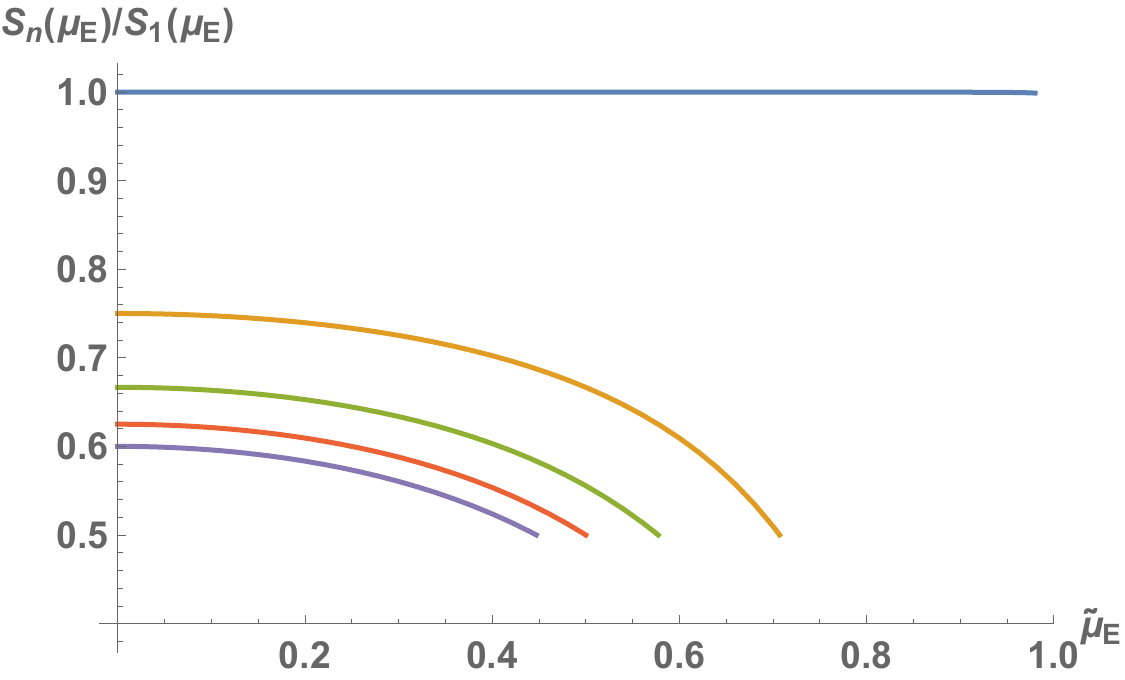}
  \captionof{figure}{Corrected Renyi entropy with imaginary chemical potential for Coulomb sources. The UV cutoff is chosen such that $\V = 10^{40}$~. From bottom to top, the curves corresponds to $n=1,2,3,4,5$~.}
  \label{fig:c3}
  \end{center}
\end{figure}

Notice that now $\tmu_{\rm E}\partial_{\tmu_{\rm E}}S_n$ remains positive if $\C \geq 0$ but now depends on $n$~. As can be seen in \autoref{fig:c1}, the Renyi entropy approaches an asymptotic value which depends on the chemical potential. In \autoref{fig:c2}, it can be seen that the number of entangled pairs gets larger with $\tmu$~, and in \autoref{fig:c3} the Renyi entropy decreases with $\tmu_{\rm E}$ becomes valid only when $\tmu_{\rm E}n \geq 1$~, showing that the analytic continuation is valid only within a finite window for each $n$~. 

Finally, the structure coefficient of the $\langle J J \rangle$ correlator becomes
\begin{align}
    C_V = 64\pi^2\left(\frac{\cR}{\ell}\right)~,
\end{align}
which does not get a modification in terms of $\C$ as we do not consider quantum backreactions on the geometry, and the conformal weights of the holographic twist operators read
\begin{align}
    h_n = \frac{\pi\ell}{\kappa}\left[\left(n -\frac{1}{n}\right) + n\left(\frac{\mu_{\rm E}}{2\pi}\right)^4\right] - \frac{\C}{V_{H^1}}n(1-x_n)~,\qquad \partial_n h_n\rvert_{n\to1,\mu\to0} = \frac{2\pi\ell}{\kappa} - \frac{\C}{V_{H^1}}~,
\end{align}
showing that the corrections imply a decreasing on the value of the dual central charge. 

Similar results apply to the pure Maxwell theory in arbitrary dimensions (see Appendix A of \cite{Arenas-Henriquez:2022ntz}) showing universality of how the corrections play a role in the dual CFT. 
As a final comment, we would like to point out that the entanglement entropy (also in the Maxwell theory with $d>2$) becomes modified due to the logarithmic corrections of the form
\begin{align}
    S_{\rm E} = \V(1-\tmu^2_{\rm E}) - \C\log\left(\V(1-\tmu_{\rm E}^2)\right)~.
\end{align}
This behaviour has been observed in systems with a spontaneous breaking of a contiuous symmetry due to low energy excitatins of the ground state \cite{kallin2011anomalies, Metlitski:2011pr, Alba2013EntanglementSO}. The subleading divergences depend on the system size and the coefficient is related with the number of Goldstone modes $\tilde{N}$ \cite{Metlitski:2011pr} associated with the symmetry breaking as 
\begin{align}
\C = \frac{\tilde{N}(d-1)}{2}~.
\end{align}

Holographically, this imposes a constraint on how the coefficient of the logarithmic corrections appearing in the thermal and quantum fluctuations in the gravity side behave, implying that $2\C/(d-1)$ must be an integer positive number, and show how low energy fluctuations on the gravity side are mapped on the CFT. 
In \cite{Faulkner:2013ana}, the logarithmic corrections is mapped as the relative entropy between the regions disconnected by the Ryu--Takayanagi surface. They show that this relative entropy produces, in the large $N$ limit of the Klebanov--Strassler theory \cite{Klebanov:2000hb}, the same logarithmic corrections, which in the field theory side correspond a spontaneously broken global symmetry produces by massless bulk excitations. This indicates that the corrections that are zero-order in the gravitatinal coupling constant are mapped to this phenomena. Similar corrections can also be found in \cite{Fujita:2009kw} for several holographic theories, and appear by considering $1$-loop contributions to the gravity partition function. 

\newpage
%%%%%%%%%%%%%%%%%%%%%%%%%%%%%%%%%%%%%%%%%%%%%%%%%
%%%%%%% Acceleration and Entanglement %%%%%%%%%%%
%%%%%%%%%%%%%%%%%%%%%%%%%%%%%%%%%%%%%%%%%%%%%%%%%
\section{Acceleration and Entanglement}\label{Sec:Acceleration}

Three-dimensional gravity is trivial at the classical level, in the sense that there are no gravitational waves, i.e., propagating degrees of freedom. Furthermore, as in the four-dimensional case, the theory is non-renormalizable by power counting. However, this does not imply that the theory can not be quantized as the theory can be viewed as a topological field theory \cite{Witten:1988hc}.  Nonetheless, in \cite{Maloney:2007ud} compute the sum over all known contribution to the three-dimensional AdS gravity partition function, i.e., all three-dimensional Euclidean geometries that at the boundary are conformal to a two-torus with Brown--Henneaux excitations \cite{Brown:1986nw}, and shown that the resulting path integral does not have a physical interpretation as the partition function contains an infinite tower of negative states breaking modular invariance at the boundary\footnote{In \cite{Keller:2014xba, Benjamin:2019stq}, using modular boostrap techniques, similar results have been obtained.} (see also \cite{Yin:2007gv} for a similar result). The negative result also extends to the supersymmetric case.  This indicates that the quantum theory of pure three-dimensional AdS gravity may not exist. There have been several attempts to cure this problem \cite{Alday:2019vdr,Maxfield:2020ale, Benjamin:2020mfz, DiUbaldo:2023hkc, Abajian:2023bqv, Collier:2023fwi}. A straightforward way out of the puzzle is the fact that there may be some contributions that were not considered in the sum. The authors argue that solutions that include cosmic strings, such as the string-based models of three-dimensional gravity \cite{Dijkgraaf:2000fq} that always contain such topological defects. Other possibility is to admit that the minimal pure gravitational theory does not include as a quantum theory, and one must include matter fields. The minimal extension would correspond to couple the theory with a scalar field.

In this section, we present a novel family of solutions of three-dimensional gravity conformally coupled with a scalar field. The solutions have acceleration, and a subset of them are accelerating black holes. The acceleration is due to a cosmic string in the spacetime which split the spacetime in half and fields around the wall obey generalized Israel conditions \cite{Israel:1966rt, Aviles:2019xae} that we treat as dynamical field equations following \cite{Charmousis:2006pn}. We show that, due to acceleration, the geometries contribute non-trivially to the gravitational path integral already at the Euclidean saddle. We follow to compute the RT entanglement entropy of a particular accelerating black hole using the Dong's Wald-like formulation of holographic entanglement entropy for bulk theories with matter fields \cite{Dong:2013qoa}. We discovered that the computation becomes significantly challenging due to the absence of conservation laws. Nonetheless, in the hairless case we are able to identify the RT minimal surface and its area is obtained in a small acceleration expansion. 

We found that the entanglement entropy decreases due to acceleration indicating some loss of information in the dual field theory.

%%%%%%%%%%%%%%%%%%%%%%%%%%%%%%%%%%%%%%%%%%%%
%%%%%%%%%%%%%%%%%%%%%%%%%%%%%%%%%%%%%%%%%%%%
%%%%%%%%%%%%%%%%%%%%%%%%%%%%%%%%%%%%%%%%%%%%
\subsection{Accelerating Black holes in  Three Dimensions}
Let us consider Einstein gravity with a self-interacting conformally coupled scalar field. The action reads
\begin{align}\label{ECCs}
    I_{\rm EC} = \int_{\cal M} \d^3x\sqrt{-g}\left(\frac{R-2\Lambda}{2\kappa} - \frac12 (\partial\phi)^2 - \frac{1}{16}R\phi^2 - \lambda \phi^6\right)~,
\end{align}
where the cosmological constant is arbitrary and $\kappa = 8\pi G_3$. Conformally coupled scalar fields have been used to model quantum effects within gravitational theories on black hole backgrounds \cite{Phillips:1996em}, such that they serve to study quantum gravity non-perturbatively. 
Morevoer, the theory is better behave than minimally coupled scalar as proffer the description of wave propagation \cite{Sonego:1993fw,Noakes:1983xd} and has a well-defined Cauchy problem \cite{Noakes:1983xd}.
The theory can be mapped to a minimally coupled scalar field with Higgs-like potentials by field redefinition \cite{Henneaux:2002wm, Hertog:2004dr, Papadimitriou:2007sj}. In the four-dimensional case, the theory can be embedded as a maximally supersymmetric supergravity, which can be uplifted to M-theory, where domain walls \cite{Papadimitriou:2006dr} and instantons \cite{deHaro:2006wy} solutions have been found\footnote{For extensions of this theories in the presence of torsion see \cite{Aviles:2024muk}.}.

Remarkably, the first black hole solution found by Ruffini \cite{Ruffini:1971bza} and Bekenstein \cite{Bekenstein} in four dimensions are the first examples in the literature that circumvent the no-hair theorem. In three dimensions, the first black hole solution in AdS gravity was found in \cite{Martinez:1996gn} when there is no hexic self-interacting term, i.e., $\lambda = 0$~, and extended to the full theory in \cite{Henneaux:2002wm}. The authors shown that the scalar field decay non-trivially contributing non-trivially to boundary variations and, therefore, to Hamiltonian charges, such that the asymptotic symmetry group differs from the one of the BTZ solution. Nonetheless, relaxing consistently the Brown--Henneaux boundary conditions \cite{Brown:1986nw}, the canonical generators of asymptotic symmetries form the Virasoro algebra with the Brown--Henneaux central charge \cite{Henneaux:2002wm}.
A striking feature of the theory, is that the non-trivial coupling between the scalar field and the Ricci tensor can be seen on-shell as a massive term of the scalar as the matter sector posses conformal symmetry, i.e.,
\begin{align}
    T^\mu{}_\mu = 0 \Longrightarrow R = 6\Lambda~, 
\end{align}
such that the scalar field has a mass $ \tilde {m}^2 = 3\Lambda/4$~, which in the AdS case, lies within the Breitenlohner--Freedman bound \cite{Breitenlohner:1982bm} and implies the stability of the solutions when different boundary conditions are considered \cite{Klebanov:1999tb}. Therefore, exact solutions to this theory admit mixed boundary conditions and are suitable to study multi-trace deformations in the dual CFT. The conformal dimension of the dual scalar operator can be $\Delta_\pm = 1 \pm 1/2$~, and the Lorentzian bulk-to-bulk propagators have support only on the light cone \cite{Breitenlohner:1982bm}. 
These properties have been used to study the decay of the conformal vacuum in the dual theory using bulk instantons \cite{deHaro:2006wy, Papadimitriou:2007sj}.

The field equations are
\begin{align}\label{EOMccs}
    R_{\mu\nu} - \tfrac12 R g_{\mu\nu} + \Lambda g_{\mu\nu} ={}& \kappa T^{\rm EC}_{\mu\nu}~, \nonumber \\ \Box \phi-\tfrac18 R \phi - 6 \lambda \phi^5 ={}& 0~,
\end{align}
where
\begin{align}
    T_{\mu\nu}^{\rm EC} := \partial_\mu\phi\partial_\nu\phi - \frac12 g_{\mu\nu}(\partial \phi)^2 + \frac18\left(R_{\mu\nu} - \frac12 R g_{\mu\nu} +g_{\mu\nu}\Box-\nabla_\mu\nabla_\nu \right)\phi^2 - g_{\mu\nu}\lambda \phi^{6}~,
\end{align}
is the stress tensor of the theory. 
Considernig statical symmetry solutions
\begin{align}\label{staticmetric}
    ds^2 = -f(r)\d t^2 + \frac{\d r^2}{f(r)} + r^2\d\theta^2~,
\end{align}
due to the conformal nature of the matter sector, the trace of the field equations implies that the scalar field backreacts in the metric function $f(r)$ with a term of order $1/r$. In higher dimensions, the new term decays as $\sim 1/r^{4D+2}$, where $D$ is the bulk dimension, and also apply to Maxwell matter fields \cite{Hassaine:2007py} (see \autoref{SubSec:ConfBosons}).
We have found that such backreaction also holds even when we go beyond the ansatz \eqref{staticmetric} and introduce acceleration that breaks the spherical symmetry. The metric in polar-like coordinates
\begin{align}\label{Hacc}
    ds^2 = \frac{1}{\Omega(r,\theta)^2}\left( -P(r)\d t^2 + \frac{\d r^2}{P(r)} + r^2 \frac{\d \theta^2}{Q(\theta)} \right)~,
\end{align}
with
\begin{align}
    \Omega ={}& \sigma(\cA r\cos(m\sqrt{\sigma}\theta)-1)~, \\ P(r) ={}& -\Lambda r^2 + \sigma m^2(1-\cA^2 r^2)\left(1-\frac{q}{r}\right)~, \\ Q(\theta) ={}& 1-\cA q \cos(m\sqrt{\sigma}\theta)~,
\end{align}
together with the scalar field
\begin{align}
    \phi(r,\theta) =\sqrt{\frac{8}{\kappa}}\sqrt{\frac{\Omega(r,\theta)}{\sigma(s r -1)}}~, 
\end{align}
where $\sigma$ is a sign and $q, m$, and $s$ are integration constants, are solutions to the field equations \eqref{EOMccs} provided that
\begin{align}\label{constraintsacch}
    q = \frac{2s}{3s^2-\cA^2}~,\qquad \lambda =-\frac{\kappa^2}{512}\left(\Lambda - \frac{\sigma m^2(s^2-\cA^2)^2}{3s^2-\cA^2}\right)~.
\end{align}
The weak field limit of the solution, i.e., $m = 0 = q$~, renders the geometry into an accelerated Rindler observers with acceleration equal to $\cA/m$, indicating that the solution corresponds to a three-dimensional accelerating black hole \cite{Anber:2008qu, Anber:2008zz, Astorino:2011mw, Arenas-Henriquez:2022www}, see \cite{Arenas-Henriquez:2023vqp} for further details.

Accelerating black hole have this atypical property that conformal infinity is located by a surface parametrized by the zero locus of ~$\Omega(r,\theta)$~, henceforth referred to as the conformal function, rather than a single point, as usually given by radial infinity for AlAdS solutions. Nonetheless, one could change coordinates such and render the conformal function to be a function of a single new coordinate, but then the metric function $f(r)$ acquires a new dependence and the horizon radii are now given by a surface rather than a single point of the radial coordinate. This peculiarity makes the holographic analysis of the accelerating solutions rather different than for non-accelerating solutions. We will further comment on this issue on \autoref{Sec:Conc}. 

The solution \eqref{Hacc}, has a non-trivial Cotton tensor, therefore is a non-conformally flat accelerating black hole with a curvature singularity showed by the Kreschmann scalar, viz.,
\begin{align}
R_{\mu\nu\rho\sigma}R^{\mu\nu\rho\sigma} = 12\Lambda^2 + \frac{6m^4q^2\Omega(r,\theta)^6}{r^6}~,
\end{align}
which diverges at $r=0$~. Indicating a black hole singularity at this point.
Moreover, the solutions are either asymptotically locally (A)dS or flat, depending on the value of the cosmological constant\footnote{We use conventions where $T_{[ab]} = \frac12(T_{ab} - T_{ba})$ and $T_{(ab)} = \frac12(T_{ab} + T_{ba})$ for any tensor $T$~.}
\begin{align}
    \lim_{\Omega\to 0}R^{\mu\nu}{}_{\rho\sigma} = 2\Lambda \delta^{[\mu}{}_{\rho}\delta^{\nu]}{}_{\sigma}~.
\end{align}

Nonetheless, in order to have a proper interpretation of the solution to be a black hole, we need at least one compact Killing horizon that encloses the singularity. We first notice that the horizon locations are given by the vanishing of the metric function $f(r)$, as the Killing vector field $\chi^2=(\partial_t)^\mu$ has norm
\begin{align}
    \chi^2 = -\frac{f(r)}{\Omega(r,\theta)^2}~,
\end{align}
which changes from being light-like to space-like and the vice versa each time that the metric function vanishes. 
To analyse the horizon radii let us first introduce a more suitable set of coordinates
\begin{align}
    r = \frac{m}{A y}~,\qquad t = A m \tau~,\qquad x = -\cos\left(m\theta \sqrt{\sigma}\right) ~,
\end{align}
where we have also consider the rescalings
\begin{align}
    A = \cA m~,\qquad \xi = \cA q~, \qquad s = - \cA \alpha~,
\end{align}
that we shall refer to as \emph{canonical coordinates}. The line element \eqref{Hacc} renders
\begin{align}
    ds^2 = \frac{1}{\tilde \Omega(x,y)^2}\left( -\tilde P(y)\d\tau^2 + \frac{\d y^2}{\tilde P(y)} + \frac{\d x^2}{\tilde Q(x)}\right)~,
\end{align}
with
\begin{align}
\tilde \Omega ={}& A(x+y)~, \\ \tilde P(y) ={}& -\sigma(1-y^2)(1-\xi y) - \frac{\Lambda}{A^2}~, \\ \tilde Q(x) ={}& \sigma(1-x^2)(1+\xi x)~, 
\end{align}
and the scalar field
\begin{align}
    \phi(x,y) = \sqrt{\frac{8}{\kappa}}\sqrt{\frac{x+y}{y+\alpha}}~.
\end{align}

We first notice that the scalar field has a pole at $y = -\alpha$~. Therefore, we need to impose the restriction $y > -\alpha$ in order to have a well-defined causal structure. The restriction \eqref{constraintsacch} in canonical coordinates now become
\begin{align}
    \xi = \frac{2\alpha}{1-3\alpha^2}~,\qquad \lambda = -\frac{\kappa^2}{512}\left[\Lambda - \sigma A^2\left(1-\frac{\alpha^2-2\alpha\xi}{3}\right)\right]~,
\end{align}
such that solving for $\alpha$ in terms of $\xi$, we get two possible branches
\begin{align}
    \alpha_\pm = \frac{-1\pm\sqrt{1+3\xi^2}}{3\xi}~.
\end{align}
Depending on the values of $\xi$~, the two branches take possible different domains. For the sake of simplicity, let us consider $\xi>0$ from now on, such that the two branches have each two possible domains, shown in \autoref{table1}.
\begin{table}[h]
\centering
  \begin{tabular}{|c||c||c|}
    \hline
     & $|\xi|<1$ & $|\xi|>1$ \\
    \hline\hline
     $\alpha_+ \in$  & $\left(0,\frac{1}{3}\right)$ & $\left(\frac{1}{3},\frac{\sqrt{3}}{3}\right)$\\
    \hline\hline
    $\alpha_- \in$  & $\left(-\infty,-1\right)$ & $\left(-1,-\frac{\sqrt{3}}{3}\right)$\\
 \hline
  \end{tabular}
\caption{Admissible domains for the two different branches $\alpha_\pm$ depending on the value of $\xi$~.} 
\label{table1}
\end{table}

Another important aspect before studying the solutions causal structure, it is crucial to acknowledge that the function $\tilde Q(x)$ needs to be positive in order to maintain the Lorentzian structure of the metric. Consequently, we need to constrain the possible values of the transverse coordinate $x$, such that we need to analyse each class $\sigma = \pm$ individually. The permissible domains are presented in \autoref{table2}.

\begin{table}[h]
\centering
  \begin{tabular}{|c||c||c|}
    \hline
    Class & $|\xi|>1$ & $|\xi|<1$ \\
    \hline\hline
    \multirow{2}{*}{$\sigma = 1$} & $x<-1$ & $x<-1/\xi$ \\
    \cline{2-3}
    & $-1/\xi < x < 1$ & $-1<x<1 $\\
    \hline\hline
    \multirow{2}{*}{$\sigma=-1$} & $-1<x<-1/\xi$ & $-1/\xi < x < -1$\\
    \cline{2-3}
    & $x>1$ & $x>1$ \\ \hline
  \end{tabular}
\caption{Possible domains for the transverse coordinate depending on the different values of $\xi$. The different possibilities are given by imposing that $\tilde Q(x) = \sigma(1-x^2)(1+\xi x) > 0$~, which imposes two different regimes for each class and for each region $|\xi| \lessgtr 1$~, such that there are 8 possible different domains if we consider $\xi>0$~, if we consider also $\xi<0$~, the possibilities grown by an extra factor of 2.} 
\label{table2}
\end{table}

As can be seen in \autoref{table2},  the transverse coordinates is not necessarily compact and we still lack of a black hole horizon interpretation. In order to achieve such, we follow the method developed in \cite{Arenas-Henriquez:2022www} to construct three-dimensional accelerating black holes. As we already notice that we have non-compact domains for $x$, we first restrict the possible minimum and maximum values such that $x \in [x_0,x_{\rm brane}]$~. We then, take a mirror copy of the spacetime, and glue with the original spacetime along some of this two surfaces, say $x_{0}$ and then identify the two surfaces $x = x_{\rm brane}$~. The resulting spacetime has a compact direction $x$ but contains two domains walls/cosmic strings\footnote{In three dimensions, a domain wall has the same world-volume as a cosmic string such that their names can be used interchangeably. Therefore, we will also sometimes refer to the domain wall energy density to as \emph{tension}.} produced by the gluing procedure. 

The presence of the domain wall splits the spacetime equally along the transverse coordinate such that there is a $\mathbb{Z}_2$ symmetry. Therefore, we shall write all quantities assuming this dihedral split of the geometry. 
Taking $x_i = \{x_0,x_{\rm brane}\}$~, the domain walls line element read
\begin{align}
    ds^2 = \gamma_{MN}dy^M dy^N = \frac{1}{A^2(x_i+y)^2}\left(-\tilde P(y)\d \tau^2 + \frac{\d y^2}{\tilde P(y)}\right)~,
\end{align}
such that the outward pointing normal to the wall reads
\begin{align}
    n^M = \frac{\tilde Q(x_i)^{-\frac12}}{\tilde \Omega(x_i,y)}\left(\frac{\partial}{\partial x}\right)^{M}~,
\end{align}
and the associated extrinsic curvature
\begin{align}
    {\cal K}_{MN} := \frac12 {\cal L}_n \gamma_{MN} = A\gamma_{MN}\sqrt{\tilde Q(x_i)}~.
\end{align}

\begin{figure}[htbp]
  \begin{minipage}[t]{0.45\linewidth}
    \centering
\includegraphics[scale=0.42]{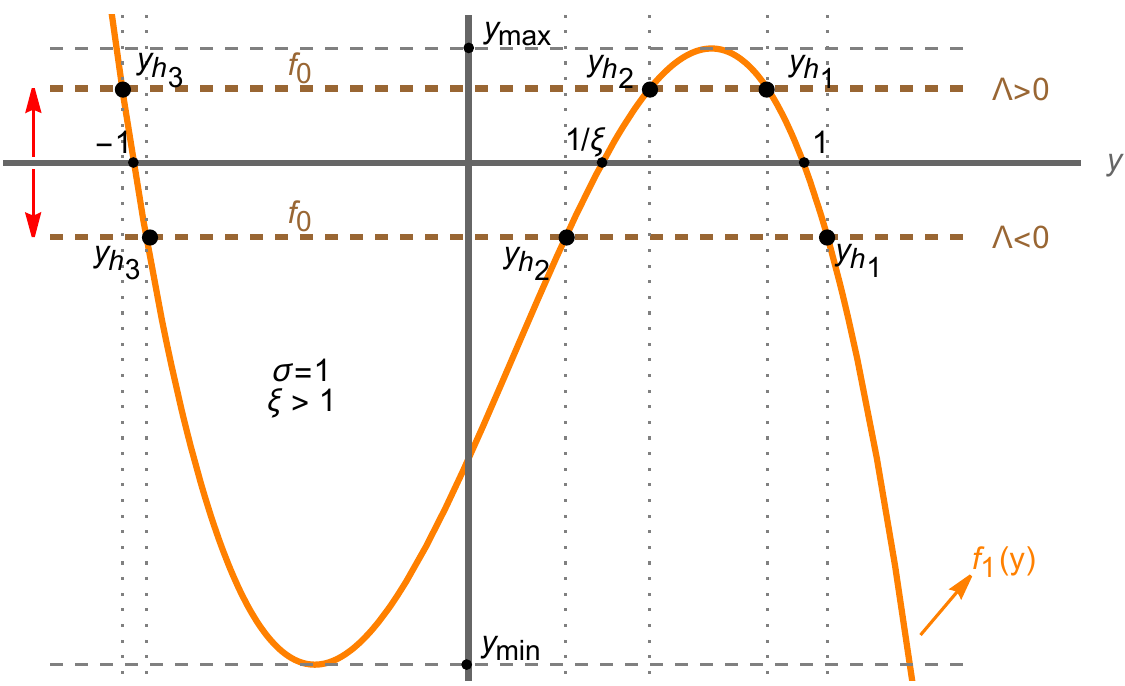}
  \end{minipage}
  \hfill
  \begin{minipage}[t]{0.45\linewidth}
    \centering
   \includegraphics[scale =0.42]{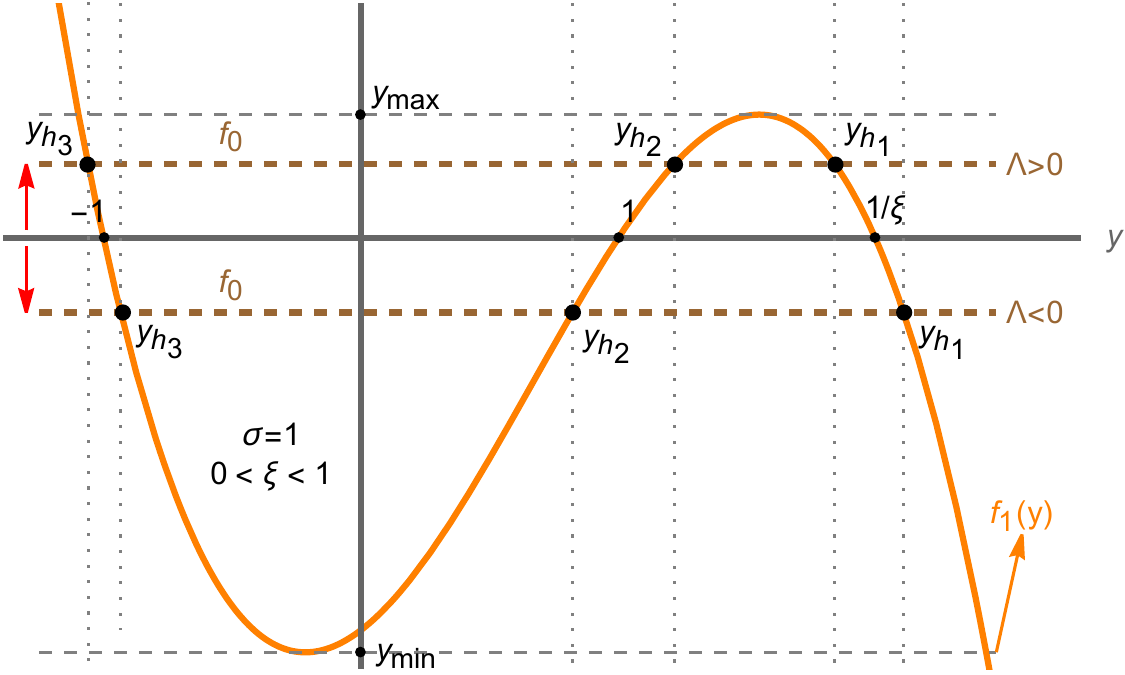}
  \end{minipage}
  \caption{Horizon structure of the class $\sigma=1$~.}
  \label{Fig:sm1hor}
\end{figure}

\begin{figure}[htbp]
  \begin{minipage}[t]{0.45\linewidth}
    \centering
\includegraphics[scale=0.42]{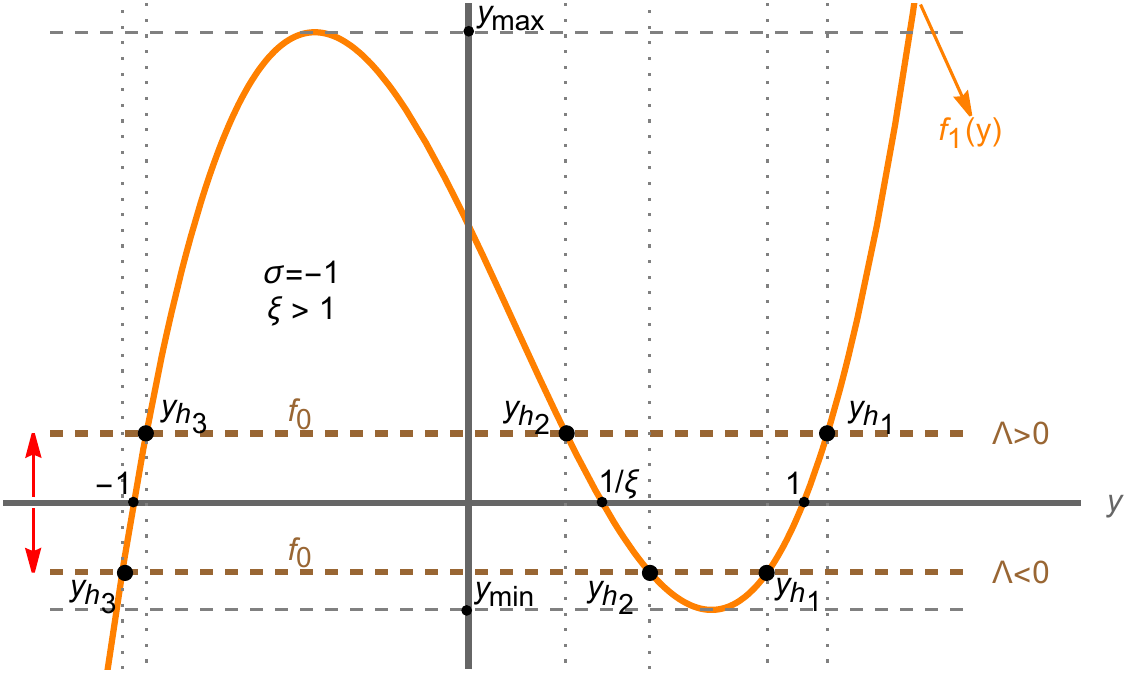}
  \end{minipage}
  \hfill
  \begin{minipage}[t]{0.45\linewidth}
    \centering
   \includegraphics[scale =0.42]{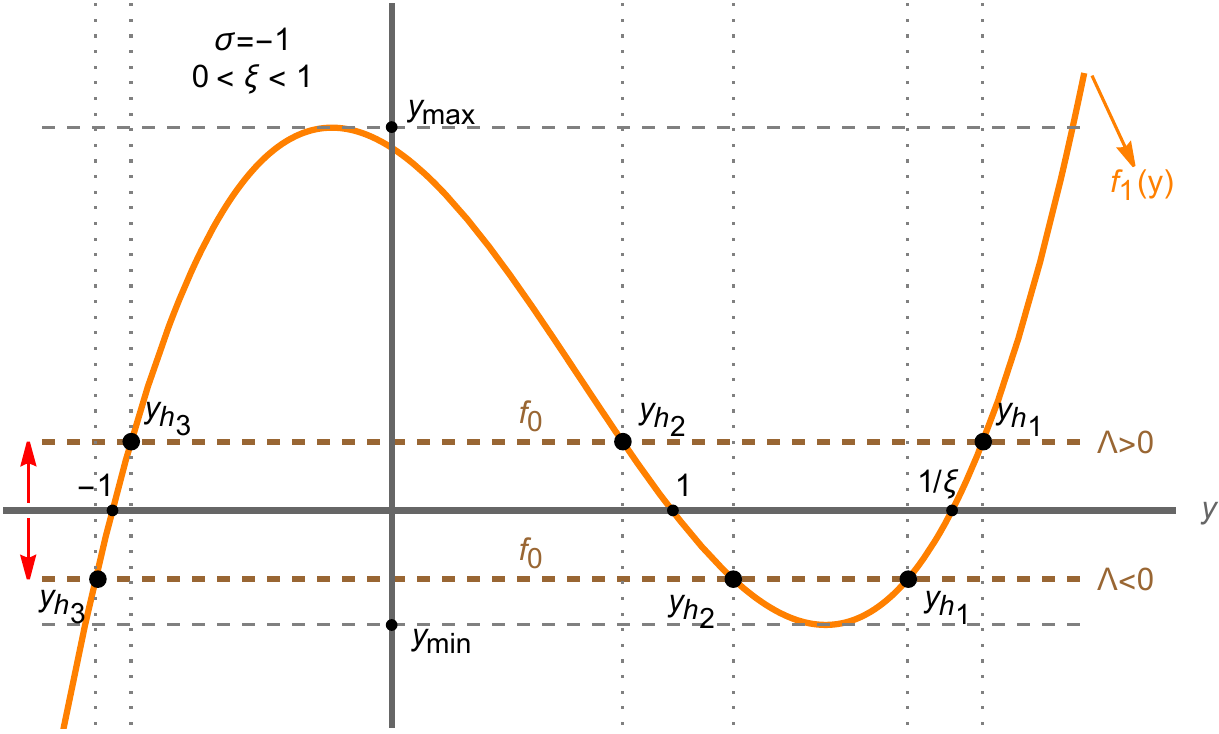}
  \end{minipage}
  \caption{Horizon structure of the class $\sigma=-1$~.}
  \label{Fig:sp1hor}
\end{figure}

At the domainw wall, the canonical momenta associated with dynamical fields jump along the surface producing a non-trivial energy density \cite{Israel:1966rt}, which in the case at hand is given by the generalized Israel junction conditions found in \cite{Aviles:2019xae} for scalar-tensor theories. The domain wall stress tensor reads
\begin{align}
    T_{MN} = -2f(\phi)\left([{\cal K}_{MN}] - \gamma_{MN}[{\cal K}]\right)+2n^M[\partial_M \phi]f'(\phi)\gamma_{MN}~,
\end{align}
subject to
\begin{align}
  n^M [\partial_M \phi] = 2f'(\phi)[{\cal K}]~.
\end{align}
Here $[Y]:=Y^+ - Y^-$ corresponds to the difference of some quantity $Y$ evaluated on each side of the wall, which in our case $[Y] = 2Y|_{x_i}$ due to the $\mathbb{Z}_2$ symmetry. Finally, $f(\phi) = \frac{1}{2\kappa}\left(1-\frac{\kappa}{8}\phi^2\right)$ is the coupling between the scalar field and the Ricci scalar in the Lagrangian \eqref{ECCs}. 
Therefore, the string energy density reads
\begin{align}\label{mudw}
   \mu = \gamma^{MN}\int_-^+ \d x~ T_{MN} = \frac{2A}{\kappa}\sqrt{\tilde Q(x_i)} = \frac{2m\cA}{\kappa}\sin(m\theta\sqrt{\sigma})\sqrt{Q(\theta_i)}~.
\end{align}
In order to have a single domain wall configuration, we will consider the point of gluing $x_0$ to be a \emph{tensionless} points, which are given by $x_0 = \{-1,1,1/\xi\}$ as $\tilde Q$ vanish at this points. Therefore, the spacetime has a curvature singularity with compact horizon's with a single domain wall located at some $x_{\rm brane}$~. In polar-like coordinates, we use the free parameter $m$ as
\begin{align}
    x_{\rm brane} = -\cos(m\pi\sqrt{\sigma})~,
\end{align}
such that $\theta \in [-\pi,\pi]$ is a compact coordinate, and the location of the string is controlled by $m$~. 
In order to understand the Killing horizons, let us split the metric function as
\begin{align}
    \tilde P(y) = -f_0 + f_1~,\qquad f_1 :=-\sigma(1-y^2)(1-\xi y)~,\qquad f_0:=\Lambda/A^2~,
\end{align}
such that the intersection of $f_0$ and $f_1$ occurs when $\tilde P$ vanishes corresponding to an horizon location. 
Clearly, the different values of the radii depend on the value of the cosmological constant\footnote{We assume during the whole section that the acceleration is positive.} and the value of $\sigma$. As can be seen in \autoref{Fig:sm1hor} and \autoref{Fig:sp1hor}, for both classes the solution has from one to three Killing horizons. In the case of $\Lambda = 0$~, the solution has three horizons at $y = \{-1,1,1/\xi\}$~. 

\begin{figure}[h!]
    \centering
    \includegraphics[scale=1.8]{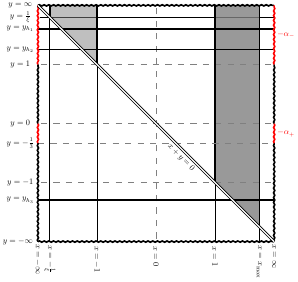}
    \caption{Constant time diagram of class $\sigma=-1$ with $|\xi|<1$ black hole. The maximum value of $x$ is restricted to be $x_{\rm max}$~. The diagonal doubled line corresponds to the conformal boundary and the grey area is the initially allowed domain of the transverse coordinate. On the vertical axes, the red region correspond to the two possible branches $\alpha_\pm$~, such that choosing one of the branches, any point in the red sector is allowed, and the $y$ coordinate must be chosen such that $y > -\alpha_\pm$~. }
    \label{fig:example1}
\end{figure}

Finally, in order to understand the physical properties of the horizons, let us consider the class $\sigma = -1$ with   $|\xi| <1$ as a particular example. As aforementioned, we need to restrict the possible values of the $y$ coordinate such that the scalar field is well-defined everywhere. The causal structure of this case can be seen in \autoref{fig:example1}. 

\begin{figure}[h!]
    \centering
    \includegraphics[scale=1.2]{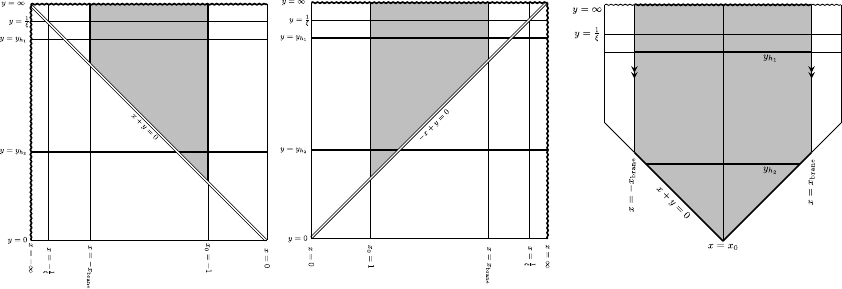}
    \caption{Causal structure of Hairy accelerating black hole with $\sigma = -1$ and $|\xi|<1$ on the branch $\alpha_+ \in (0,\tfrac13)$~. The solution has two Killing horizons labeled by $y_{h_1}$ and $y_{h_2}$~. The second horizon $y_{h_2}$ corresponds to a Rindler horizon.}
    \label{fig:example1a}
\end{figure}

We can now consider the $\alpha_-$ branch, and focus on the upper left corner, given by the domain $x \in (-1,-1/\xi)$~. In this case, we insert a domain wall at $x = x_{\rm brane}$ and let the tensionless point $x_0 = -1$ where the gluing surface. The procedure is shown in \autoref{fig:example1a}.  

\begin{figure}[h!]
    \centering
    \includegraphics[scale=1.2]{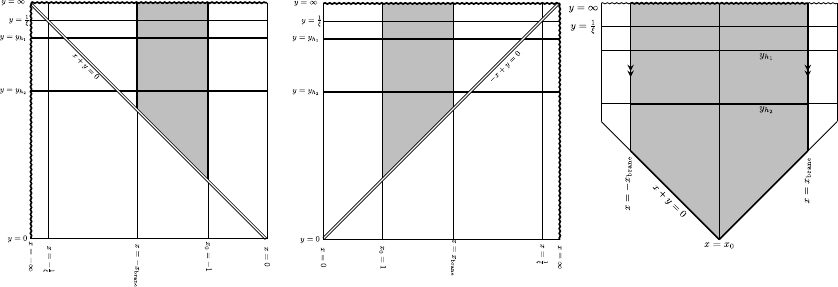}
    \caption{Causal structure of Hairy accelerating black hole with $\sigma = -1$ and $|\xi|<1$ on the branch $\alpha_+ \in (0,\tfrac13)$~. The solution has two Killing horizons labeled by $y_{h_1}$ and $y_{h_2}$~.}
    \label{fig:example1b}
\end{figure}

The black hole contains two Killing horizons $y_{h_1}$ and $y_{h_2}$~. The second horizon is not compact, and touches conformal infinity. This horizons are termed \emph{Rindler horizons} or \emph{Accelerating horizons} and appear naturally in accelerating systems. Accelerating black holes with at least one Rindler horizon are referred to as \emph{rapidly accelerating black holes} or to be in a \emph{rapid phase}. 
In order to avoid the accelerating horizon, we can move the location of the cosmic string as shown in \eqref{fig:example1b} where the domain is restricted further such that the second horizon does not touch conformal infinity. Accelerating black holes with no Rindler horizons are refereed to as \emph{slowly accelerating black holes} or to be in a \emph{slow phase}. This phenomena also occurs in four dimensions when a negative cosmological constant is considered \cite{Dias:2002mi}, in our case always is possible to have such structure. The slow phases are suitable to study thermodynamics of accelerating black holes as in principle can be seen as system in thermal equilibrium eventhough are accelerating, see for instance \cite{Gregory:2017ogk}, and reference therein. 
As can be seen from both cases, the singularity lies at $y = \infty$, which correspond to $r = 0$~, and is enclosed by the first horizon $y_{h_1}$~. As previously mentioned, there is always at least one horizon (which can be either Rindler or compact) so there is no possible naked singularity as a solution.
This is just one instance of construction; various other possibilities exist, as outlined in \cite{Cisterna:2023qhh}.

All the aforementioned properties appear in the four-dimensional avatar (see for instance \cite{Griffiths_Podolsky_2009}, and reference therein), showing that the inclusion of a conformally coupled scalar is suitable, not just due to the ease in which backreacts at the metric, but also because the study of horizons structure is now resembled at three dimensions. 

As a final comment before moving to the holographic analysis of the solutions, we will now focus on the different limits of \eqref{Hacc}.
Let us first focus on the $\sigma = -1$ class. As can be seen from \eqref{constraintsacch}, there are two possible different limits on $s$ that make the scalar contribution to vanish. The first case correspond to $s\to \infty$~, in which the scalar field vanishes and the metric becomes the one of the accelerating BTZ \cite{Arenas-Henriquez:2022www} when $\Lambda <0$. The field equations and action principle become simply the ones of pure Einstein gravity in three dimensions. This solution is continuously connected to the BTZ solution in the zero-acceleration limit $\cA \to 0$~. Then, class $\sigma=-1$ is dubbed the \emph{hairy accelerating BTZ}.
Another interesting limit, is to consider $s\to 0$~, which also makes $q$ to vanish, but not the scalar profile. Nonetheless, the stress tensor vanishes in this limit
\begin{align}
    \lim_{s\to 0}T_{\mu\nu} = 0~.
\end{align}
Such solutions are referred to as stealth configurations. In this case, we have a stealth over the accelerating BTZ. 
Let us finish with the zero-acceleration limit $\cA\to 0$~, which recovers the HMTZ solution \cite{Henneaux:2002wm} when $\Lambda = -1/\ell^2$. This can be seen by using the redefinition's
\begin{align}
    q = -\frac{2B}{3}~,\qquad m^2= \frac{3B^2}{\ell^2}(1+\nu)~,\qquad s = -B^{-1}~,
\end{align}
such that in the limit $\cA\to0$~, the metric, scalar field, and coupling constant becomes
\begin{align}
    ds^2 = -F(r)\d t^2+\frac{\d r^2}{F(r)} + r^2\d \theta^2~,\\ 
    F(r) = \frac{r^2}{\ell^2}-(1+\nu)\left(\frac{3B^2}{\ell^2}+\frac{2B^3}{\ell^2r}\right)~,
\end{align} 
\begin{align}
    \phi(r) = \sqrt{\frac{8}{\kappa}}\sqrt{\frac{B}{B+r}}~,\qquad \lambda = -\frac{\kappa^2\nu}{512\ell^2}~,
\end{align}
respectively.
The solution is not connected with the BTZ black hole, as the limit $B\to0$ makes the scalar profile to vanish, but the metric function becomes simply the one of the \emph{massless} BTZ, i.e., $F(r) = r^2/\ell^2$~.

\begin{figure}[h!]
    \centering
    \includegraphics{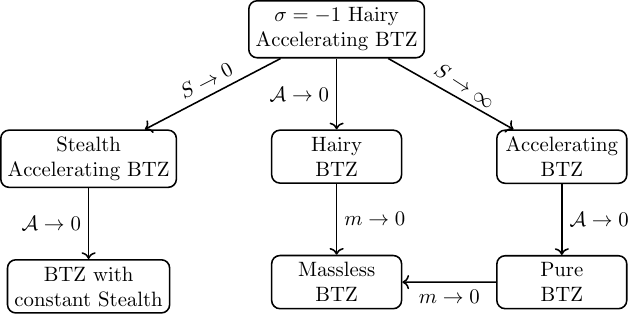}
    \caption{Limits of the $\sigma = -1$ hairy accelerating black hole.}
    \label{fig:hierm1}
\end{figure}

The full hierarchies of solutions contained in \eqref{Hacc} for the $\sigma=-1$ class with negative cosmological constant can be seen in \autoref{fig:hierm1}.

The class $\sigma = 1$~, has similar limits. The limit $s\to\infty$ makes the scalar field to vanish and the solution becomes the class I$_{\rm C}$ found in \cite{Arenas-Henriquez:2022www}. This solution is an accelerating black hole that is not connected with the BTZ solution as in order for the solution to exists, the acceleration must be finite. The limit $s\to0$ becomes a stealth over the I$_{\rm C}$ black hole, and the zero acceleration limit $\cA\to 0$ recovers the HMTZ solution as well. 

\begin{figure}[h!]
    \centering
    \includegraphics{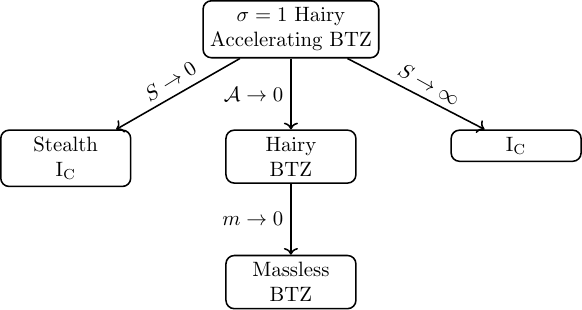}
    \caption{Limits of the $\sigma = -1$ hairy accelerating black hole.}
    \label{fig:hierp1}
\end{figure}

The full hierarchies of solutions contained in \eqref{Hacc} for the $\sigma=1$ class with negative cosmological constant can be seen in \autoref{fig:hierp1}.

%%%%%%%%%%%%%%%%%%%%%%%%%%%%%%%%%%%%%%%%%%%%%
%%%%%%%%%%%%%%%%%%%%%%%%%%%%%%%%%%%%%%%%%%%%%
%%%%%%%%%%%%%%%%%%%%%%%%%%%%%%%%%%%%%%%%%%%%%
%%%%%%%%%%%%%%%%%%%%%%%%%%%%%%%%%%%%%%%%%%%%%
\subsection{Holography and Acceleration}

In this section, we will study the holographic properties of the hairy accelerating black holes, and see how acceleration modify the dual theory. In order to explore the implications of the AdS/CFT correspondence in this solution, let us analyse the class $\sigma =-1$ with negative cosmological constant. 
As explained in the previous section, accelerating black holes have a conformal infinity at non-constant surface, which in the case at hand is given by
\begin{align}
    r = \frac{1}{\cA \cosh(m\theta)}~.
\end{align}
This striking feature, makes hard to identify the holographic coordinate. A possible resolution, is to cast the metric in the Fefferman--Graham (FG) gauge \cite{Fefferman:1984asd} 
\begin{align}\label{FG}
    ds^2 = \frac{\ell^2\d\rho^2}{4\rho^2} + \frac{1}{\rho}g_{ij}(\rho,x^i)\d x^i \d x^j~,
\end{align}
where the metric at constant $\rho$ admits a power expansion 
\begin{align}
    g_{ij} = g_{(0)ij} + \rho\left( g_{(2)ij} + \tilde{g}_{(2)ij}\log \rho \right) + \dots~,
\end{align}
by applying the coordinate transformation proposed in \cite{Anabalon:2018ydc} 
\begin{align}
    \frac{1}{\cA r} ={}&  \xi + \sum_{n=1}^\infty F_n(\xi)\rho^n~, \\ \cosh(m\theta) ={}& -\xi + \sum_{n=1}^\infty G_n(\xi)\rho^n~,
\end{align}
and solve order by order the functions $F_n(\xi)$ and $G_n(\xi)$ by imposing the FG fall-off \eqref{FG}, that contains no cross terms $g_{\rho i}$. This procedure allows to identify the holographic radial coordinate $\rho$, but the first order of one of the functions, $F_1(\xi)$ can not be obtained and appears as a conformal factor of the boundary metric, showing explicitly the conformal degree of freedom in the boundary metric\footnote{The AdS spacetime does not have a boundary in the usual sense, but rather a conformal boundary, i.e., a boundary obtained after applying a conformal compactification \cite{Penrose:1985bww, Skenderis:2002wp}. Therefore, any bulk field does not induce a buondary field, but rather a conformal class of it, i.e., a field up to Weyl rescaling.}.  
In the three dimensions, the dual CFT suffers from a Weyl anomaly (see \cite{Duff:1993wm} and reference therein) such that the generating functional is not conformal invariant due to the non-trivial scalar curvature of the theory. Morevoer, in two dimensions, the stress tensor of a CFT is a quasi-primary operator transforming non-trivially under conformal transformations, such that conserved quantities are not invariant under Weyl rescaling. Therefore, in order to study the holographic properties of the three-dimensional accelerating black hole, we follow \cite{Hubeny:2009kz, Cassani:2021dwa} and introduce a new coordinate
\begin{align}
    z := \frac{1}{r} - \cA \cosh(m\theta) \Longleftrightarrow y=z-x~,
\end{align}
such that the conformal boundary is now located at $z = 0$ and the metric takes the form
\begin{align}\label{HaccZcoord}
    ds^2 ={}& \left(N \d z - N_i \d x^i\right)^2 + h_{ij}\d x^i \d x^j =  N^2\d z^2 + h_{ij}(\d x^i - N^i\d z)(\d x^j - N^j\d z)~,
\end{align}
where the shift function
\begin{align}
    N_i = -\frac{\partial_i r}{\Omega(z,x) \sqrt{P(z,x)}}~,
\end{align}
The outward-pointing normal to the $z={\rm const}$ hypersurface is
\begin{align}
    n = \frac{1}{N}\left(N^i \partial_i -\partial_z\right)~,
\end{align}
and the associated extrinsic curvature
\begin{align}
    K_{ij} := \frac12 {\cal L}_n h_{ij} = -\frac{1}{2N}\left(\partial_z h_{ij} - \nabla_i N_j - \nabla_j N_i\right)~,
\end{align}
where the covariant derivative $\nabla_i$ is respect to the induced metric $h_{ij}$~. 
Although the metric \eqref{HaccZcoord} contains cross terms between $z$ and the boundary coordinates $x^i$ and only matches the FG gauge at leading order, the extrinsic curvature near the boundary behaves as
\begin{align}
    K^i{}_j \sim \frac{1}{\ell}\delta^i{}_j + \dots~,
\end{align}
which is enough to obtain holographic data through the holographic stress tensor \cite{Miskovic:2006tm}. We will provide additional commentary on this issue at the conclusion of the section.
The renormalization procedure consist in first considring a cut-off $z = \delta$ such that $h_{ij}$ becomes the induced metric at this hypersurface, and then add boundary terms that preserve general covariance, such that taking the limit $\delta\to0$ gives a renormalized on-shell action and a well-defined variational principle for the boundary metric $g_{(0)ij}$\footnote{As the boundary of AdS is not endowed with a metric structure, but rather a conformal class of metrics, boundary conditions must be imposed on a quantity that has a well-defined transformation under Weyl rescaling. See \cite{Papadimitriou:2007sj} for further details.} as well. 
Then, we can compute holographic $n$-point functions using the renormalized action by first taking variations respect to the induced metric $h_{ij}$ and then take the limit to the boundary as 
\begin{align}
    g_{(0)ij} = \lim_{z\to 0}z^2 h_{ij}~.
\end{align}
Then, we follow to obtain the vacuum expextation value of the holographic stress tensor by computing the quasi-local energy tensor
\begin{align}
    T_{ij}[h] = \frac{2}{\sqrt{-h}}\frac{\delta I_{\rm ren}}{\delta h^{ij}}~,
\end{align}
where $I_{\rm ren}$ is a renormalized gravitational action, which is related with the holographic stress tensor as
\begin{align}
    \langle T^{ij}\rangle = \frac{2}{\sqrt{-g_{(0)}}}\frac{\delta I_{\rm ren}}{\delta g_{(0)ij}} = \lim_{z \to 0}\frac{2}{\sqrt{-g}}\frac{\delta I_{\rm ren}}{\delta g_{ij}} = \lim_{z\to 0}T^{ij}[h]~.
\end{align}
The action \eqref{ECCs} must be supplemented with a generalized Gibbons--Hawking--York (GHY) boundary term of the form
\begin{align}
    I_{\rm GHY} = \frac{1}{\kappa}\int_{\partial \cal M} \d^2x \sqrt{-h}\left(1 - \frac{\kappa}{8}\phi^2\right)K~,
\end{align}
in order to have a well-defined variational principle for the bulk metric $g_{\mu\nu}$~. 
In order to get a renormalized action, let us first consider the case with no acceleration, i.e., $\cA \to 0$~. Following the holographic renormalization procedure \cite{Skenderis:2002wp}, we found that 
\begin{align}\label{IrenHacc}
    I_{\rm ren} = I_{\rm EC} + I_{\rm GHY} + I_{\rm ct} + I_{\rm log}~,
\end{align}
where
\begin{align}\label{Ict}
    I_{\rm ct} = \frac{1}{\kappa}\int_{\partial \cal M}\d^2x\sqrt{-h}\left[\frac{1}{\ell} - \frac{\kappa}{4}\left(\frac{\phi^2}{2\ell} - \phi n^\mu\partial_\mu\phi\right)\right]~,\qquad I_{\rm log} = \frac{\log z}{24\pi}\int_{\partial \cal M}\d^2x\sqrt{-h}{\cal R}~,
\end{align}
with ${\cal R}$ the scalar curvature of the boundary metric $g_{(0)}$, is on-shell finite. 
The first term in $I_{\rm ct}$ corresponds to the counterterm associated with pure three-dimensional gravity \cite{Balasubramanian:1999re} while the remaining terms are associated with divergences of the scalar field. The last counterterm $I_{\rm log}$ appears to cancel logarithmic divergences \cite{Emparan:1999pm} associated with $\tilde g_{(2)}$ in \eqref{FG}~.

The holographic stress tensor associated with \eqref{IrenHacc} reads
\begin{align}\label{TijHacc}
    \langle T_{ij}\rangle = \lim_{z\to 0}T_{ij}[h] = -\lim_{z\to0}\left(1-\frac{\kappa}{8}\phi^2\right)\left(K_{ij} - K h_{ij} + \frac{1}{\ell}h_{ij}\right)~,
\end{align}
which on-shell
\begin{align}
    \langle T^i{}_j\rangle = \frac{3B^2}{2\kappa\ell}(1+\nu){\rm diag}(-1,1)~,
\end{align}
which shows no conformal anomaly as the boundary metric is flat, i.e.,
\begin{align}
    \ell^2g_{(0)ij} = \lim_{z\to0}\ell^2z^2 h_{ij} = {\rm diag}\left(-1,\ell^2\right)~.
\end{align}
The holographic mass can be now obtained using the holographic stress tensor using the Killing vector $\xi^t = \partial_t$ on the Noether charge
\begin{align}
    Q[\xi] = \int_\Gamma \d\Sigma_i\langle T^i_{}{j}\rangle\xi^j ~,
\end{align}
c v  for some space-like hypersurface $\Gamma$ and normal $d\Sigma_i$~. For the energy in our case we get
\begin{align}
    M = \int \d\theta \sqrt{-g_{(0)}}\langle T^t{}_t\rangle = \frac{3\pi}{\kappa\ell^2}B^2(1+\nu) = \frac{\pi m^2}{\kappa}~.
\end{align}

We found that the Euclideanized version of the action \eqref{IrenHacc} becomes the free energy of the thermodynamic system, i.e.,
\begin{align}
    I^{\rm E}_{\rm ren} = \beta M - S = \beta F~,
\end{align}
where 
\begin{align}
    S = \left(\frac{2\pi r_h}{B+r_h}\right)\frac{2\pi r_h}{\kappa}~,
\end{align}
is the black hole entropy in terms of the horizon radius $r_h$, in agreement with the Wald entropy formula \cite{Wald:1993nt}, and 
\begin{align}
    \beta = \frac{1}{T} = \frac{B^3 (1+\nu)+r_h^3}{2 \pi  r_h^2 \ell ^2}~,
\end{align}
is Euclidean time period that equals the inverse black hole temperature.

Moving forward, we now consider the accelerating hairy black hole. The solution now contains a domain wall splitting the spacetime, with a non-trivial energy density \eqref{mudw}. In \cite{Charmousis:2006pn, Aviles:2019xae} it is shown that in order to have a well posed variational principle for the string metric, and obtain the Israel equations as field equations, one need to consider a new term supported at the string, which in the case at hand reads\footnote{In principle, one needs to split the spacetime manifold from each side of the domain wall and consider a gravitational action for each manifold and the action \eqref{Idw} at the wall. Nonetheless, we use the fact that the spacetime split the spacetime symmetrically such that one can write everything in terms of a single manifold with some extra factors of 2 in the bulk quantities that we have already considered from the beginning.}
\begin{align}\label{Idw}
    I_{\rm dw} = -\frac{1}{\kappa}\int_\Sigma \d^2y \sqrt{\gamma}\left[\left(1-\frac{\kappa}{8}[\phi]^2\right)[{\cal K}] - \m\right]~.
\end{align}
This extra term not only gives a well-defined variational principle, but moreover cancels extra subleading divergences of ${\cal O}\left(\cA/z\right)$ that appear due to acceleration. Therefore, the action
\begin{align}\label{IrenHaccdw}
    I_{\rm ren}^{\rm E} = I_{\rm EC}^{\rm E} + I_{\rm GHY}^{\rm E} + I_{\rm ct}^{\rm E} + I_{\rm log}^{\rm E} + I_{\rm dw}^{\rm E}~,
\end{align}
is on-shell finite, and the associated holographic stress tensor, which is also given by \eqref{TijHacc} as the domain wall does no contribute to the quasi-local energy tensor, is finite and suffers from a Weyl anomaly, i.e.,
\begin{align}
\langle T^i{}_i\rangle = \frac{c}{24\pi}{\cal R}[g_{(0)}]~,\qquad c = \frac{12\pi\ell}{\kappa}~,   
\end{align}
where ${\cal R}[g_{(0)}]$ is the scalar curvature of the boundary metric and $c$ is the Brown--Henneaux central charge \cite{Brown:1986nw}.
The free energy and holographic energy can be computed explicitly and are given by a elliptic integrals in the $\theta$ coordinate. The quantities are rather lenghty and we shall not present them here.
This procedure is also well-defined in the hairless limit $s\to\infty$, where the action becomes the one of \cite{Balasubramanian:1999re} for pure three-dimensional AdS gravity with the extra wall contribution and the action recover the accelerating BTZ free energy. 
This extra terms shows that the acceleration modifies the Euclidean partition function already at the saddle point.  

As a final comment regarding to holographic aspects of the hairy accelerating black hole before delving into the holographic entanglement entropy, let us discuss boundary conditions and the holographic dictionary. 
AS aforementioned, the boundary conditions must be imposed on a particular coefficients in the asymptotic expansion of the bulk fields once the generating function becomes well-defined. The action \eqref{IrenHacc} is identified to the generating function of the dual CFT corresponding to Dirichlet boundary conditions, i.e., fixing the leading modes in the asymptotic expansion, which for the scalar field takes the form\footnote{If $\Delta=d/2$ one should include logarithmic modes in the expansion.}
\begin{align}
 \phi \sim z^\Delta_- (\phi_{(0)}+\dots) + z^\Delta_+ (\phi_{(1)} + \dots)~,
\end{align}
such that for the scalar field $\phi_{(1)}$ becomes the vacuum expectation value (vev) of the dual scalar operator with conformal dimension $\Delta_+$ sourced by $\phi_{(0)}$~. The Dirichlet 1-point functions and local Ward identities are obtained using the renormalized action and are given in terms of the coefficients in the FG expansion, but depend on the dimension and details of each theory. One could further consider Neumann boundary conditions which fixed the next-to-leading order in the asymptotic expansion, such that for the scalar field $\phi_{(0)}$ represents the 1-point function of the dual scalar operator with conformal dimension $\Delta_-$ sourced by $\phi_{(1)}$~. In order to have a well-defined variational principle, one needs to supplement the renormalized action with a term proportional to the normal derivative of the bulk field (see for instance \cite{Minces:1999eg}). Another option, is to consider mixed boundary conditions which corresponds to a deformation of the Neumann theory and fixes a linear combination between the source $\phi_{(1)}$ and a polynomial of $\phi_{(0)}$~. Then, the source of the dual operator with vev $\phi_{(0)}$ and conformal dimension $\Delta_-$~, is given by a linear combination dictated by the relation beteween the first coefficients in the expansion. The Neumann theory and its deformation is only suitable in the AdS/CFT correspondence if both modes are normalizable and a different scheme of quantization that allows having a conformal scalar operator of dimensions $\Delta_-$ becomes admissible. In order for both modes to be normalizable, the mass of the scalar field must lies in the window  \cite{Witten:2001ua, Papadimitriou:2007sj} 
\begin{align}\label{window}
    -\left(\frac{d}{2}\right)^2 < \tilde{m}^2\ell^2 < -\left(\frac{d}{2}\right)^2+1~,
\end{align}
where the lower bound is the usual BF bound that accounts for the stability of the scalar field, and the new upper bound ensures that the boundary theory admits a new quantization procedure, such that Neumann and mixed boundary conditions are admissible. 
Just as the Neumann boundary conditions modify the action principle by new boundary terms, the mixed boundary conditions also impose a new boundary term in the action given by the linear combination of the modes. The new term added is actually finite, i.e., does not contribute with new divergences to the bulk action, such that implies a deformation of the dual generating function. Indeed, the finite new term added is given by a polynomial of the $\phi_{(0)}$ such that the dual theory is multi-trace deformed. Such deformation can be marginal, i.e., do not break scale invariance, if the deformation is proportional to $\phi_{(0)}^{d/\Delta_-}$~, otherwise it will contribute with non-trivial terms to the trace of the stress tensor, producing a Weyl anomaly and drive the theory away from the conformal fixed point and lead to emergence's of new scales in the RG group. 

Returning to the harry accelerated BTZ, the asymptotic expansion of the scalar field   
\begin{align}
    \phi \sim z^{\Delta_-}(\phi_{(0)}(x^i) +\dots) + z^{\Delta_+}(\phi_{(1)}(x^i)+\dots)~,
\end{align}
where 
\begin{align}
    \phi_{(0)} = \sqrt{\frac{8}{\kappa}}(\cA \cosh(m\theta)-s)^{-\Delta-}~,\qquad \phi_{(1)} = -\sqrt{\frac{2}{\kappa}}(\cA \cosh(m\theta)-s)^{-\Delta_+}~,
\end{align}
and $\Delta_\pm = 1\pm\frac12$ are two solutions of 
\begin{align}
    \Delta(\Delta-2) = \tilde m^2 \ell^2~,
\end{align}
which appears as the first order in the asymptotic expansion of the scalar field equation. As aforementioned, the effective mass of the scalar field is $\tilde{m}^2\ell^2 = -3/4$~, which lies window \eqref{window}. Moreover, the scalar field satisfies mixed boundary conditions 
\begin{align}
    \phi_{(1)} = -\frac{\kappa}{16}\phi_{(0)}^3~,
\end{align}
such that, to have a well posed variational principle if we consider mixed boundary conditions we need to supplement the action \eqref{IrenHaccdw} with an extra boundary term \cite{Witten:2001ua, Papadimitriou:2007sj} of the form
\begin{align}
    I_{\rm mbc} = -\frac{\kappa}{64}\int_{\partial \cal M}\d^2x \sqrt{-g_{(0)}}\phi_{(0)}^4 = -\frac{\kappa}{64}\int_{\partial \cal M}\d^2x\sqrt{-h}\phi^4~,
\end{align}
which is, indeed, a marginal multi-trace deformation of the dual CFT.

%%%%%%%%%%%%%%%%%%%%%%%%%%%%%%%%%%%%%%%%%%%%%
%%%%%%%%%%%%%%%%%%%%%%%%%%%%%%%%%%%%%%%%%%%%%
%%%%%%%%%%%%%%%%%%%%%%%%%%%%%%%%%%%%%%%%%%%%%
%%%%%%%%%%%%%%%%%%%%%%%%%%%%%%%%%%%%%%%%%%%%%
\subsection{Entanglement entropy}

Let us now analyse the holographic entanglement entropy of the accelerating solutions. In order to do so, one needs to consider a more general prescription than the RT proposal, as now the theory has non-trivial couplings between the fields. In \cite{Dong:2013qoa}, a Wald-like formula has been proposed for the holographic entanglement entropy of a generic theory of gravity. The dual entanglement entropy now reads
\begin{align}
    S_{\rm E} = -2\pi\int \d^{d-1}y\sqrt{\gamma}\left\{ \frac{\partial L}{\partial R_{\mu\nu\rho\sigma}}\delta_{\mu\nu}\delta_{\rho\sigma} + {\rm higher~derivative~terms}\right\}~,
\end{align}
Here $\delta_{\mu\nu}$ is the Levi-Civita tensor in the two orthogonal direction to the RT minimal surface, and the second terms appear only for higher derivative gravity theories, which for \eqref{ECCs} vanish. Then, considering a constant time slice, the entanglement entropy becomes
\begin{align}
    S_{\rm E} = \frac{2\pi}{\kappa}\int \d\theta \sqrt{\gamma} \left(1-\frac{\kappa}{8}\phi^2\right)~.
\end{align}
Using $\theta$ as the Hamiltonian time of the minimization problem, as the scalar field now depends on the time coordinate as well, leads to a highly non-trivial equation as the functional has no conservation laws and we have not being able to obtain an analytical expression for the RT surface. Nonetheless, the issue can be avoided in the hairless case. 

The accelerating BTZ solution (the $s\to\infty$ limit of \eqref{Hacc} with $\sigma=-1$) becomes easier to handle as it can be mapped to Rindler-AdS by the coordinate transformation
\begin{align}
    \frac{\tilde{R}^2}{\ell^2} - 1 = \frac{f(r)}{\alpha^2m^2\Omega^2}~,\qquad \tilde{R}\sinh\Theta = \frac{r\sinh(m\theta)}{m\Omega}~,
\end{align}
becomes
\begin{align}
    ds^2 = -\left(\frac{\tilde{R}^2}{\ell^2} - 1\right)\d T^2 + \frac{\d \tilde{R}^2}{\frac{\tilde{R}^2}{\ell^2} - 1} + \tilde{R}^2\d\Theta^2~,
\end{align}
provided that
\begin{align}
    T = mt~, \qquad \alpha = \sqrt{1+m^2\ell^2\cA^2}~.
\end{align}
Then, we can find the RT surface in this coordinate system. Let us consider a slice bounded by a line of latitude $\theta_0$ and treat $\phi$ as the Hamiltonian time in the minimization problem. The minimal surface is then parametrized by
\begin{align}
    \tilde R_{\rm e} = \ell\left(1-\frac{\cosh^2\Theta}{\cosh^2\Theta_0}\right)^{-\frac12}~,
\end{align}
where $\Theta_0$ is fixed such that at the endpoints of the surfaces the radial coordinate goes to the conformal boundary $\tilde R\to\infty$~. Returning to the accelerating black hole, the condition is mapped to $r(\theta_0) = -(\cA \cosh(m\theta_0))^{-1}$ with $\theta_0 = m^{-1}\Theta_0$~, such that the surface is anchored to the conformal boundary. In this coordinates the surface becomes
\begin{align}
    r_{\rm e} = \frac{\alpha \cA m^2\ell^2\cosh(m\theta) + m\ell\cosh(m\theta_0)\sqrt{\sinh^2(m\theta_0) - \alpha^2\sinh^2(m\theta)}}{\alpha(\alpha^2\cosh^2(m\theta) - \cosh^2(m\theta_0))}~,
\end{align}
which is, indeed, a minimal surface of the original problem, and has an expansion for small acceleration
\begin{align}
    r_{\rm e}(\theta) = \frac{m\ell}{\sqrt{1-\frac{\cosh^2\Theta}{\cosh^2\Theta_0}}} - \frac{m^2\ell^2\cA\cosh(m\theta)}{\cosh^2(m\theta_0) - \cosh^2(m\theta)} + {\cal O}(\cA^2)~.
\end{align}
Although we have found a rather simpler analytical expression for minimal surface, to find the associated area becomes quite involved. The integration becomes divergent at $\theta = \theta_0$ so we introduce a cutoff $\delta$ and integrate from $\delta$ to $\theta - \delta$~. Expanding the integral around $\delta = 0$ and considering an expansion in a small acceleration we get that the entanglement entropy becomes
\begin{align}
    S_{\rm E} = \frac{c}{3}\log \left[\frac{\beta}{\pi\delta}\sinh\left(\frac{\pi L}{\beta}\right)\right] - 2\cA\ell\left(\frac{2\pi\ell^2}{\beta\delta}\sinh\left(\frac{\pi L}{\beta}\right)\right)^{\frac12}\tanh\left(\frac{\pi L}{2\beta}\right) - \frac{\cA^2\ell^4\pi}{\beta \delta}\sinh\left(\frac{\pi L}{\beta}\right)-\dots~,
\end{align}
where $L :=3\ell\theta_0$ can be related with the length of the entangling region. Notice that the temperature is independent of the acceleration so the zero-th order in the expansion correspond to the entanglement entropy of a thermal CFT dual to the BTZ black hole. However the next to leading order contain subleading divergences which decrease the amount of entangled pairs with the acceleration. This subleading divergences show how acceleration modifies the entanglement entropy of the dual CFT indicating some loss of information due to acceleration, and they could in principle correspond to a breaking of scaling symmetry. Nonetheless, we are not in position do such claim as we have use an expansion in the acceleration so the result can not be trust to all order. However, this is still a strong evidence that there is a modification of entanglement entropy of the dual theory due to the presence of the cosmic string responsible for the acceleration.

%%%%%%%%%%%%%%%%%%%%%%%%%%%%%%%%%%%%%%%%%%%%%%%%%%%%%%%%%%
%%%%%%%%%%%%%%%%%%%%%%%%%%%%%%%%%%%%%%%%%%%%%%%%%%%%%%%%%%
\newpage
\section{Conclusion and Future Directions}\label{Sec:Conc}
%%%%%%%%%%%%%%%%%%%%%%%%%%%%%%%%%%%%%%%%%%%%%%%%%%%%%%%%%%
%%%%%%%%%%%%%%%%%%%%%%%%%%%%%%%%%%%%%%%%%%%%%%%%%%%%%%%%%%

In this thesis, we explore three distinct scenarios wherein modifications of entanglement provide insights into the underlying Quantum Gravity theory. We start by first considering the geometric picture of \cite{Arias:2019pzy, Arias:2019zug} where a computation at leading order of a $1/G_{4}$ expansion of Renyi entropy of a Quantum Gravity theory on a dS$_4$ background is proposed. This computation involves extending the spacetime of a Rindler observer maximally and considering an orbifold generated by a $\mathbb{Z}_n$ action. This action induces a pair of Nambu--Goto terms localized at the fixed points of the orbifold, facilitating the use of the Replica trick. Our findings reveal a flat entanglement spectrum expressed in terms of the GH entropy, indicating entanglement between two disconnected Rindler observers. Consequently, the reduced density matrix corresponds to that of a single observer in dS$_4$. Leveraging two-dimensional CFT methods, we characterize the horizon states and introduce 1-loop corrections to the CFT partition function, thereby breaking the flat entanglement spectrum. We observe that the dimension of the Hilbert space of a Rindler observer on dS$_4$ corresponds to the exponential of the GH entropy, suggesting potential non-existence of the Quantum theory of dS gravity. Additionally, our setup facilitates the discovery of logarithmic corrections to the entropy of dS$_3$ spacetime under a specific limit of the orbifold parameter, extending the results of codimension-two dS holography proposed in \cite{Arias:2019zug}. The main results of that section depend on the applicability of Cardy entropy, which is still a debate as cosmological horizon symmetry algebra corresponds to a single copy of a Virasoro algebra. It would be interesting to explore if it possible to generate a second copy of the Virasoro algebra by relaxing the boundary conditions along the lines of \cite{Perez:2015jxn}. Furthermore, we would like to study the extension of the scheme to consider dS black hole solutions and matter fields. 

In the next section, we study corrections of entanglement entropy using the AdS/CFT correspondence. We extend the holographic proposal to compute charged Renyi entropy of \cite{Belin:2013uta} and couple gravity with a generic NLED Lagrangian density. We consider thermal fluctuations on the bulk path integral and find that the dual theory seems to suffer from a spontaneous symmetry breaking of a global symmetry. Moreover, the dual central charge is modified by the fluctuations indicating some RG flow appears due to bulk energy fluctuations. The spontaneous symmetry-breaking contribution to the entanglement entropy does not modify the UV structure, as it appears as a subleading logarithmic divergence, and therefore it cannot be seen holographically from the AdS classical saddle points \cite{Jeong:2022zea, Park:2022oek}. As a particular example, we use three-dimensional conformal electrodynamics coupled to AdS$_3$ gravity, finding that the black hole solution of \cite{Cataldo:2000we} can be used to study holographic charged free bosons in two dimensions. Notably, our findings extend to the uncharged scenario, demonstrating a universality in the results. However, while our analysis showcases this universality, it appears that symmetry breaking on the Conformal Field Theory (CFT) side does not correspond to a breaking of isometries on the string theory side within our results. To explore such discrepancies, one must consider the backreaction of energy fluctuations in the classical geometry. This involves solving a modified Einstein equation incorporating a quantum stress tensor that accounts for these fluctuations. For instance, prior works \cite{Solodukhin:1994yz, Frolov:1996hd} have demonstrated how effective field theory for Quantum Gravity can elucidate quantum gravitational corrections to Schwarzschild black hole geometries. Similarly, our thesis examines similar corrections, revealing a logarithmic adjustment to the associated entropy (see also \cite{Pourhassan:2022auo}). These new contributions to the geometry disrupt the isometry group of the solution\footnote{One can explore quantum fluctuations that adhere to the spherical symmetry of the solution \cite{Kazakov:1993ha, Konoplya:2019xmn}. However, despite these deformations of the geometry, there are no induced logarithmic terms in the entropy. Consequently, no symmetry breaking would manifest in the dual CFT in the AdS case.}. Similarly, this symmetry breaking should hold for AlAdS black holes. We leave a detailed study of this detail for future work. Another intriguing avenue involves defining the symmetry-resolved entanglement entropy (see \eqref{SRSn}) holographically. This endeavour would require considering the gravity theory with a fixed charge rather than a fixed chemical potential. Subsequently, one would need to consider an enhanced action principle \cite{Caldarelli:1999xj, Chamblin:1999tk}. We hope to come back to this interesting idea in the near future.  

Finally, we show a new modification of the entanglement of a holographic two-dimensional thermal CFT appearing due to the presence of a cosmic string in the AdS bulk solution which generates acceleration. Using an expansion in the acceleration parameter, we find that as the acceleration increases the accessible region of the conformal boundary decreases and also the entanglement entropy, indicating a loss of information in the dual theory due to acceleration. We have explored the hairless solution, which presents a simpler minimization problem and allows the identification of the minimal RT surface. However, the computation of its area becomes significantly intricate. To obtain an algebraic expression, we introduce a small acceleration expansion, revealing additional subleading divergences that may signify a breakdown of scaling symmetry. Since this expansion merely serves as a preliminary step, the results cannot be deemed final. An intriguing avenue for future investigation lies in employing the prescription proposed by \cite{Bakhmatov:2017ihw} to reduce the problem's dimensionality by investigating dual AdS geometries.

As a concluding remark, it's noteworthy that for the computation of holographic quantities of accelerating black holes, we adopt the framework of \cite{Hubeny:2009kz,Cassani:2021dwa}. This framework entails a coordinate transformation that effectively maps the holographic boundary surface to a parametrization based on a single coordinate. Nonetheless, such a map does not correspond to a Penrose--Brown--Henneaux (PBH) transformation \cite{Penrose:1985bww, Brown:1986nw}  and the resulting line element deviates from the FG gauge. Despite the solution being AlAdS, it is not clear what is the meaning of the transformation from the CFT perspective. For instance, is possible to induce a boundary Weyl transformation by relaxing the FG gauge to the Weyl-FG (WFG) gauge \cite{Ciambelli:2019bzz}
\begin{align}
    ds^2 = \left(\frac{\d z^2}{z^2} - k_i(z,x)\d x^i\right)^2 + h_{ij}(z,x)\d x^i \d x^j~,
\end{align}
where the quantities $k_i$ and $h_{ij}$ can be asymptotically radially expanded as
\begin{align}
    h_{ij}(z,x) = \frac{1}{z^2}\sum_{p \geq 0} z^{2p}h_{ij}^{(2p)}(x)~,\qquad k_i(z,x) = \sum_{p\geq 0}z^{2p}k_i^{(2p)}(x)~,
\end{align}
where $h_{ij}^{(0)}$ correspond to the boundary metric and $k_{i}^{(0)}$ is a boundary Weyl connection. This new ingredient modifies non-trivially the boundary theory \cite{Ciambelli:2019bzz,Jia:2021hgy, Ciambelli:2023ott} as it contains a new gauge current which modifies the boundary Weyl anomaly and the diffeomorphism mapping the WFG to the FG is not always non-trivially charged. 
The form of the WFG gauge resembles the form of the line element of an accelerating black hole when the new transformation of \cite{Hubeny:2009kz, Cassani:2021dwa} is considered. Nonetheless, is easy to check that for accelerating black holes, the quantity $k_i$ has a subleading divergence in contrast to the one of the WFG gauge, viz
\begin{align}
    k_i^{(0)} = \sum_{p\geq -1}z^{p}k_i^{(0)}(x)~.
\end{align}
The emergence of this new decay imposes the necessity of exploring more generalized boundary conditions beyond the WFG and FG gauges. This kind of geometries has been studied in detail for three-dimensional AdS gravity in \cite{Grumiller:2016pqb} generating non-trivial canonical boundary charged and the asymptotic symmetry algebra contains two extra affine $\mathfrak{sl}(2)$ current algebras (see \cite{Avery:2013dja, Troessaert:2013fma,Ojeda:2019xih, Ojeda:2020bgz,Cardenas:2021vwo, Lara:2024cie} for holographic studies of general boundary conditions in AdS$_3$~). It suggests that a more appropriate holographic description of accelerating black holes may be achievable within the framework of a generalized FG gauge. Such an approach would complement \cite{Grumiller:2016pqb} by extending them to higher dimensions and computing obstruction tensors arising from these generalized boundary conditions and finding the new renormalized Euclidean on-shell action \cite{Arenas-Henriquez:2024ypo}.

\vspace*{\fill}
\begin{center}
\pgfornament[width=1cm]{9}
\end{center}
\newpage
%%%%BIBLIOGRAPHY%%%%%%%%%%%%%
\bibliographystyle{JHEP}
\bibliography{bibliography}
%%%%%%%%%%%%%%%%%%%%%%%%%

\end{document}